\documentclass[prl, twocolumn]{revtex4-2}
\usepackage{amssymb}
\usepackage{amsmath}
\usepackage{epsfig}
\usepackage{bm}

\begin{document}

\title{  Highly correlated two-dimensional viscous electron fluid    in moderate magnetic fields }
\author{  P. S. Alekseev and A. P. Alekseeva
}
\affiliation{
 Ioffe  Institute,  194021  St.~Petersburg, Russia
}

\begin{abstract}

Magnetotransport phenomena often provide critically important information about two-dimensional (2D) electron  systems. For example, the independence of magneto-photo-resistance of 2D electrons in best-quality  quantum wells on the polarization helicity of incident
radio-frequency radiation have been treated as a puzzling effect, which is important for characterization of these systems, but had
no well-established explanation up to now. Here we develop a phenomenological model  of dynamics of a highly correlated  2D electron fluid
 in moderate magnetic fields, in which shear viscosity and the memory effects in inter-particle interaction are crucial.
 In this system, successive collisions of electrons joined in pairs (that is, the pair correlations in time) turn out  to be as important
 as uncorrelated collisions  of statistically independent electrons.  The resulting photoresistance exhibits
an irregular shape  of magnetooscillations,  the absence of the dependence on the helicity of the circular polarization of radiation,
 and a giant peak near the doubled cyclotron   frequency. All these effects were observed  in experiments on best-quality GaAs quantum wells
in moderate magnetic fields at low temperatures.   Although the most general conditions of applicability of the developed phenomenological model
  is not fully clarified at now,  this coincidence can point out that  2D electrons in such systems form
 the  highly correlated viscous fluid.
\end{abstract}

\maketitle

\section{1. Introduction}
Microwave-induced  resistance oscillations (MIRO) of 2D electrons in moderate magnetic  fields were initially
observed~\cite{exp_1,exp_2,exp_3} and then extensively studied~\cite{rev_M} in  GaAs/AlGaAs quantum wells and than in other
2D nanostructures. This is an intriguing phenomena,  whose correct explanation is, apparently, crucial for understanding of
physics of  2D electrons. Originally, MIRO were attributed to  transitions of non-interacting  2D electrons at high Landau levels
in external dc and ac electric fields~\cite{theor_1,theor_1_1,theor_2,joint}. Such ``displacement mechanism''~\cite{Ryzhii,Ryzhii_Suris}
is based on taking into account the radiation-assisted scattering of 2D electrons in quantized states on disorder,  that result
in  unequal probabilities of electron transitions in the opposite directions at a dc field. Theories~\cite{theor_1,theor_1_1} yield
the correct profile of the magnetooscillations of photoresistance in not too  clean samples.   In Ref.~\cite{theor_2} the ``inelastic mechanism''
 was proposed in which the crucial role is played by radiation-induced redistribution   of electrons by energy. Its contribution
  to photoresistance explains  the temperature dependence of MIRO.  In Ref.~\cite{Polyakov_et_al_classical_mem} it was shown that
 the classical memory effects in scattering of   2D electrons  on disorder also can lead to  MIRO. In Ref.~\cite{Beltukov_Dyakonov} was
developed a different powerful  approach to theory of MIRO, which is based on a phenomenological description  of memory effects
 in classical dynamics of non-interacting 2D electrons, scattered on disk defects.   This approach describes
the essence and properties of MIRO in a lucid transparent manner.

Theories~\cite{theor_1,theor_1_1,theor_2,joint,Polyakov_et_al_classical_mem,Beltukov_Dyakonov} have a substantial problem  which is
the inconsistency between an almost full lack of dependence of MIRO on the sign of the circular polarization of radiation
in some experiments~\cite{Smet_1,Ganichev_1} and the need to involve special factors of experimental setup~\cite{rev_M,Savchenko} or
very special models of electron dynamics~\cite{Mikhailov,Chepelianskii_Shepelyansky} to explain this lack in theory.  Moreover,
in best-quality GaAs quantum well and graphene samples were observed peculiar  features in transport effects, in particular,
a non-sinusoidal profile of  MIRO and similar oscillating  photo-magneto-conductivity as well as a giant peak near the doubled cyclotron
frequency in photoresistance~\cite{Smet_1,Ganichev_1,exp_GaAs_ac_1,exp_GaAs_ac_2,exp_GaAs_ac_3,recentest_,peak_gr,peak_gr2}.
All these effects  has been remained unexplained within disorder-based models up to now.

In parallel with the studies of MIRO,  the giant negative magnetoresistance was observed  in the same best-mobility GaAs quantum
wells (see, for example, Refs.~\cite{exps_neg_1,exp_GaAs_ac_1,exps_neg_2,exps_neg_3,exps_neg_4,Gusev_1,recentest_}).
In Ref.~\cite{je_visc}  it was explained as the result  of   formation of a viscous  fluid from 2D electrons in
the quantum wells~\cite{Gurzhi_Shevchenko}. Later, a very similar huge negative magnetoresistance was detected in  other
 high-quality conductors: 2D metal PdCoO$_2$~\cite{Moll}, the  3D Weyl semimetal WP$_2$~\cite{Gooth}, and single-layer
 graphene~\cite{graphene_4,graphene_5}, for which other bright evidences of the hydrodynamic electron transport   were also
 discovered~\cite{Levitov_et_al,graphene_1,graphene_3,graphene_4,graphene_5,Polini_Geim,Kr}. In Ref.~\cite{vis_res_1}
it was pointed out that the high-frequency viscosity coefficients of 2D electrons exhibit a single resonance at the twice  cyclotron
frequency,  being a characteristic sign of electron hydrodynamics. In the case of a strongly non-ideal fluid, such ``viscoelastic''
resonance manifests itself  via the excitation  of the  transverse magnetosonic waves~\cite{vis_res_2,Semiconductors}.  Later,
 many other hydrodynamic effects were observed in high-quality GaAs quantum wells and graphene samples.  For example, in Ref.~\cite{Kr}
 it was demonstrated that in graphene samples with constrictions  THz radiation induces a strong heating of electrons leading to the
hydrodynamic regime.

In view of these results,  a question arises: whether  the giant peak   and the polarization-sign-independent  magnetooscillations
in photoresistance of high-purity 2D electron systems are explained within a model of a viscous  electron fluid?

In this work  we propose a phenomenological model of a highly correlated viscous 2D electron fluid in defectless samples  in moderate
magnetic fields. We start from the Landau Fermi-liquid model with the strong Landau interparticle interaction.  The two key points
of our model  are:  the viscoelasticity effect leading to excitation of the shear-stress waves   and  the magnetic-field-induced memory effects
in the  inter-particle scattering.   Based on the approaches and ideas of Refs.~\cite{Beltukov_Dyakonov,vis_res_2,Semiconductors,el,el2,el3,AAGS},
we formulate the viscoelastic motion equations of the fluid  containing  the  memory terms. These equations accounts for multiple
repeated collisions of the electrons joined in pairs [see Fig.~1($a$,$b$)], 
 that is, for the strong pair correlations in time.    The strong interparticle
interaction  and presence of moderate  magnetic fields  are the sufficient conditions of applicability of formulated model.
Resulting  photoconductivity exhibits magnetooscillations of an irregular shape, a large peak at the doubled cyclotron frequency,
and is independent of the helicity  of polarization of radiation. All these properties qualitatively  coincide with the ones observed
in experiments~\cite{Smet_1,Ganichev_1,exp_GaAs_ac_1,exp_GaAs_ac_2,exp_GaAs_ac_3,recentest_} on magnetotransport in ultra-high-quality
 GaAs  quantum wells and graphene~\cite{r}.

Apparently, the strength of interparticle interaction is not too large  in the structures examined
 in works~\cite{Smet_1,Ganichev_1,exp_GaAs_ac_1,exp_GaAs_ac_2,exp_GaAs_ac_3,recentest_},
      and the above sufficient condition of applicability
  of our model  (the strong Landau interaction) is not fulfilled in them.
   However, we discuss that  there can be more general applicability    conditions;
    they  can depend on particular electron flow
parameters and be much wider.
  The demonstrated good qualitative agreement of  our theory and the experiments can be
a promising  evidence
  of  formation of a highly correlated 2D electron fluid in such  systems and  of suitability of our model for  their description.

  \begin{figure}[t!]
\centerline{\includegraphics[width=1.0 \linewidth]{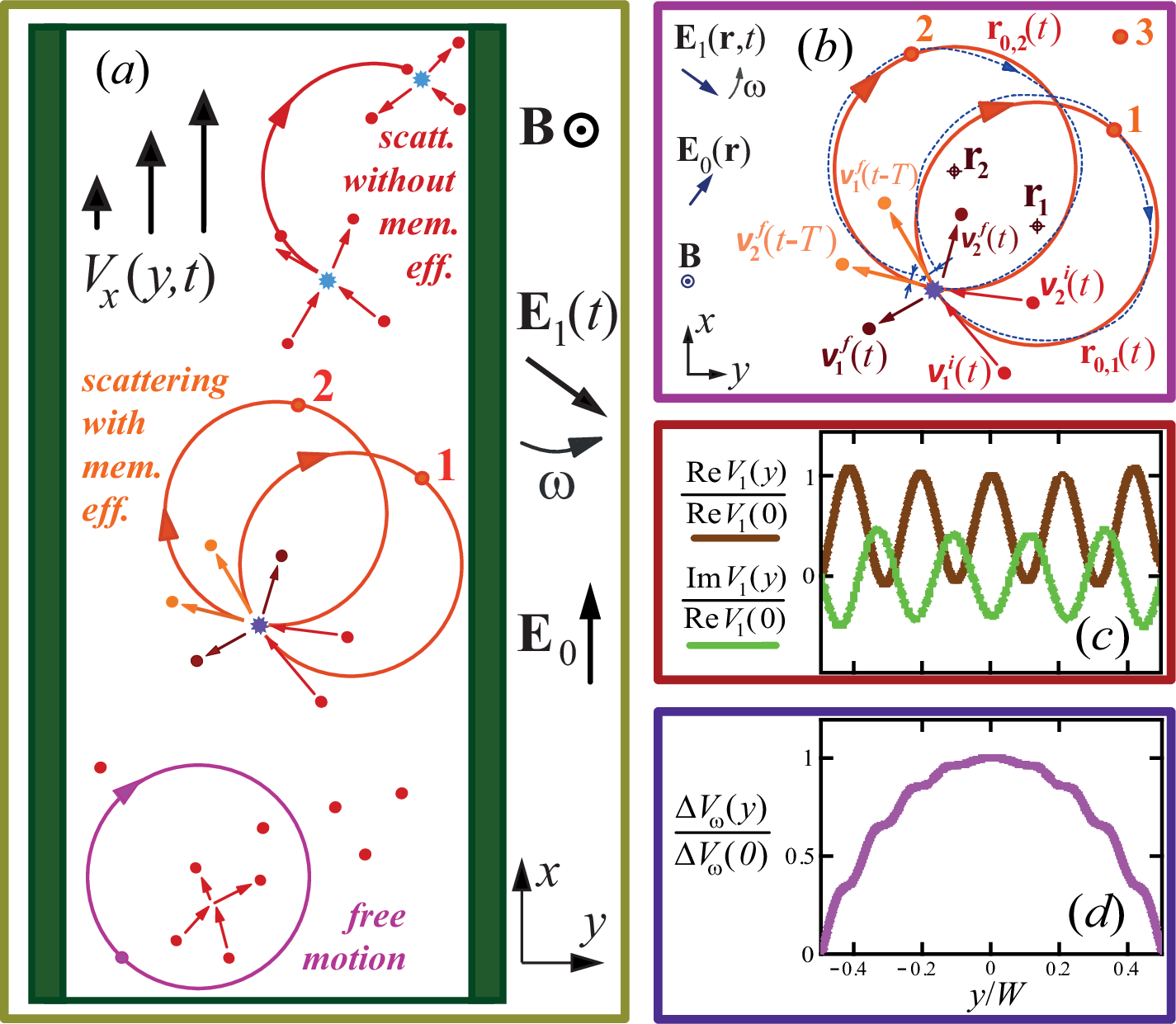}}
\caption{
        ($a$): Flow of a highly correlated 2D electron fluid in a long defectless sample with rough edges in a magnetic field~$\mathbf{B}$,
a dc electric field~$\mathbf{E}_0$,  and a circularly  polarized  radiation field~$\mathbf{E}_1(t)$. Trajectories of two Fermi
liquid electron-like quasiparticles  in an extended collision, leading to the memory effect, as well as trajectories of a collisionless
motion and of two successive   collisions without the memory  effect are drawn.
      ($b$):~Sketch of the microscopic mechanism of the memory effect in collisions of electron-like quasiparticles in a 2D  electron
Fermi-liquid in a classical magnetic field $\mathbf{B} = B \mathbf{e}_z$.  Red solid lines are the quasiparticle trajectories
 at  $\mathbf{E}_0, \mathbf{E}_1 , \mathbf{V }\equiv 0 $, which are  exact cyclotron circles. Blue dashed lines are the trajectories
 accounting the elastic force,  related to  space-dependent perturbations of   the quasiparticle energy spectrum,  as well as to
 the forces from  the full (sums of external and internal) dc and the ac electric fields.
       ($c$):~Ac component $V_1 (y) \sim E_1 $ of the  fluid velocity  in the flow in a long sample shown in panel~(a).
Shear stress waves are seen.
  ($d$):~Profile of the radiation-induced correction (relaxational and elastic contributions)
 $ V_\omega  (y) \propto  V_\omega  ^{rel,\,el} (y)  \sim E_0 E_1^2 $
 to the dc flow component $ V_0 (y) \sim E_0  $ in the same sample.
 All curves  curves in panels ($c$),$(d)$ are
plotted for the parameters:   $\omega \tau_2 = 40$, $\tau_q / \tau_2 = 0.7$,   $W/L_s =1$, and  $\omega_c / \omega = 0.34 $.
  ($\tau_q$ is the interparticle departure time and  $L_s  $ is the decay length of shear stress waves).
 }
\end{figure}

\section{ 2. Model of correlated electron fluid  in moderate magnetic fields }
We develop a phenomenological  hydrodynamic theory of a highly-correlated viscous 2D electron fluid
 in classical magnetic fields  starting from the Landau
Fermi-liquid model. Our theory is surely applicable for 2D electrons with low densities~$n_0$ and corresponding interparticle interaction
parameter $r_s \gg 1$ those form dc and ac flows in sufficiently wide (not ballistic) samples.
 Such systems were  realized  in inversion layers on Si crystals, ZnO-based structures, low-density
GaAs quantum wells, where  metal-insulator transition, renormalization of the quasiparticle effective mass,  and other correlation effects
were observed~\cite{Shashkin0,Shashkin,222}.

    As for the systems  with moderate interaction, $r_s \lesssim  1$,  the widest conditions of applicability of our
    phenomenological model  are not fully clarified at now,
  but can include also the systems with  $r_s \lesssim 1$  for particular flow parameters~\cite{f,KFS,KFS2}.
 Indeed, for example,   electron hydrodynamics  for stationary flows
   in moderate magnetic fields is realized in bulk samples at any inter-particle scattering rates~\cite{a,Holder}.
          Note also that if there are no sources of small-scale disturbances in the system,
          then kinetic  and hydrodynamic approaches
          often give qualitatively similar results for the flow distributions
           with a minimum spacescale equal to the microscopic length~\cite{c}.

At high frequencies, $\omega   \gg 1 / \tau_2 $, and long wavelengths,  $\Delta x  \gg v_F/\omega $,
 a flow of a strongly non-ideal viscous 2D electron fluid in magnetic field
 consists of the two components [here $\tau_2$ is the shear stress relaxation time of electron-like
 quasiparticles;
   $ \Delta x $~is the characteristic spacescale  of inhomogeneities, $v_F$ is the Fermi velocity].
The first one is formed by of  magnetoplasmon modes related to perturbations of the electron  density, $n(\mathbf{r},t)$, and
 the hydrodynamic velocity $\mathbf{V} (\mathbf{r},t) $.  The second one is formed by   the transversal   magnetosonic waves related to
perturbations of the shear stress $\sigma_{ij} (\mathbf{r},t)$ and the velocity $\mathbf{V} (\mathbf{r},t) $~\cite{vis_res_2}.
Such  waves can be excited in Fermi-liquids  at large dissipativeless inter-particle  interaction,
described  by the Landau interaction term~[57,58,43,41,42,59,46].
The inter-particle scattering leads to  weak relaxation of both the components.

The Navier-Stokes-like non-stationary equations of the 2D electron Fermi-liquid at a strong interparticle interaction
 were derived in Refs.~\cite{el,el2,el3} for zero magnetic field and  in Ref.~\cite{Semiconductors} for a nonzero
classical magnetic fields.  The sufficient criteria of applicability of those viscoelastic equations  can be formulated
as follows~\cite{vis_res_2,AAGS,Semiconductors}:
\begin{equation}
 \label{crit}
 \begin{array}{c}
 \displaystyle
 F_{1} \gg 1 \: ,  \;
\, \;\;   \; F_m \ll F_1  \;\; \mathrm{at   } \;\; m \geq 3
 \, ;
\\ \displaystyle
R_c \ll \Delta x
\quad  \mathrm{or}  \quad
v_F/ \omega \ll \Delta x
  \, ,
  \end{array}
\end{equation}
 where $F_m $ are the Landau interaction  parameters, $F_2$ can be arbitrary,
 $R_c = v_F/ \omega_c $ is the cyclotron radius
of electron-like quasiparticles,     and $\omega_c  = eB/(mc)$ is the cyclotron frequency. These criteria
do not imply, generally speaking, a thermodynamical equilibrium   of quasiparticles   in a frame moving
with the flow velocity $\mathbf{V}(\mathbf{r},t)$ (see discussion  in Refs.~\cite{el2,el3,vis_res_2}).

Memory  effects in an electron fluid in magnetic field,
 from the classical-mechanical point of view,  are  induced by  subsequent collisions  of  some electrons
joined in pairs due to the cyclotron rotation [see Fig.~1(a,b)].
The probability for electron to pass a cyclotron circle without collisions with other electrons within this picture is:
\begin{equation}
\label{P0}
   P (B) =e^{-T/\tau_q} \,,
\end{equation}
where  $\tau_q$ is the interparticle  scattering departure  time   and  $ T=2\pi/\omega_c $ is the cyclotron period.
The value $P$ increases up from zero to unity with the increase of magnetic field.
  Herewith we do not study the regimes of too high magnetic fields when
 quantization of the density of states can become substantial
and   high-order memory effects, related to
  many recollisions, $N_r \gg 1$, within an extended collision, appears.
 Apparently, these simplifications takes place when $ \omega_c \lesssim 2\pi/\tau_q $.
   In Fermi systems, the time $\tau_q$ is  much shorter than the shear stress relaxation
 time  $\tau_2$~\cite{Alekseev_Dmitriev,Novikov}.
So we consider the diapason of magnetic fields: $    1/ \tau_2 \ll \omega_c  \lesssim 2\pi/\tau_q $,
 when the electron dynamics remain classical
and~$N_r \sim 1$.

Due to extended collisions, the dynamic of the fluid in the current moment~$t$ is determined by the statistical
 properties of the system in the moments~$t$ and
\begin{equation}
\label{t}
   t'=t-N_r T\
 ,  \quad
N_r=1,2,3, ...
\end{equation}
of previous collisions of paired particles, which are being  scattered in the moment  $t$
 \{see Fig.~1(a,b) and details
in Supplemental material~(SM)~\cite{SI}\}.
  As a result, paired electrons leads to the appearance of some retarded  non-linear terms in the fluid dynamic equations of
 Refs.~\cite{je_visc,vis_res_2,Semiconductors}.  Such ``memory terms'' can be of the different types.  The terms of one type are
the retarded relaxation terms due to the interparticle scattering,  being partly similar to the retarded relaxation terms due to
the  extended collisions of non-interacting electrons with  localized defects~\cite{Beltukov_Dyakonov}. The contributions of another type
are related to  retarded corrections of  the Landau parameters $F_m$, reflecting non-local in time perturbation
of the elastic part  of the interparticle interaction.


\section{ 3. Retarded non-linear equations\\ of fluid dynamics }
  We formulate the following motion equations for the particle density~$n$,
the hydrodynamic velocity  $\mathbf{V}$,  and the inequilibrium momentum flux~$ \hat{\Pi} = -\hat{ \sigma } $,
accounting the described above viscoelastic and memory effects:
  \begin{equation}
\label{main_eq_gen}
\begin{array}{l}
\displaystyle
 \partial n / \partial t     \,   +  \, n_0 \, \mathrm{div }\mathbf{V} \,= \,  0
  \,,
\\
\\
\displaystyle
     \partial V_i   / \partial t   \, = \,
          e   E_i /m +    \omega_c  \epsilon _{ikz} V_k - (1/m) \, \partial \Pi_{ij}  /\partial x_j
 \,,
\\
\\
\displaystyle
     (1 - \widetilde{\delta F } _{2,\,ij} )   \, \partial \Pi_{ij} / \partial t
        \, = \,  2 \omega_c  \epsilon _{ikz} \Pi_{kj}   \, - \, \Pi_{ij} / \tau_{2} \, -
\\
\\
\displaystyle
\; \;  \;\; \;  \;\; \; \; \;  \;\;   \;\;  \; \;  \;\;   -  \, m(v_F^{\eta})^2 \,  V_{ij}/4 -
         \Gamma _{ijkl} \,  \Pi_{kl} (\mathbf{r},t-T) \, .
\end{array}
\end{equation}
These equations are an extension of  (i)~the Navier-Stokes-like equations
of highly viscous electron fluid~\cite{vis_res_2,Semiconductors} and
 (ii)~the retarded relaxation equations for scattering of independent electrons on disk defects~\cite{Beltukov_Dyakonov}.
In Eqs.~(\ref{main_eq_gen}), the electric field  $\mathbf{E} = \mathbf{E}(\mathbf{r},t) $
consists of the applied field and the internal field induced by a non-equilibrium charge density~$ e \,\delta n (\mathbf{r},t)$:
$ \mathbf{E} (\mathbf{r},t) = \mathbf{E} ^{ext} (t) + \mathbf{E} ^{int} (\mathbf{r}, t)$; $e$ and $m$ are the electron charge
and the quasiparticle effective  mass; $\epsilon _{ikm}$~is the antisymmetric unit tensor; $V_{ij} = \partial V_i/\partial x_j
+ \partial V_j /\partial x_i  $; $\tau_{2}$~is  the shear stress relaxation time without the memory effects; $v^{\eta}_F $ is
the parameter defining the magnitude of the viscosity $\eta$ of a highly viscous fluid at zero magnetic field~\cite{vis_res_2,Semiconductors}:
$ \eta_0 = ( v_F^\eta ) ^2  \tau_2  / 4 $; and the values $\widetilde{\delta F} _{2,\,ij}$~are proportional to the retarded
perturbations of the Landau parameter~$F_2$.   The values   $m$, $\tau_2$,  $ v_F^\eta$,
and $\omega_c$ (in the equation for $\partial \hat{\Pi }/ \partial  t $)  are expressed via the Landau
 parameters $F_0$, $F_1$, and $F_2$~\cite{Semiconductors}.

It is noteworthy that in Eqs.~(\ref{main_eq_gen}) the parameter $v_F^{\eta} $ in a strongly non-ideal Fermi liquid
 is renormalized by the electron-electron interaction and becomes much greater than its value in a Fermi gas,
  the Fermi velocity $v_F$~\cite{vis_res_2,Semiconductors}.      This is an example of the
  effects
   of the Fermi-liquid renormalizations   of the values characterizing quaiparticles as compared with the values
   for  the ``bare electrons'' (or other ``bare particles''; see, for example book~\cite{LP}).

The term $ - \Gamma_{ijkl}(\mathbf{r},t ) \Pi_{kl}(\mathbf{r},t-T)$ in Eq.~(\ref{main_eq_gen}) describes the retarded relaxation
 of $\Pi_{ij}$ due to the  extended collisions of pairs of quasiparticles with one return to the same point [see Fig.~1($b$)].
Such extended collisions are sensitive to the macroscopic motion of the fluid via the forces acting on quasiparticles from the internal
field~$ \mathbf{E} ^{int}( \mathbf{r} , t ) $  and  from the elastic tension
 $ \hat{ \sigma} (\mathbf{r},t) =  - \hat{\Pi} (\mathbf{r},t) $.
 The latter one is induced  by perturbations of the quasiparticle energy spectrum  by nonzero $\mathbf{V}$, which is expressed via
the term $- m(v_F^\eta) ^2 V_{ij} /4 $~\cite{el,Semiconductors}. As a result, the tensor $ \hat{\Gamma}  (\mathbf{r},t )  $ depends on
the fluid variables in the moments $t'  \leq t $.

At rare interparticle collisions, when $\omega _c \tau_2 \gg 1 $, the tensor  $ \Gamma _{ijkl} (\mathbf{r},t )  $  depends mainly
on the shift
\begin{equation}
 \Delta_{mn} (\mathbf{r},t) = \varepsilon _{mn} (\mathbf{r},t)
-  \varepsilon _{mn} (\mathbf{r}, t-T)
\end{equation}
of the  strain tensor, $\varepsilon _{mn}  = \partial u_m/\partial x_n + \partial u_n /\partial x_m  $, in
one cyclotron period \{here $\mathbf{u}(\mathbf{r},t)$ is the displacement of fluid elements; see Fig.~1(b)
 and qualitative justification of this property of $\Gamma _{ijkl} $ in~\cite{SI}\}.
Far from the resonance frequency $\omega= 2 \omega_c$, the fluid dynamics is almost elastic  (combination of magnetosonic waves)
 and the value $\mathbf{u}(\mathbf{r},t)$ is the proper value fully describing states of the system (as it was described in book~\cite{LL7}
  for conventional elastic media).
 For the frequencies near
the resonance,   the value $\mathbf{u}(\mathbf{r},t)$ characterizes the inelastic displacement of fluid elements.
 In both this cases     the ``mismatch'' of the strain tensor is expressed via the velocity gradient~$V_{mn} (\mathbf{r},t')$:
\begin{equation}
\label{Delta_ij}
 \Delta _{mn} (\mathbf{r},t) =
    \int _{t-T} ^t dt' \; V_{mn} (\mathbf{r}, t'  \, )
 \:.
\end{equation}
For a small-amplitude flow,  the retarded relaxation tensor $ \hat{\Gamma } ( \hat{\Delta} ) $ is expanded  into power
series by~$ \hat{\Delta} = \hat{\Delta} (\mathbf{r},t) $:
\begin{equation}
\label{Gamma_ex}
\begin{array}{c}
\displaystyle
   \Gamma _{ijkl}    [ \mathbf{u}(\mathbf{r}',t')]
    = \Gamma ^{(0)}_{ijkl}
        + \alpha _{ijkl}^{ mnop} \, \Delta _{mn} \, \Delta_{op}
     \:,
\end{array}
\end{equation}
where  the coefficients $\Gamma ^{(0)}_{ijkl} $ and $\alpha _{ijkl}^{ mnop}$ are proportional to the probability
 $P(B)$~(\ref{P0}) for a particle to make a cyclotron rotation without collisions.

When the inter-particle interaction is sufficiently strong [$F_1 \gg 1$, see Eq.~(\ref{crit})] and
all memory and nonlinear effects are neglected,
 the kinetic equation for quasiparticles leads
 to the viscoelastic equations containing the Landau parameters~$F_{0,1,2}$~\cite{el,Semiconductors}.
They are equations~(\ref{main_eq_gen})
 with $\widetilde{\delta F} _{2,ij}  \equiv 0 $ and $\Gamma _{ijkl} \equiv 0$.
 The perturbation of the magnitude of the parameter~$ F_2 $ on  the values~$ \delta F_{2,ij}  $
 due to the fluid motion is a nonlinear effect, leading to
  the factor  $(1 - \widetilde{\delta F} _{2,ij} ) $,
   $\widetilde{\delta F} _{2,ij}= \delta F_{2,ij} /(1+F_2) $, in the third of  Eqs.~(\ref{main_eq_gen}).
  The correction $\widetilde{\delta F} _{2,ij}$ can  depend
on the fluid dynamic  in the present and in the past, being some bilinear
by $ \mathbf{V} (\mathbf{r},t') $ and  $ \hat{\Pi} (\mathbf{r},t') $
expression.  The physical origin of these effects is a perturbation  of  the interaction energy in an inhomogeneous flow
 at the moment~$t$
due to
 the correlations in the  distances between electrons at the moments~$t'<t$
  (see details in~\cite{SI}). Owing to a small probability
 of interparticle scattering, the values $\delta F_{2,ij}$ depend   mainly
 on the variables $\mathbf{V}$ and $\hat{\Pi}$
 in the current moment $t$  and in the retarded moments $t'=t-N_r T$~[see Eq.~(\ref{t})].
 Indeed, in those moments $t'$, the two quasiparticles, which travel
by cyclotron trajectories and are scattered at the current moment $t$,
 were located near their current  positions at $ t $
with a large probability [see Fig.~1($b$)].    We again account only one cyclotron
rotation:~$ \widetilde{\delta F} _ {2,ij}  (\mathbf{r},t) = \widetilde{\delta F}_ {2,ij} ^{(1)}
[ \, \mathbf{V} (\mathbf{r},t-T) \,  ] $.

The magnitude of the retarded correction  $\widetilde{ \delta F }_{2,ij} ^{(1)}$ to the second order Landau parameter   is considered
to be proportional  to the power of the energy absorbed  near the moment  $t'=t-T$  by the flow from the external field,
 $ \mathcal{W} (\mathbf{r},t) = (V_{kl} \Pi_{kl})|_{\mathbf{r},\,t'}$, as well as to the all other linear combinations
 of the tensors $\hat{V}$  and  $\hat{\Pi}$:
\begin{equation}
 \label{Gamma_ex2}
 \begin{array}{c}
  \displaystyle
     \widetilde{\delta F} _{2,ij}^{(1)} [\mathbf{V}(\mathbf{r}',t')]
          = \beta_{ij}^{klps}
          V_{kl} (\mathbf{r},t-T)  \, \Pi_{ps} (\mathbf{r},t-T)
   \,  ,
\end{array}
\end{equation}
where $ \beta_{ij}^{klmn} $ are the coefficients, which are proportional to the probability  $P(B)$~(\ref{P0}),  like as the
 coefficients $\Gamma_{ijkl}$~\cite{SI}. We note that Eq.~(\ref{Gamma_ex2}) at fixes ac frequency $\omega $ leads to the elastic terms
 in equations~(\ref{main_eq_gen})
   playing the role of
 the non-local and non-linear  by $V_i$ and $\Pi_{ij}$ corrections
to the cyclotron terms~$ 2 \omega_c  \epsilon _{ikz} \Pi_{kj} $.

The necessary  condition of applicability of the above memory terms has the form~\cite{SI}:
$V_{char} T R_c / \Delta x \, \ll \, a_B$,
where $V_{char} $ is the characteristic flow amplitude and $\Delta x$ is the characteristic  spacescale of its inhomogeneity.
This is the condition on the the electric field magnitudes $E_{0,1}$ determining the magnitude of $V_{char} $ in a given flow geometry.

\section{ 4. Fluid flow in long sample}
We consider a Poiseuille flow   in a defectless long sample  as a minimal model to study magneto-photo-transport
 in a highly correlated   electron fluid~[see Fig.~1($a$)].

Note that the high-quality GaAs quantum wells often contain macroscopic defects, those can appear during the growth process~\cite{d1}
 or can  be made artificially~\cite{d2}. Their appearance leads to a complication of the flow of the 2D electron fluid
  and an effective reduction in the flow
width~\cite{je_visc,disks}.  So, when describing a real flow in a sample by our model, the value $W$ is to be considered
as a certain effective width $W_{eff}$ of the conducting channel, which can  be significantly less than the actual sample width.

The external electric field $\mathbf{E} ^{ext} (t) = \mathbf{E}_0 +\mathbf{E}_1(t)$ consist of  the  dc field $\mathbf{E}_0 =
E_0 \, \mathbf{e}_x$ from  an applied bias  and the radiation field $\mathbf{E}_1(t) = \mathbf{E}_1 ^\pm
e^{-i\omega t } + c.c.$  with the left or the right  circular polarizations: $ \mathbf{E}_1^\pm = E_1\,
( \, \mathbf{e}_x \pm  i \mathbf{e}_y \, ) / 2 $.
We use the simplest diffusive boundary conditions:
\begin{equation}
   \mathbf{V}| _ {y=\pm W/2} = 0
\:.
\end{equation}
 The internal field $\mathbf{E}^{int} (y,t )  $   in this geometry is directed
 along the $y$ axis, being the Hall field. For simplicity, the sample is considered to be sufficiently narrow: $W \ll l_p $,  where
 $l_p $ is the characteristic plasmon wavelength.  In such case, the field $\mathbf{E}^{int}  (y,t) $ screens the $y$ component
of the incident ac field $\mathbf{E}_1(t)$. As a result, the ac plasmonic flow component, related to
$ e \, \delta n (y,t)  $,
is  suppressed and the ac flow is formed mainly by the viscoelastic eigenmodes~\cite{vis_res_2,SI}.   Herewith
only the $x$ component of the hydrodynamic velocity $ \mathbf{V} $ and the $xy$  components  of  the tensors $V_{ij}$ and
$ \varepsilon_{ij}$ are present.

In order to find the non-linear dc-ac conductivity $\sigma _{\omega} = j_0 [E_0, \mathbf{E}_1(t) ] /E_0 $, first, we calculate the linear
responses $ \mathbf{V} _0 $ and  $\mathbf{V} _1$  of the fluid on the dc $ \mathbf{E}_0  $  and  the ac $\mathbf{E}_1(t) $
fields \{here $j_0 [E_0,\mathbf{E}_1(t)] $ is the dc current density at incident radiation,
 averaged over the sample\}. The dc response
$\mathbf{V}_0=V_0\mathbf{e}_x$ is  a conventional 2D  Poiseuille flow:
\begin{equation}
 \label{V0}
    V_0 (y)  = eE_0[\,(W/2)^2  - y^2\, ] / (2m\eta_{xx})
    \: .
\end{equation}
Here   $\eta_{xx} = [(v^\eta_F)^2\tau_2/4 ] / (1+4 \omega_c ^2) $  is the dc diagonal viscosity. The velocity $V_0$
  depends  on magnetic field via the dc diagonal viscosity $\eta_{xx}  \sim 1/B^2$~\cite{je_visc}. The linear
responses~$ \mathbf{V}_1^\pm (y,t) =V_1^\pm (y,t)\, \mathbf{e}_x$  on the ac field of the both  polarizations~$\mathbf{E}_1^\pm$
are identical that reflects the screening of the  field component $E_{1,y}(t)$.  The velocity takes
the form~\cite{vis_res_2}: $ V _1 (y,t) = V_1 (y)  \, e^{ -i \omega t } + c.c. $,
\begin{equation}
\label{V1}
          V_1 (y)   = i eE_1 \,  [\,1  - f_\lambda  (y)\, ]/(2m\omega)
          \:,
\end{equation}
where   $ \lambda  = \sqrt{ - i \omega / \eta_{xx} (\omega ) } $ is the eigenvalue of the transverse magnetosonic waves,
\begin{equation}
\label{eta_xx_omega}
\eta_{xx} (\omega )   =    \frac{[(v_F^\eta)^2 \tau_2/4] (1-i\omega \tau_2) }
{ 1+(4 \omega _c^2 - \omega^2) \tau_2^2 -2i\omega \tau_2}
\end{equation}
 is the ac diagonal viscosity~\cite{vis_res_1,vis_res_2}, and the function
\begin{equation}
\label{cosh}
 f _\lambda  (y) = \cosh(\lambda y) / \cosh(\lambda  W/2)
\end{equation}
describes the flow profile. In the viscoelastic regime, $ \omega  \sim \omega_c \gg 1/\tau_2 $,  the imaginary  part dominates
 in $\eta_{xx} $ far from   the resonance at $\omega = 2 \omega_c$,    therefore Navier-Stokes equations~(\ref{main_eq_gen})
 turns out into Hooke's equations~[42,43]. The     wavelength and the decay length of the shear waves, related to the value $\lambda$,
 at $ \omega  > 2 \omega_c $  are estimated as  $ l_s  = v_F^{\eta} /\omega $ and     $ L_s  =  v_F^{\eta} \tau_2 $~(see details
in Refs.~\cite{vis_res_2} and~\cite{SI}; we consider here that $l_s,L_s \ll l_p$).  The parameter $v_F^{\eta} $
within the model of a strongly non-ideal Fermi liquid
is much greater than    the Fermi velocity $v_F$,  thus  a  flow with characteristic spacescales $ \Delta x  \gtrsim  l_s = v_F^{\eta}
/ \omega \gg  v_F/\omega $ can be  described    hydrodynamically~\cite{vis_res_2,Semiconductors}.

Second, to calculate the   photoconductivity in the second by $E_1$ order, $ \sim E_1^2$,  we account
   the following contributions in the hydrodynamic velocity: $
    V = V_0+ V_1   + V_ \omega $,
where $V_0= V_0(y) $ and $V_1 = V_1(y,t)$  are the linear responses presented above   and   $V_\omega  = V_ \omega (y) $ is
the nonlinear dc  component being proportional  to $ E_0 E_1^2 $. The latter one is calculated in Supplemental material~\cite{SI}
 on the base of  Eqs.~(\ref{main_eq_gen})-(\ref{Gamma_ex2})   by the perturbation   theory by the nonlinear memory inelastic,
$\sim \Gamma_{ijkl}$,    and elastic, $\delta F_{2,ij}$, terms.
 In this calculation, to obtain the value $V_\omega (y) $ in the lowest approximation,
one should account  in the nonlinear memory terms  the flow characteristics    $V$, $V_{xy}$,
 $\Delta_{xy}$, and  $\Pi_{xy}$  corresponding to the linear responses~$ V_0 $ and~$  V_1 $.

Based on the nonlinear dc   velocity  $V_{\omega}(y)$,     we find the mean sample photoconductivity
$ \Delta \sigma _ \omega  = \sigma _ \omega  -  \sigma _ 0   $ (here $\sigma _0 $ is the dc mean conductivity,
that is~$\sigma_\omega$ at $E_1=0$):
\begin{equation}
    \Delta \sigma _{\omega}  =  e n_0
   \int _{-W/2} ^{W/2} V_\omega(y)\,dy \,/(WE_0)
   \:.
\end{equation}
The non-linear values $V_\omega $ and  $\Delta \sigma _{\omega}$ within our model consists
 of  the inelastic  (relaxational)
  and the elastic
contributions:
$ V_{\omega}   = V _{\omega}  ^{ rel }  + V_{\omega}  ^{ el }$,
 $   \Delta \sigma _{\omega}   = \Delta \sigma _{\omega}  ^{ rel }  + \Delta \sigma _{\omega}  ^{ el }$.
We will see below that the relaxational and the elastic contribution have different magnetic field dependencies.

The result for the relaxational contribution to the nonlinear dc velocity can be presented in the form:
\begin{equation}
\label{result_SI0_0}
  V_{\omega} ^ {rel } (y)  \, = \, \frac{e ^3 E_0 E_1 ^2  \alpha }{ m^3 (v^\eta_F)^2}
  \,
   \frac{u + 2 \, \mathrm{Re} \,v }{\omega^4}
   \:  J(y) \, ,
\end{equation}
where $\alpha $ is a linear combination of the coefficients $ \alpha_{ijkl}^{mnop}  $ in Eq.~(\ref{Gamma_ex})
corresponding to the Poiseuelle flow geometry, $J(y)$ is the dimensionless factor determining the profile of
the nonlinear flow component:
\begin{equation}
  \label{J_y00}
  J(y) =
   \frac{| \lambda   |^2   }{|\cosh (\lambda W/2  ) |^2}
   \int  _{|y|} ^{W/2}   d \tilde{y}  \;  \tilde{y}
     \;|  \sinh (\lambda \tilde{y}  ) | ^2 \
     \, ,
\end{equation}
and the dimensionless  values $u$ and $v$ contain the dependence on the frequencies $\omega$ and $\omega_c$:
\begin{equation}
\label{u}
 \begin{array}{c}
 \displaystyle
  u =
  4 \sin ^2 ( \pi \omega / \omega _c )
 \, ( -1 + 4 \omega_c ^2 \tau_2^2 )
 \:,
 \\
 \\
\displaystyle
 v =
  ( 2\pi \omega / \omega _c)
   \Big[\sin( 2\pi \omega/\omega _c )
  -
   2i  \sin^2 ( \pi \omega/ \omega _c )  \Big]
\times
\\
\\
\displaystyle
  \times
    \frac{(-1 + i \omega \tau_2 + 4 \omega_c ^2 \tau_2^2 )(1+ 4 \omega_c^2 \tau_2^2)}
    {1+ ( 4 \omega_c^2-\omega^2 ) \tau_2^2 -2 i \omega \tau_2 }
\:.
 \end{array}
\end{equation}
For the elastic contribution to the nonlinear dc velocity it was obtained~\cite{SI}:
\begin{equation}
\label{result_SI_0}
   V ^{el}_{\omega}(y)  = \frac{e ^3 E_0 E_1 ^2   }{ m^3 (v^\eta_F)^2  }
   \: \frac{   |\eta_{xx} (\omega)| ^2   }{ \omega \, \eta_{xx} } \:  \Phi
   \,  J(y ) \:,
\end{equation}
where the factor $ J(y)  $ is the same   as for the inelastic contribution
[Eq.~(\ref{J_y00})] and  the factor $ \Phi = \Phi ( \omega , \omega_c ) $ has the form:
\begin{equation}
\label{u20}
\begin{array}{c}
  \displaystyle
\Phi  = 4 \, \{
\,
 [  - a \,\omega_c \tau_2   +    2 d \,    ( \omega_c /\omega  )^2 ] \,
 \sin ( 2 \pi \omega /\omega_c ) +
 \\
 \\
   \displaystyle
 + \,
  [-  2  b \, \omega_c \tau_2  +   c\, ] \, (\omega_c / \omega )  \,
 \cos ( 2 \pi \omega / \omega_c )
 \,
  \}
  \:.
  \end{array}
\end{equation}
Here $a,b,c,d$ are combinations of the coefficient $ \beta_{ij}^{klps}  $ from Eq.~(\ref{Gamma_ex2})
 for the case of the Poiseuelle flow geometry [see Eq.~(S37) in SM]. We remind that the coefficients $\alpha$ in  Eq.~(\ref{result_SI0_0})
as well as $ a, b, c, d $
 in  Eq.~(\ref{result_SI_0}) are proportional to the probability $P = e^{-T/\tau_q} $ for two quasiparticles
in a pair to make a complete
cyclotron rotation without a collision with  a third quasiparticle.

\begin{figure}[t!]
\centerline{\includegraphics[width=0.98\linewidth]{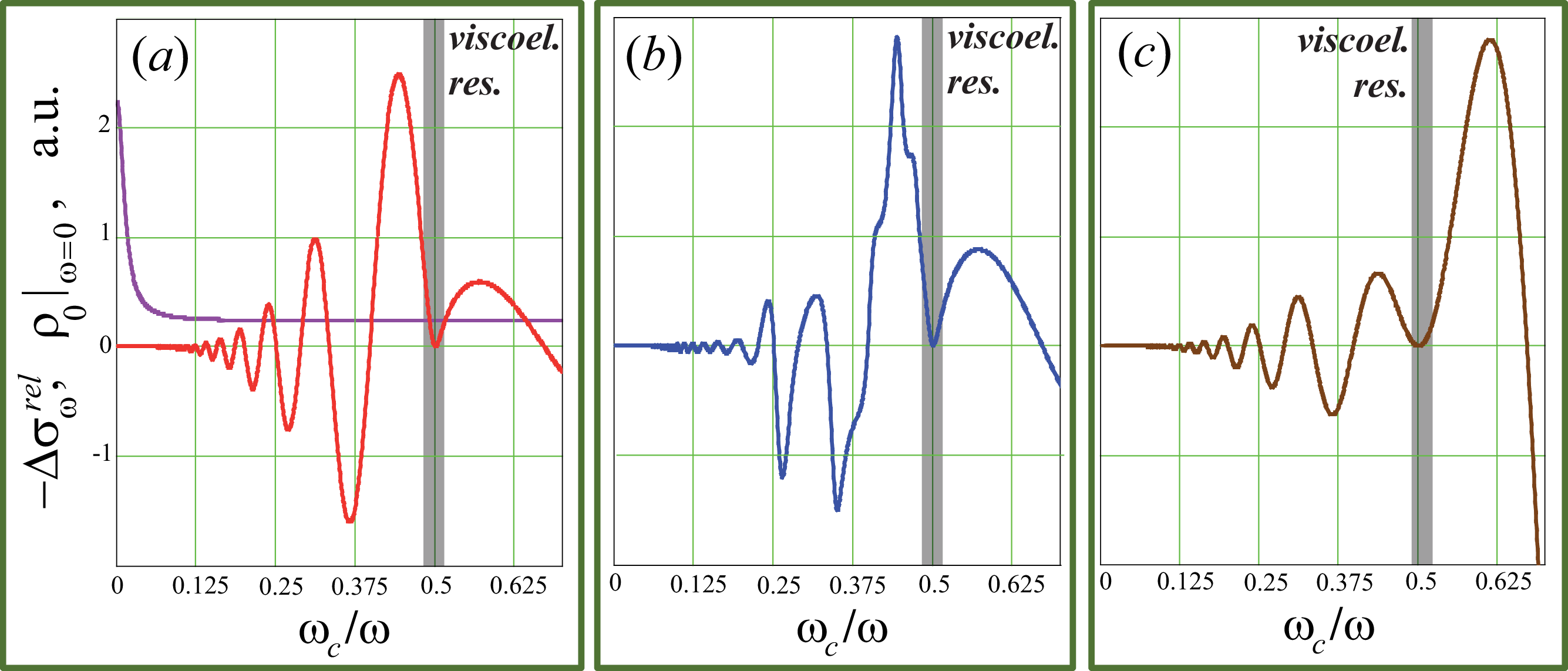}}
\caption{
Photoconductivity (relaxational contribution) with the inverse sign,  $  - \Delta \sigma_\omega ^{rel}$, for Poiseuille flows of a highly correlated  electron fluid
 in long defectless samples. The parameters of the system are $ \omega \tau_{2} = 40 $; $\tau_q / \tau_2 = 0.7$;     $  W / L_s =6   ,\,
 1.19   ,\, 0.03 $   [for red, blue, and brown curves in panels ($a$-$c$), respectively].  Grey stripes schematically denote the region
near  the viscoelastic resonance, $\omega = 2 \omega_c$, where the  fluid dynamics is mainly  dissipative  and
 the elastic contribution in photoconductivity [related to the perturbations of the Landau parameters
 $ \delta F_{2,ij} \sim V_{kl} \Pi _{mn} \sim \eta_{xx}(\omega) $] becomes large.
  In panel~($a$) we also plot  the dc magnetoresistance $\varrho_0|_{\omega = 0 }
(\omega _c)$ (violet curve) for  a sample with the same parameters of the electron   fluid and some density of defects.   Such
magnetoresistance consists of the two contributions~\cite{je_visc}:  the hydrodynamic
 one $\varrho_{xx,0} ^ {hydr} (B)$, being  proportional to $\eta_{xx} \sim (1+4\omega_c^2 \tau_2^2)^{-1}$ and dominating at $\omega_c \tau_2
 \lesssim 1 $  ($ \omega_c /  \omega \lesssim 0.1  $), and the Ohmic (Drude) one,  $\varrho_{xx} ^{D}(B) = const$, being proportional to
the rate of  the scattering of quasiparticles on impurities,  $1/\tau_{imp}$,   and dominating at $\omega_c \tau_2 \gg 1 $.
 }
\end{figure}

The profiles of the linear and nonlinear parts of the  dc responses,   $V_1(y)$ and
 $ V_{\omega}(y) \propto  V_{\omega}^{rel,el} (y) $,
for considered relatively
narrow samples are plotted in Fig.~1($c$,$d$). The ac-field-induced correction to the dc velocity, $V_{\omega}(y)$, inherits both
the oscillations of  $V_1(y)$ by~$y$, related to standing shear waves~[see Eq.~(\ref{cosh}) and Fig.~1($c$)], as well as
the parabolic profile of  the dc  Poiseuille flow~$V_0(y) $~(\ref{V0}).

In Fig.~2 we plot the relaxational contribution to photoconductivity, $\Delta   \sigma ^{rel} _{\omega} \sim V_\omega ^{rel}
\sim \Gamma_{ijkl} $, for the cases of wide, medium, and narrow samples. It is seen that the shape of magnetooscillations of
$\Delta \sigma_\omega ^{rel} (B)$ is regular  for the wide and the narrow samples, while it is irregular for  medium-size samples
with the widths  $W\sim L_s$, $W > l_s$. This property of $\Delta \sigma_\omega ^{rel} (B)$   reflects the formation of smeared standing
 waves inside the sample with relatively big amplitudes, which takes place  when:
\begin{equation}
 \label{cond0}
   \lambda( B ) \, W  \, \approx \, i \pi   ( 1+2N_w )
   \,,\quad
 N  _w = 0,\pm1,\pm2,... \:.
\end{equation}
 At such conditions  the amplitude of the ac response $V_1(y)$ exhibits the resonance [see Eq.~(\ref{cosh})], and consequently
 $V_\omega (y)$ and $ \Delta \sigma _ \omega $ also exhibits positive or negative peaks [see Fig.~2(b)]. Note that
the value $ \Delta \sigma _ \omega ^{rel}$ is not proportional just to the absorbed power $\mathcal{W}(B)$  for wide 
 or for narrow samples (see the plot of $\mathcal{W}(B)$ in Fig.~S3 of Supplemental material~\cite{SI}),  but is related
 with both the real and imaginary parts of $V_1(y)$.

The magnitude of the calculated response strongly depends on magnetic field, frequency, and sample width.
For example, let us make the estimate for the photoconductivity magnitude at the magnetic field and frequency corresponding
to the regime: $\omega_c \sim \omega \gg 1/\tau_2 $, $|\mathrm{Im}(\lambda )  | \gg | \mathrm{Re}(\lambda ) | $,
and the sample widths in the interval, $| \mathrm{Im}(\lambda ) | ^{-1}  \ll W \ll | \mathrm{Re}(\lambda ) | ^{-1} $.
In this case, the standing shear waves are well formed. For the relaxational contribution in the limiting case
$ F_1 \sim 1 $ ($ v_F \sim v_F^\eta $) and  far from the resonances appearing at condition~(\ref{cond0}), the formulas
in Supplemental material~\cite{SI} yield:
\begin{equation}
\label{est}
   \frac{  \Delta \sigma _{\omega} ^{rel}}{ \sigma _0 }
     \, \sim \,
   P(\omega_c)  \, \Big( \frac{eE_1}{m} \Big)^2 \frac{  \tau_2 }{a_B^2 \,  \omega^3 }
   \:.
\end{equation}
Here we imply that the magnitude $E_1$ satisfies the linear by $E_1^2$  regime, when $ |\Delta \sigma _{\omega} ^{rel} | \ll \sigma _0 $.
This result is based on consideration  in~\cite{SI} of the process of an extended collision in detail, leading to  the estimate
for the parameters $\alpha_{ijkl}$  in Eqs.~(\ref{Gamma_ex}) via the screening radius $\sim a_B $.  Note that there is no a direct dependence
on the sample width in~Eq.~(\ref{est}).

The estimate of the relative photoresistance of  the Ohmic samples, being analogous to  the value
 $ \Delta \sigma _{\omega} ^{rel} /  \sigma _0 $ in a Poiseuille flow,
within  the model of Ref.~\cite{Beltukov_Dyakonov}
of magnetotransport due to the memory effects at scattering of non-interacting electrons in smooth defects
 yields:
\begin{equation}
\label{estOhm}
   \frac{  \Delta \varrho _{ \, xx, \, \omega} }{ \varrho ^D _{  0} }
   \, \sim \,
   P_{imp}(\omega_c)  \, \Big( \frac{eE_1}{m} \Big)^2 \frac{  1 }{r_0^2  \, \omega^4 }
   \:,
\end{equation}
where $ \varrho ^D _{  0} \equiv \varrho ^D _{  xx,0} $ is the Drude dc resistivity, $r_0$ is the defect radius,
 $P_{imp} =e^{-2\pi/(\omega_c\tau_q^{imp})} $  is the probability
for electron make a rotation without scattering  on defects, being similar to $P(B)$~(\ref{P0}), and $\tau_q^{imp}$ is the impurity
 scattering departure time. We see from Eqs.~(\ref{est}) and~(\ref{estOhm}) that, for reasonable size of the defect radius, $a_B \sim r_0 $,
 and at comparable scattering times, $\tau_q^{imp} \sim \tau_q  (T_e)$, in the hydrodynamic regime  a much larger magnitude of
the relative photoconductivity, $ \Delta \sigma _\omega ^{rel}/ \sigma_0$,  is attained than 
the similar value, $ \Delta \varrho _{ \, xx, \, \omega} / \varrho ^D _{  0} $,   
within the theory~\cite{Beltukov_Dyakonov} for Ohmic flows. Finally, we note that in the hydrodynamic samples
 with a very low defect density  at sufficiently large magnetic fields,
when $\omega _c \tau_2 $ is larger than some threshold value depending on sample width and the defect density,  the scattering
 of quasiparticles on these defects becomes important, and the flow acquires a mixed hydrodynamic-Ohmic  character~\cite{SI}.

The resulting  mean sample photoconductivity  $  \Delta \sigma _{\omega} $ does not depend on the sign ``$\pm$'' of the circular  polarization
of the ac field   $\mathbf{E}_1 (t)$. Indeed, as we mentioned above, in narrow samples, $W\ll l_p $, the ac linear response  $ V_1(y,t) $,
which is mainly formed by the viscoelastic contribution, is  independent  of the sign~$\pm$
due to the screening of the $y$ component of the incident field    $ \mathbf{E}_1(t) $ by the internal field
 $\mathbf{E}^{int}\sim (d \delta n/dy) \, \mathbf{e}_y $. Thus the nonlinear
 components of the velocity,  $V_\omega ^{rel}$ and $V_\omega ^{el}$,  stemming
from the retarded relaxation term, $\sim \Gamma_{ijkl}$,   and from the retarded  elastic terms,
$ \sim\delta F_{2,ij}$, are  also independent of the polarization sign~$ \pm $.
  So the independence of $\Delta \sigma _\omega $ of the sign $\pm$ is a necessary (but not sufficient)  evidence
of the formation of a high-frequency hydrodynamic flow.

  \begin{figure}[t!]
\centerline{\includegraphics[width=1.0 \linewidth]{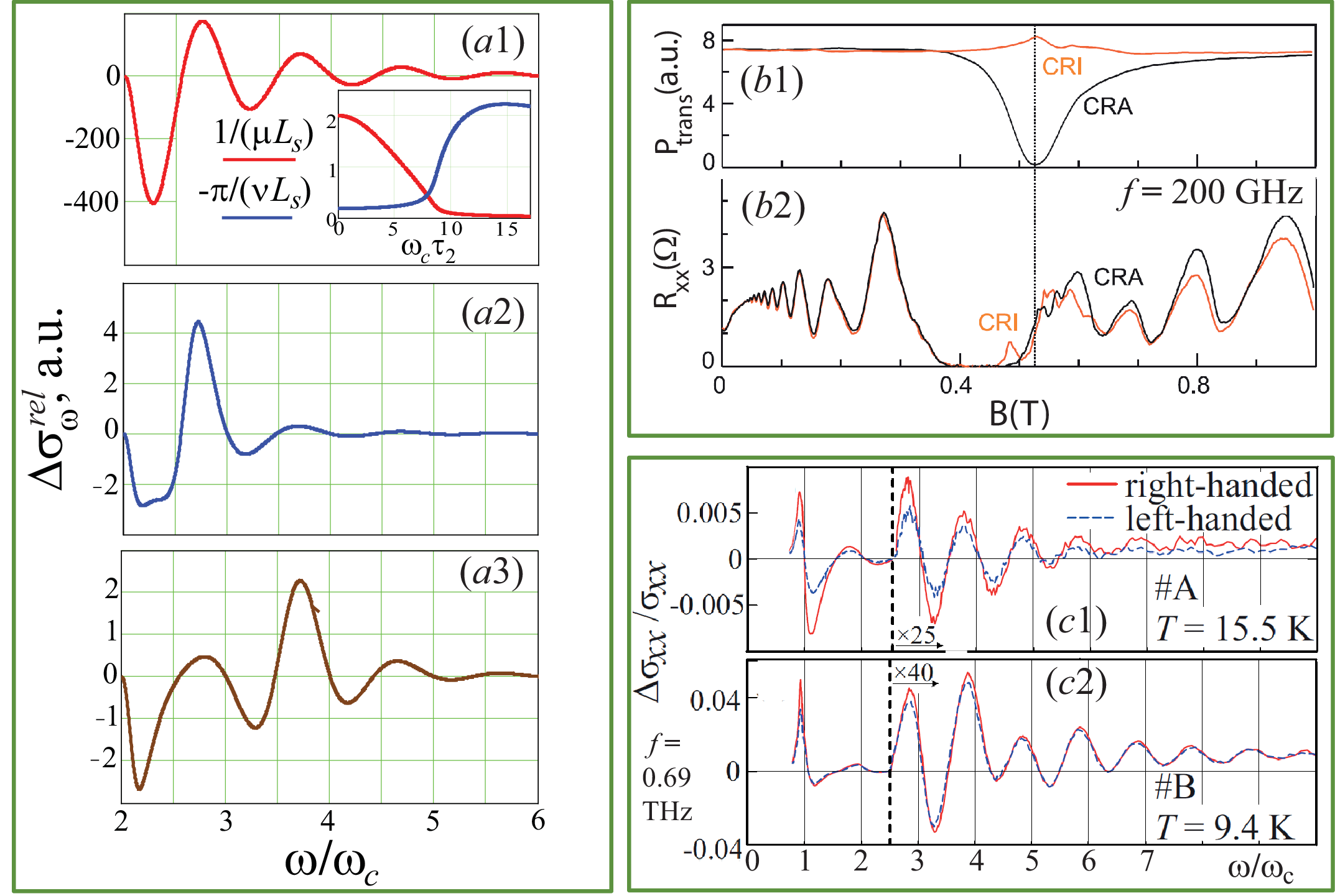}}
\caption{
($a1$)-($a3$):
Calculated relaxational contribution to   photoconductivity,~$\Delta \sigma_\omega ^{rel}$,
due to  the memory effect  in the interparticle scattering for the Poiseuille flows with
the parameters  $\omega \tau_2 = 17 $;  $\tau_q / \tau_2 = 0.7$;
 $  W/L_s =5.5, \, 0.85, \, 0.66  $ [for the upper, center, and lower subpanels, respectively].
    Inset: the dependencies of the decay length
($1/\mu$) and one half of the wavelength ($-\pi/\nu$)    of  magnetosonic  waves
on the value $\omega _c \tau _2 \sim B $ at $ \omega \tau_2
 = 17 $   in the diapason $ 0<  \omega_c  <\omega $.
  ($b1$)-($b2$):
Experimental data from Ref.~\cite{Smet_1}  on the cyclotron resonance in absorption (upper subpanel) and  MIRO (lower  subpanel)
 in an
ultra-high quality GaAs quantum  well     with the experimental value of ``mean  sample mobility''  $\mu _{exp}= 18 \cdot 10^6
 $~cm$^2/$V$\cdot$s.     Red and black curves refer to the inactive and the  active circular polarizations of radiation.
  The value $P_{trans}$ in upper subpanel is the power of the radiation transmitted through the layer with 2D electrons in arbitrary units;
  the value $R_{xx}$ in lower   subpanel is the longitudinal resistance of 2D electrons at incident radiation in  magnetic fields
   perpendicular to 2D layer (that is,  magneto-photo-resistance).
      The radiation  frequencies $f$ are labeled on both panels ($b$) and ($c$).
  ($c1$)-($c2$):
Experimental data  from Ref.~\cite{Ganichev_1}
  on oscillating magneto-photo-conductivity~$\Delta \sigma_\omega$, analogous to MIRO, for two GaAs quantum wells:
 with a moderate experimental mobility
(sample $\#$A  with  $\mu  _{exp} = 0.82 \cdot 10^6  $~cm$^2/$V$\cdot$s, upper subpanel)
and  with a  higher experimental  mobility
(sample $\#$B  with  $\mu _{exp}= 1.8 \cdot 10^6  $~cm$^2/$V$\cdot$s,  lower subpanel).
Red and blue curves refer to the right and the left
circular polarizations of radiation.
  {\em Comment on  comparison of panels ($a$)-($c$)}:
   It is seen that the theoretical curves in panels~($a$2) and~($a$3) are similar to the experimental curves
      in panels~($b$2) and~($c$2) by the presence of irregular non-sinusoidal oscillations,
   while  the theoretical curve in panel~($a$1)
   is similar to the experimental curves in panel~($c$1) by the presence of regular sinusoidal
   smoothly decreasing with $\omega/\omega_c $  oscillations
   (see a detailed explanation in~Section~6.1).
  }
\end{figure}

In the regime linear by $ E_1^2 $ and $ E_0$,  the photoresistance  $\Delta  \varrho_{\omega} =  \varrho_{\omega} - \varrho_0 $
is proportional to $-\Delta \sigma _{\omega}$  (here $\varrho_{\omega}   = 1 / \sigma_\omega  $ and $  \varrho_0 = 1/\sigma_0$
are the sample mean resistivities at nonzero and at zero $E_1$). Therefore in Fig.~2 and Fig.~4 below we have presented
the value~$-\Delta \sigma _{\omega}$.

In Fig.~3($a$) we plot the  relaxational contribution  to photoconductivity  $\Delta   \sigma ^{rel}  _{\omega} $ for smaller values
of the parameters $\omega \tau_2$ and $W/L_s$ than in Fig.~2 (that is, for a ``smoother'' viscoelastic
system; the used value of $\tau_q$
  lies at the edge of the model applicability
condition~$    1/ \tau_2 \ll \omega_c  \lesssim 2\pi/\tau_q $).
  It is seen that the magnetooscillations of $\Delta   \sigma _{\omega}  ^{rel}  $  are regular (sinusoidal with damping)
 in relatively wide samples: $W \gg L_s$, and become  irregular  (some  oscillations have peculiar values of  amplitudes) in
 the narrower samples, whose widths $W$ are  comparable with the transversal wave decay length~$L_s$. As like for the curves
$ \Delta   \sigma ^{rel}  _{\omega}  (B)$ in Fig.~2, these irregularities   are the manifestations   of the ``geometric'' resonances
 related to coincidence of $W$ with a half-integer numbers of the magnetosonic  wavelengthes [see Eq.~(\ref{cond0})].   For the samples
shown in  Fig.~3($a$) these resonances are much more smeared    as compared with the ones shown in Fig.~2
  because of not too large value of $\omega \tau_2$.

In  Fig.~4($a$) we plot the elastic  contribution   $\Delta   \sigma ^{el} _{\omega} \sim  V_\omega ^{el} \sim  \widetilde{\delta F }
^{(1)} _ {2,ij} $  to the photoconductivity  $\Delta   \sigma  _{\omega} $ originating from the $F_2$-parameter memory  term, 
 $ \sim  \widetilde{ \delta F } _{2,ij} [V(y,t')] $, 
 in Eqs.~(\ref{main_eq_gen}).   It is seen that this contribution, along with magneto-oscillations, exhibits
a strong resonance at  the doubled cyclotron frequency $\omega = 2\omega_c$. This is the viscoelastic resonance stemming from
the resonance
 in~$\eta_{xx}(\omega) $~(\ref{eta_xx_omega}) [see Eq.~(18)]
  and being reflected in the dispersion law  of magnetosonic waves~\cite{vis_res_2}. In the calculated
photoconductivity   $\Delta   \sigma ^{el} _{\omega} $ the shape of this resonance originates from the frequency dependence 
 of the linear response $V_1(y)$
 and from the form of the elastic memory term~$-\widetilde{ \delta F }_{2,ij}^{(1)} [V (y,t-T) ] \, \partial \Pi_{ij} / \partial t $.

  \begin{figure}[t!]
\centerline{\includegraphics[width=0.97 \linewidth]{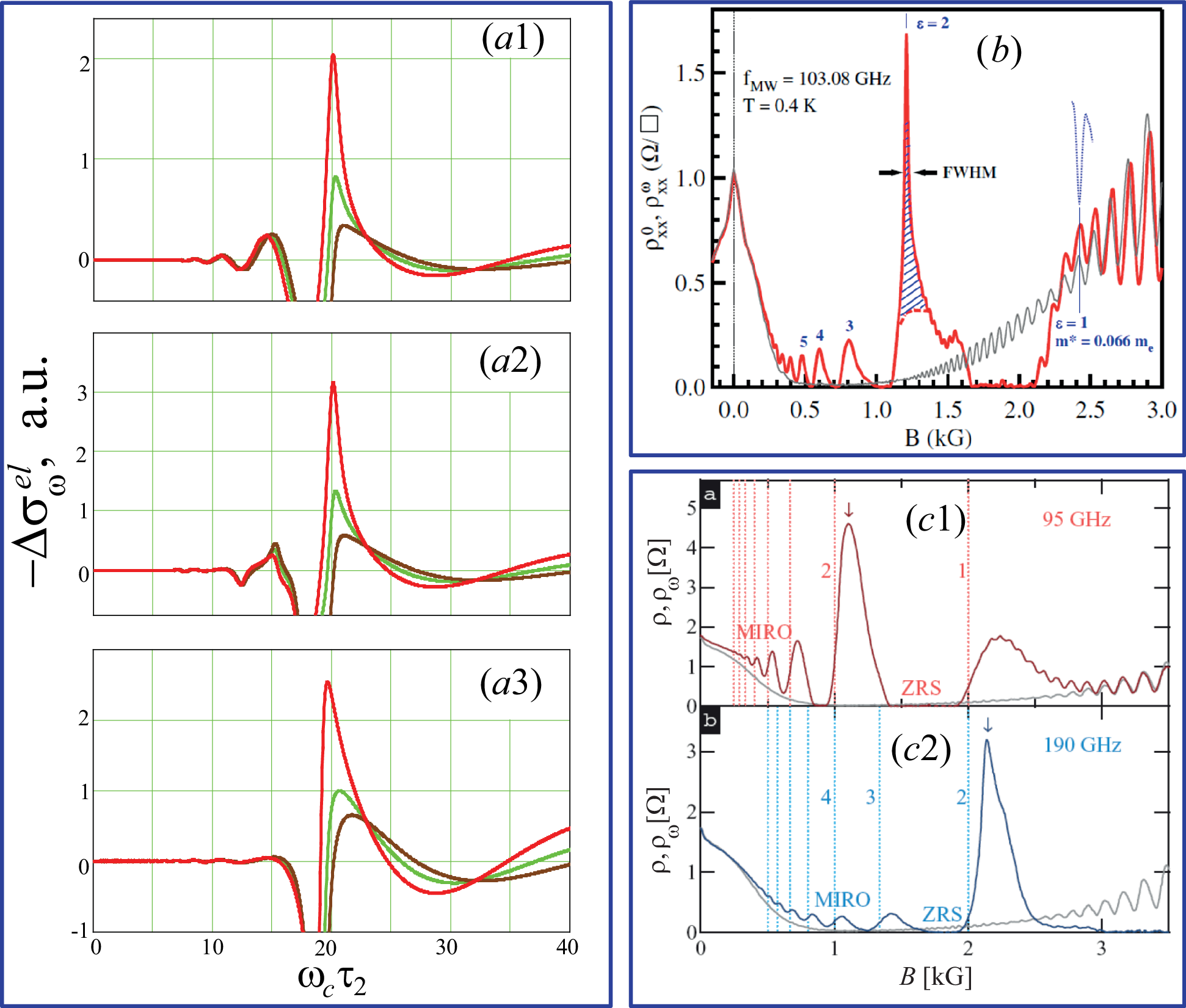}}
\caption{
 ($a$1)-($a$3):
Calculated   elastic contribution to   photoconductivity  due to the memory effect in the interparticle interaction
energy   for the Poiseuille
 flows with  the parameters  $\omega \tau_2 =  40 $;  $\tau_q / \tau_2 = 0.5 $;   $W/L_s = 4 $  (top), $W/L_s = 1.1$ (middle),
$W/L_s = 0.24$ (bottom). Curves of different colors on each of the subpanels  correspond to the following coefficients   $a,b,c,d$
 in equation~(\ref{u20}) [$a,b,c,d$ describe
 the form of the elastic memory term $ \widetilde{\delta F}_{2,ij} $ defined by Eq.~(S37) in~SM]: $b/a = -0.8 $
 (red curves;
for these and all other curves $ c ,  d = 0 $), $ b/a = -0.28  $ (green curves), $b/a =0.08 $ (brown curves).
Negative values of
$ - \Delta \sigma _\omega ^{el} $    appear on the plotted curves as
 the value $\Delta \sigma _\omega ^{el } $ is the linear by the ac power $ P \sim E_1^2 $ correction to  $\sigma_0$.
 [($b$)~and~($c$1)-($c$2)]:
Experimental data from Refs.~\cite{exp_GaAs_ac_1} and~\cite{exp_GaAs_ac_2}, respectively,
  on longitudinal resistance of high-quality GaAs quantum wells
in perpendicular magnetic fields at low temperatures without radiation
  [the values $\varrho^0_{xx} $ and $\varrho$  in panels ($b$) and~($c$)]
 and at non-zero radiation
[the values $\varrho^\omega_{xx} $ and $\varrho_\omega $  in panels ($b$) and~($c$)].
 The radiation  frequencies $f$ are labeled on both panels.
The experimental values of mean sample mobilities are
  $\mu  _{exp} = 3 \cdot 10^7  $~cm$^2/$V$\cdot$s~($b$)
  and
   $\mu  _{exp} = 1.1 \cdot 10^7  $~cm$^2/$V$\cdot$s~($c$).
     The value~$\varepsilon=1,2,3,4,5$ in panel~($b$) and the small numbers~$1,2,3,4$ near thin vertical lines
      in panel~($c$)  denote the magnetic fields correspond
       to the integer ratios $\omega / \omega_c=1,2,3,... $,
      that is, to the frequencies of the first and higher harmonics of the cyclotron resonance.
 Both the giant negative magnetoresistance   (thin grey lines), evidencing the hydrodynamic transport
regime~\cite{je_visc},   and the $ \omega = 2 \omega_c $-peak in photoresistance (red and blue lines) are seen.
    {\em Comment on comparison of panels $(a)$-$(c)$}:
   It is seen that the red theoretical curves in panels~($a$1)-($a$3), especially
  the red curve in panel~($a$1), are similar to the experimental curves
  in panels~($b$), ($c$1), and~($c$2) by the presence of a large, more or less symmetrical, peak (``spike'')
   at the magnetic fields being
   close to the doubled cyclotron resonance, $\omega_c =\omega/2$.
        Oscillations to the left side of the large peak are also seen
   on both theoretical and experimental curves in panels~($a$) and~($b$),($c$).
  }
\end{figure}

\section{ 5. Discussion on conditions of applicability of  developed theory}
  In above consideration, we have considered  the Landau Fermi-liquid model with the large parameter $F_{1}\gg1$~[see Eq.(\ref{crit})] as
the  simplest model of   a highly correlated electron liquid where the formulated equations~(\ref{main_eq_gen}) are surely applicable.
    In GaAs quantum wells samples, which we will discuss
below,  such strong interaction is not apparently realized, therefore the simplest Fermi-liquid based
 model  of Ref.~\cite{Semiconductors} is not directly applicable to them.

   However, the above phenomenological consideration of flow dynamics, based on solution of Eqs.~(\ref{main_eq_gen}),
  leads to reasonable results even in the limit $F_{1} \ll 1$ (corresponding to $r_s \lesssim 1$ or $r_s \ll 1$), when
the fluid  parameters tends to their Fermi-gas values, in particular, $v_F ^\eta \approx v_F $,  $ l_s \sim R_c \sim v_F/\omega  $,
  and  $ L_s \sim v_F \tau_2 \gg R_c  $.
Although shear magnetosonic waves are no longer formed at such~$r_s$,  equations~(\ref{V0})-(\ref{cosh})
and   (\ref{result_SI0_0})-(\ref{result_SI_0}) in this limit yield finite reasonable flow characteristics in the bulk region,
  $W/2-|y| \gg R_c $ (at $ W \gg R_c $).
Indeed, provided that there is no source of small-scale disturbances of sizes $\Delta x \ll R_c$ in a sample,
     the kinetic equation and the hydrodynamic approaches can yield similar results for non-peculiar regimes~\cite{SI,c}.
This situation corresponds to the edge of applicability of any hydrodynamic-like equations  for the flow with
  the minimal spacescale  of inhomogeneities:~$\Delta x \sim R_c$.

A generalized model, being valid for any $F_m$,
should be based on the kinetic equation for time-dependent distributions of quasiparticles, in which
  the hydrodynamic contribution
  and the ballistic contribution,
  containing the angular harmonics by the quasiparticle velocity of the third and higher orders,
  can be comparable.
Herewith non-linear memory terms  in the collision integral as well as  in the Landau interaction terms (with any unperturbed
parameters~$F_m$), apparently, survive and should be accounted in the same manner,  because condition~$F_1 \gg 1$~(\ref{crit})
 is substantial only for the viscoelastic, but not for the memory effects.
 In such system, instead of the standing shear
waves, some quasi-periodical by $2R_c$ distribution of the flow characteristics  can be formed due to
 the ballistic effects of commensurability
of~$W$ and~$2R_c$     (see, for example, consideration of Ref.~\cite{Holder} for stationary flows).
 Such  generalized model of the 2D electron fluid dynamics, based on the non-linear retarded kinetic equation,
   may lead to the results for characteristics of the electron flow being very similar
   to the results of the current purely hydrodynamic model
   in both  the diapasons of~$B$: the vicinity of the resonance magnetic $\omega_c = \omega/2 $
and the lower magnetic fields $\omega_c <\omega/2 $.

We also note that, as it was mentioned above~\cite{f},
 a quantum  reconstruction of the ground and excited Fermi-liquid states of 2D electrons in magnetic field
also can be very important for the possible form and the most general  conditions
of applicability of the macroscopic equations   of sort of~Eqs.~(\ref{main_eq_gen}) for the systems with  $r_s \lesssim 1 $.
Beside this,
for particular flow parameters, the widest conditions of literal applicability of our phenomenological model,
based on solution of the balance equations~(\ref{main_eq_gen}),  can include also the systems with $r_s \lesssim 1$.
 For example, it was demonstrated in Refs.~\cite{a,Holder} that the electron hydrodynamics   for stationary flows
 in pure samples at any  inter-particle scattering rates, $1/\tau_2$, is eventually being formed with
the increase of magnetic field, when $W > 2R_c$.

From this discussion, such conclusion can be made.
In ultra-pure samples, even at small interaction parameter, $F_1 \lesssim 1 $,
the $2\omega_c$-resonance in viscosity coefficient~(\ref{eta_xx_omega})  and
  the  memory effects in the fluid motion equations  of the sort of Eq.~(\ref{main_eq_gen})
    due to formation of quasiparticle pairs  do retain.
In bulk  samples with the widths $ W \gg R_c,v_F/\omega$
  or/and under some other conditions on the parameters $  R_c , \, \omega , \,\tau_q $,
a kinetic equation model accounting the above
 two effects as well as the ballistic commensurability effects, apparently,
can  lead to the flows and photoresponses,
 being similar to the ones derived 
 in Section~IV within the strongly non-ideal Fermi-liquid model.

\section{ 6. Discussion on relations of developed theory with   experiments }
\subsection{ 6.1. Comparison of developed theory with experiments on high-quality GaAs~quantum wells }
In Fig.~3 we compare the calculated relaxational contribution in photoconductivity $\Delta\sigma ^{rel}_{\omega} $ with
 experiments~\cite{Smet_1,Ganichev_1}  on measurements of the MIRO effect in high-quality GaAs quantum wells.

First, it is seen from panels ($b$) and ($c$)  of Fig.~3 that  for the structure with the lower ``experimental  averaged sample mobility''
[sample~$\#A$ in panel~($c$)] there is some substantial dependence of the MIRO-like photoconductivity  on the polarization sign
of radiation, whereas  this dependence is almost absent for MIRO (panel~$b$) and for the MIRO-like photoconductivity (sample~$\#B$
in panel~$C$) in the ultra-high and   the moderately high mobility samples. Second, it is also seen that for the lower-mobility sample the shape
of oscillation  is regular (sinusoidal with a slow damping with the increase of $1/B$),  whereas for the two higher-mobility samples
 the shape of oscillations is  irregular: the amplitudes of some oscillations   are too large or too small, but the other oscillations
 are  sinusoidal with a slow damping with  $1/B$ [see Figs.~3($b$2) and~3($c$2)].   The irregular shape
of magnetooscillations, as it was discussed above, is explained   within our theory
      by the formation of smeared standing magnetosonic waves inside
the sample.

The independence of $ \Delta \sigma _{\omega} $  of the polarization sign  $\pm$ and the irregularities of  oscillations  were unexplained
 within  theories~\cite{theor_1,theor_1_1,theor_2,joint,Polyakov_et_al_classical_mem,Beltukov_Dyakonov} for independent  electrons
 in bulk disordered samples.

    In the samples studied in Refs.~\cite{Smet_1}  and Ref.~\cite{Ganichev_1}  the 2D electron densities
    were $2.3\cdot 10^{11}$~cm$^{-2}$~[13]
 and $12.0\,,\:9.3\cdot 10^{11}$~cm$^{-2}$ (for samples~$\#A$, $\#B$ in~[14]).  Such densities lead to $r_s=1.1,\,0.50,\,0.56$
 and, apparently, to not large values of the Landau parameters $ F_1 $~\cite{Vignale_book}. 
    So the sufficient conditions~(\ref{crit}) of the big magnitude of the Landau interparticle interaction, which surely
 leads to the developed viscoelastic model, seems to be not fulfilled.

   However, as in was discussed in Section~5,
    a generalized model of the retarded non-linear dynamics of
    the 2D electron fluid,
    based on the kinetic equation,
   may lead to the results for characteristics of the electron flow being very similar
   to the results those  follow
   from the current purely hydrodynamic model based on the phenomenological equations~(\ref{main_eq_gen}).
   In particular,
    in the kinetic model the effects of commensurability
    of the sample width, $W$, and the trajectory size, $2 R_c$,
    can play the role
    of the resonances induced by
    the standing shear waves.
    Provided such closeness of the kinetic  and the hydrodynamic models  is actually realized,
     the explanation of the photoresponses  observed in Ref.~\cite{Smet_1,Ganichev_1} within our
theory can evidence about formation of flows of a highly correlated 2D electron fluid
 with the geometric resonant effects and  the memory effects.

In Fig.~4 we compare the calculated elastic contribution to photoconductivity, $\Delta \sigma ^{el} _{\omega}  (B)$,
 with the magneto-photo-resistance measured in the GaAs quantum wells of record quality~\cite{exp_GaAs_ac_1,exp_GaAs_ac_2}
 and exhibiting a strong peak at $\omega = 2 \omega_c$. Rather similar results on observation of  magneto-photo-resistance with
the strong $\omega = 2 \omega_c$-peak were also obtained in Ref.~\cite{peak_gr,peak_gr2} for high-quality graphene samples.

 In experimental data shown in  panels~($b$) and~($c$) of Fig.~4 a large  narrow peak
is observed near the doubled cyclotron frequency, $\omega= 2 \omega_c $, and much smaller peculiarities (or oscillations)
  at higher cyclotron harmonics. It is noteworthy that the stronger is the negative magnetoresistance, the sharper and
bigger is the peak in magneto-photo-resistance   (see discussions in \cite{vis_res_2,SI}). It is seen from the experiments as well as
from the theoretical results that,  depending on the specific parameters of the sample and  the electron fluid, the resonance changes
its shape (see Fig.~4).   In other ultra-high-quality graphene and GaAs quantum well samples  the $2\omega_c$-resonance gradually disappears,
   depending on the sample width, temperature, and the electron density~\cite{recentest_,peak_gr2}.
     In lower magnetic fields, $\omega_c <\omega/ 2  $,
magnetooscillations  appear on  the theoretical curves, qualitatively corresponding to peculiarities on the experimental curves
(see Fig.~4). Note that there are no resonance at   $\omega= 2 \omega_c $ in the relaxation contribution
$ \Delta \sigma ^{rel} _{\omega} $~[see Figs.~2 and~3($a$)].

 The magnitudes of the experimental data
on the resistance change due to radiation,  $ (\varrho_\omega - \varrho _0 $), [cited in Fig.~4] are relatively large,
 comparable with $\varrho _0$ at $B=0$.
As we have described in Section~4,   we perform the calculations of $(\varrho_\omega - \varrho _0) $  in the lowest,
$ \sim E_1^2 $, order by the ac field amplitude.
So it is important for comparison of theory and experiments that in Ref.~\cite{exp_GaAs_ac_2}
   an eventual
   appearance of the $2\omega_c$-resonance in photoresistance with  the increase of radiation power $ \mathcal{P} \sim E_1^2 $
 was observed and the measurements of the dependance of its amplitude   on $E_1^2$ were performed.
The resonance appear already at  relatively small $E_1$, when $ | \varrho_\omega - \varrho _0 |\ll\varrho _0  $, and the value
$(\varrho_\omega - \varrho _0)$ turns out linear in  $E_1^2$ (see Fig.~3 in Ref.~\cite{exp_GaAs_ac_2}).

In recent experiment~\cite{recentest_}   on ultra-pure GaAs quantum well samples of various widths
 the giant negative magnetoresistance and the peak in photoresistance near $\omega = 2\omega  _c $,
similar to the ones presented at Fig.~4, were observed.
For medium sample widths and at low radiation powers, the irregular shape of MIRO and
the giant  $2\omega_c$-peak were seen very well \{see Fig.~4(b) in Ref.~\cite{recentest_}\}.

 In this way, with the same precaution about the strength of interparticle interaction in the discussed GaAs and graphene  samples
 (which was put above in this section, when we discussed the results of experiments~[13,14]),
 the appearance of a large distinct peak at $\omega= 2 \omega_c $  in magneto-photo-resistance of GaAs
 and graphene samples can be related to the viscoelastic resonance   in the values $\eta_{xx}(\omega)$, 
 $V_1$, and~$\Delta \sigma _ \omega ^{el} $.

\subsection{ 6.2. Properties of high-quality samples  which can lead to
realization of  hydrodynamic regime }
It is necessary to recall that for the implementation of the high-frequency hydrodynamic regime, the width of the effective channel
 in which the viscous flow is formed must be sufficiently narrow, in particular, narrower than the wavelength of the magnetoplasmon
 at the frequency of external radiation (see Section~4 and Ref.~\cite{vis_res_2}).

On the one hand, the actual widths of the samples studied in experiments~\cite{Smet_1,Ganichev_1,exp_GaAs_ac_1,exp_GaAs_ac_2}
 were much larger, hundreds of microns or several millimeters.
 Only in experiment~\cite{recentest_} the samples with small as well as large widths~$W$, from 25 to 400~$\mu$m were studied,
 with intent to trace the signs of the hydrodynamic regime with change the value~$W$. Moreover, in work~\cite{Ganichev_1},
 a part of measurements were carried out by irradiating only a  region  of the sample, a spot with the size being  significantly
  smaller than the sample size.

On the other hand, as noted above, in GaAs quantum well samples with very high mobilities,
 macroscopic ``oval'' defects  are typically present~\cite{d1,d1_new}.
 In such samples, both the low- and high-frequency fluid flows
  are formed in the regions between the defects, and the effective width of such flows,
  $W_{eff}$,  is significantly smaller than the sample width~\cite{je_visc,disks}.
Since the samples studied in works~\cite{Smet_1,Ganichev_1,exp_GaAs_ac_1,exp_GaAs_ac_2}   are similar in many parameters
and properties (electron densities, experimental values of mean mobilities, the appearance of
 the giant negative magnetoresistance) to the samples studied in works~\cite{d1,d1_new},
 it is reasonable to expect that macroscopic defects were also present in the samples studied in works~\cite{Smet_1,Ganichev_1,exp_GaAs_ac_1,exp_GaAs_ac_2}, leading the the appearance of conductive
 channels with sufficiently small effective widths $W_{eff}$.
This can explain the applicability of hydrodynamic equations of type~(\ref{main_eq_gen})
[or analogous kinetic equations] to the description of photoconductivity of sufficiently wide samples
 without taking into account the  plasmonic contribution to ac  flows.

Let us discuss the possible reasons why in some samples the relaxational contribution
     of photoconductivity due to the retarded terms in the tensor $ \Gamma $ is predominantly manifested,
while in others the elastic contribution due to the retarded terms~$\delta F _{2,ij}$
in the perturbed Landau parameter~$F_2$   dominates.
    We note that, as it is seen from Eqs.~(\ref{result_SI0_0}) and~(\ref{result_SI_0}),
 the relaxational and the elastic contributions both
are proportional to the factor $J(y)$, thus their magnitudes depends
on the   sample geometry  and flow  geometry in a similar way, provided the flow in the whole sample is fully hydrodynamic.

First, it is important that these two
 contributions differ by their dependence     on the frequencies $\omega$ and $\omega_c$
 [see Eqs.~(\ref{result_SI0_0}) and (\ref{result_SI_0})].
   It is seen from Figs.~2,~3($a$), and~4($a$)
that, within our theory,  the elastic contributions
 has maximum values  mainly around the magnetic field corresponding to $\omega=2 \omega_c$
[as this
contribution is proportional to $|\eta_{xx}(\omega)|^2$, see Eq.~(\ref{result_SI_0})],
 while the relaxation contribution   exhibits a non-trivial oscillating behavior
 mainly  in the diapason  $ \omega_c  < \omega /2 $.
 Therefore, in the cases when the hydrodynamic regime is realized only partially
and there is some substantial  plasmonic contribution in the ac flow component
 in  some diapason of magnetic fields,
 the relaxational or the elastic hydrodynamic contributions can predominantly
  control the ac flow component at the other diapason
 of magnetic fields.
     As a result,  the last two contributions
     can manifest itself
     in one of the regions $\omega_c\sim \omega/2$  or $\omega_c<\omega/2$,
    in which  the plasmonic contribution is relatively small.
    In the diapason of magnetic fields, where the plasmonic contribution dominates, 
    the non-linear photoresponses have
     the properties,
    those have not been studied in this work.

    Note that the linear photoresponses  for samples with any~$W_{eff} $,
    in which the flow consists of
 the both plasmonic and  the  hydrodynamic components,
    were theoretically explored   in Ref.~\cite{vis_res_2}.
  Studies of the properties of the ``mixed  non-linear ac-dc'' flows with
 the   ac and dc  hydrodynamic as well as the ac   plasmonic contributions
   is the subject of a further separate study.

  Second, the relaxational and the elastic contributions within our model
  are proportional to different constants, namely,
    $\alpha$ and $a$, $b$, $c$, $d$.
  A qualitative estimate of the constant $\alpha$ is made in Supplemental Materials~\cite{SI}.
The value of $\alpha$ (more precisely, the characteristic magnitude of  $\alpha_{ijkl}^{nmop}$) is determined
only by the processes of the extended collisions of quasiparticles~\cite{SI} and, thus,
is finite even in an almost ideal Fermi gas where the interaction parameter $r_s$ is small, $r_s \ll 1$.
 The retarded non-linear corrections~$\widetilde{\delta F} _{2,ij}^{(1)}$
  to the Landau constant $F _{2}$, which depends on the properties of the flow via the parameters $a$-$d$ within
  our model [see Eq.~(S37)],
  strongly varies with the strength of the interelectron interaction, expressed via the interaction
   parameter~$ r_s = ( \pi \, n_0)^{-1/2}/a_B$.
     In this way, the relaxational or the  elastic contribution in photoresistance
     can dominate,
  even if the plasmonic component is absent, in samples with different, relatively  small or relatively large,
  interaction parameter~$r_s$.

  Indeed, in experiment~\cite{Ganichev_1}, in which the studied samples had the relatively large electron densities,
   $n_0 \approx  10^{12}$~cm$^{-2}$, where the interaction parameter $r_s$ is relatively small, $r_s\approx 0.5$, 
    and the value $F_2$ and the  correction~$\widetilde{\delta F} _{2,ij}^{(1)}$ is also expected to be small,
   the magnetoconductivity exhibits only the properties corresponding
  to the relaxational contribution within our model.
   On the other hand, in experiments~\cite{Smet_1,exp_GaAs_ac_1,exp_GaAs_ac_2}, in which the giant peak
  in photoresponce near $\omega_c  = \omega /2 $ is more or less pronounced [see Figs.~3($b2$) and 4($b$,$c$)],
  the samples has the 2D electron densities in the diapason from
 $ 2\cdot 10^{11}$~cm$^{-2}$ to $ 3 \cdot 10^{11}$~cm$^{-2}$,
   those corresponds to the
 larger values of the parameter $r_s$, thus the correction~$\widetilde{\delta F} _{2,ij}^{(1)}$ is expected to be larger, 
  than in samples studied in Ref.~\cite{Ganichev_1},
   where~$n_0 \approx 10^{12}$~cm$^{-2}$.

\section{ 7. Conclusion }
We have proposed and developed a simple phenomenological theory of non-linear magnetotransport in the  highly correlated
2D electron  fluid. The $2\omega_c$-resonance in the viscosity coefficient  and the ac memory effects in the  inter-particle
 interaction are crucial in   the theory.
 The sufficient conditions of the applicability of the developed theory is the  large magnitudes of Landau interaction
parameters for the considered 2D electron systems.   The calculated photoconductivity exhibits an independence on the helicity
of the polarization  of incident radiation,  irregular  magnetooscillations, and a large peak at the doubled cyclotron frequency.
 Although the most general
 conditions of applicability of the developed theory have not been fully understood now,  observation of all these effects
 on  best-quality   GaAs quantum wells
 can be the evidence
 of the formation  of a highly correlated 2D viscous electron fluid in these~structures
 and
 of suitability of our model (or it close analog)  for  the description of such~fluid.

\begin{acknowledgments}
We sincerely thank M.~I.~Dyakonov  for numerous discussions of transport  experiments  on high-quality  2D electron systems, those led to
this work, as well as for discussions   of some of the issues raised in this work,  for reading the preliminary version 
 of the article,
for advice and   support. We are very grateful to  \mbox{Y.~M.~Beltukov}, \mbox{I.~S.~Burmistrov}, and  \mbox{A.~P.~Dmitriev} 
  for many fruitful  discussions.  One of us (\mbox{P.~S.~A.}) sincerely thanks
\mbox{A.~N.~Afanasiev}, \mbox{E.~G.~Alekseeva}, \mbox{I.~P.~Alekseeva}, \mbox{N.~S.~Averkiev}, 
 \mbox{K.~A.~Baryshnikov}, \mbox{K.~S.~Denisov}, \mbox{M.~M.~Glazov}, 
\mbox{I.~V.~Krainov}, \mbox{S.~M.~Postolov}, \mbox{M.~A.~Semina}, and~\mbox{S.~A.~Tarasenko}  for advice and~support.

This work  was partially financially supported by  the Russian Science Foundation, Grant~18-72-10111
[development of the general concept of the model,
solution of its equations for description of photo-magneto-resistance of the Poiseuille flow,
comparison of theory and experiments, those are presented
Sections~2,~4-6 of the main text and in Sections~3-8 of Supplemental material].
One of us (P.~S.~A.) is grateful to the Foundation
for the Advancement of Theoretical Physics and Mathematics
``BASIS'' (Grant~23-1-2-25-1) for financial
support of a part of this work [description of the retarded contribution in the relaxation
of the shear stress, which is presented
in Section~3 of the main text and in Section~2 of Supplemental material].

\end{acknowledgments}

\clearpage

\setcounter{equation}{0}
\setcounter{figure}{0}

\renewcommand{\thefigure}{S\arabic{figure}}
\renewcommand{\thesection}{S\Roman{section}}
\renewcommand{\theequation}{S\arabic{equation}}

\setcounter{equation}{0}
\setcounter{figure}{0}

\renewcommand{\thefigure}{S\arabic{figure}}
\renewcommand{\thesection}{S\Roman{section}}
\renewcommand{\theequation}{S\arabic{equation}}

\onecolumngrid
\begin{center}

{\Large  {\bf Supplemental material   to the article
 ``Highly correlated two-dimensional viscous electron fluid    in moderate magnetic fields'' }
\linebreak
}

{
\large P. S. Alekseev and A. P. Alekseeva
\linebreak \linebreak}
{\small
 Ioffe  Institute, Politekhnicheskaya 26,
  194021,   St.~Petersburg,   Russia
\linebreak
}
\end{center}

{\small
Here we present the details of our theoretical model of the highly correlated electron fluid, of the solution of its equations, and of derivation of  the properties of the
resulting  photoconductivity.  We discuss possible extensions of our theoretical model to more general hydrodynamic-like  systems
 in disordered samples. We also  compare our results with preceding theoretical works and the related experiments of photo-magneto-transport in the best-quality
 GaAs quantum wells and graphene.
\linebreak
\linebreak
\linebreak}
\twocolumngrid

\subsection{ 1. Memory effects in scattering
 of 2D non-interacting electrons on localized defects }
Memory effects in classical ac magnetotransport of non-interacting 2D electrons  in disordered samples were considered,  for example,
in~Refs.~\cite{S1,S2,S3,Polyakov_et_al_classical_mem,S12,Beltukov_Dyakonov}.

For  the scattering of 2D electrons on defects with a localized potential (``impurities''), the memory effects in a perpendicular
 magnetic  field $\mathbf{B}$ are due to the appearance of  (i)   the electrons  not scattering  on  defects (their  trajectories are located
between the impurities) and of (ii) the so-called ``extended collisions''.    The last events consist of several returns of an electron
to the same impurity after a first scattering on it because of  cyclotron rotation (see Fig.~2 in Ref.~\cite{Beltukov_Dyakonov}).
Both these effects (i) and (ii) become substantial in the classically strong magnetic fields, when the electron mean free path relative
to the scattering   on defects becomes comparable with or   longer than the length of the cyclotron  circle~\cite{S1,S12}.

Within the approach of Ref.~\cite{Beltukov_Dyakonov},  such events are accounted in a phenomenological physically transparent way by
the retardation term~$-\hat{\Gamma } (t) \mathbf{V}(t-T)$  in the Drude-like  equation for the mean velocity $\mathbf{V}(t)$:
\begin{equation}
 \label{main_motion_eq_imp}
\begin{array}{c}
\displaystyle
  \frac{ \partial \mathbf{V}  }{ \partial t }
  = \frac{e }{m}  \, \mathbf{E} (t)   +  [\, \mathbf{V}  \times  \boldsymbol{ \omega_c}  \, ]
   -\qquad \qquad
   \\
   \\\displaystyle
   \qquad \qquad -
(1-P )\, \Big[ \,    \frac{\mathbf{V}}{\tau_{tr}}    +  \hat{\Gamma } (t)   \, \mathbf{V}(t-T) \, \Big]
    \:,
\end{array}
\end{equation}
where $e$ and $m$ are the electron charge and the electron mass,  the electric field $\mathbf{E}(t)$ can contain dc and ac components;
 $\boldsymbol{ \omega_c} =  \omega_c \mathbf{e}_z $ is the vector along the magnetic field, $\omega_c  = eB/(mc)$ is
the electron cyclotron frequency,   $\tau_{tr}$ is the momentum relaxation time due to the scattering on disorder in the absence
of the memory effect,
\begin{equation}
 \label{P}
 P = e^{-T/\tau_{q}^{def}}
\end{equation}
is the probability for an electron to make a full cyclotron rotation without collisions with defects, $\tau_{q}^{\,def}$ is the departure
scattering time due to collisions with defects,   $T=2 \pi /\omega_c$ is the cyclotron period,   and  $\hat{\Gamma } (t)$  is
the retarded relaxation tensor due to double extended collisions.

The tensor $\hat{\Gamma } (t)$   depends on the dynamics of individual electrons in the past, $t'<t$. At sufficiently weak  flows,
such dependence is accounted by a nonlinear contributions in $ \hat{ \Gamma } $  [being additional to the unperturbed value
$ \hat{ \Gamma }^{(0)} $ corresponding to the limit $E,V \to 0 $].  This contribution is  proportional
to the vector~$ \boldsymbol{  \Delta }$ characterizing  the deviation of the electron trajectories $\mathbf{r}_0(t)$
from the exact cyclotron circles:
\begin{equation}
 \label{Gamma_imp}
   \Gamma _{ij}  (\boldsymbol{\Delta}) = \Gamma _{ij}  ^{(0)}
   +
    \alpha \Delta^2 \delta _{ij} + \beta \Delta_i \Delta_j\:.
\end{equation}
Here  the coefficients  $\hat{\Gamma }  ^{(0)}$, $ \alpha$, and $\beta $ are the microscopic characteristics  of the electron gas
and the defects,  being proportional to the probability  $P$~(\ref{P}). The vector $\Delta $ for an Ohmic flow is the mismatch $\boldsymbol{\Delta}(t)
= \boldsymbol{ \varrho }(t) -  \boldsymbol{ \varrho }  (t-T)$   of the impact scattering parameters $\boldsymbol{ \varrho }  (t) =
\mathbf{r}_0(t) - \mathbf{R}_0 $ and   $\boldsymbol{ \varrho }  (t-T) =   \mathbf{r}_0(t-T) - \mathbf{R}_0 $  of an electron in
 two successive collisions with the same  defect [here $\mathbf{R}_0$ is the position of the defect center; it follows  from the above
 formulas that $\boldsymbol{\Delta}(t) =  \mathbf{r}_0(t) - \mathbf{r}_0(t-T) $]. Such  form of relation~(\ref{Gamma_imp})  of  the part
of the tensor~$ \hat{ \Gamma } $  related to the perturbation of the trajectories between collisions
is valid when   the size of defects is small:
$a \ll R_c $. Provided this inequality, the mismatch  $\boldsymbol{\Delta}(t) $ is directly related with the forces from the dc and
the ac electric fields $\mathbf{E}_0$  and $\mathbf{E}_1(t)$  acting on an electron during a cyclotron period $t-T<t'<t$.   Herewith
in the approximation linear by  $\mathbf{E}_0$  and  $\mathbf{E}_1(t)$, the corresponding  value of the mismatch:
\begin{equation}
     \boldsymbol{\Delta}(t)
     =
     \boldsymbol{\Delta}_0 +  \boldsymbol{\Delta}_1(t)
     \:,
\end{equation}
 does not depend on particular parameters of a trajectory~$\mathbf{r}_0(t)$.

Such model allowed to analytically calculate the photoconductivity of non-interacting 2D electrons in bulk samples~\cite{Beltukov_Dyakonov}.
The obtained magnetooscillations  of the photoconductivity  are induced by the described extended collisions and are similar  in
many properties to the ones of MIRO observed  in experiments.

\newpage

\subsection{ 2. Memory effects in relaxation due to interparticle scattering in highly viscous 2D electron fluid }
Below we construct  phenomenological dynamic equations for 2D interacting electrons forming a
viscous fluid in samples with no  defects (or, more generally,  in the sample regions where defects are not substantial)
  in classical magnetic fields.
    On  the one hand, these equations are based on the Landau Fermi-liquid model
 which is a simplest model accounting  strong many-particle correlation effects in electron systems.
On the other hand, the formulated  below equations contains similar memory terms to the ones presented in previous Section~1
 for non-interacting electrons in sample with localized defects.

For a viscous electron fluid, the motion equations describes  the mean hydrodynamic velocity $\mathbf{V}(\mathbf{r},t)$,
 the perturbed electron density $n(\mathbf{r},t)$ and the stress tensor
$\hat{\sigma}(\mathbf{r},t)$.  The hydrodynamic Navier-Stokes-like equations of evolution  of these values for a viscous 2D electron fluid
 without accounting for  the memory effects were formulated and derived in
 Refs.~\cite{el,el2,el3,Novikov,je_visc,vis_res_1,vis_res_2,Semiconductors,Alekseev_Dmitriev}.
      In a strongly non-ideal electron liquid, the viscoelastic effects (first of all, propagation of
  the  shear-stress waves)
become possible within those hydrodynamic equations, and such fluid can be   referred as highly viscous
 electron fluid~\cite{LL7}.
     For a high-frequency flow in such system, the Navier-Stokes-like
equations are  transformed into  Hooke's equations of an almost elastic dynamics of an electron media with a weak damping due to
the interparticle scattering~\cite{el,Semiconductors}.

In Hooke's equations, the only physical value
 that characterizes the state of the system is
the displacement vector $\mathbf{u}(\mathbf{r},t) $, related to the strain  tensor $\varepsilon_{ij} (\mathbf{r},t)$:
\begin{equation}
 \varepsilon_{ij}
   =  \frac{ \partial  u_i } {\partial x_j }  +
    \frac{ \partial  u_j }{ \partial x_i}
 \:.
 \end{equation}
The connection between Hooke's and the Navier-Stokes equations  is based on the following relation for the hydrodynamic  velocity in an
almost fully elastic  flow:
\begin{equation}
\label{conn_s}
 \mathbf{V}(\mathbf{r},t)
   =
 \frac{ \partial \mathbf{u}(\mathbf{r},t) }{\partial  t}
 \:.
\end{equation}
This formula reflects the absence of slipping between   neighbor layers of the flow.

Generally, the dynamics of a viscous  fluid with taking into account the interparticle scattering, resulting in  slipping between
fluid layers,  is described by all the three  independent variables:  $n$,   $\mathbf{V}$, and~$ \sigma _ {ij} = - \Pi_{ij} $.
 Equation~(\ref{conn_s}) in this case is a formula for calculation, in the linear by
$\mathbf{V}$ approximation, of the displacement $\mathbf{u}(\mathbf{r},t)$   of fluid elements.
Note that the value $\mathbf{u}$ for a flow with
 the slipping of neighbour layers is  not the proper variable sufficient for  the description of  the fluid dynamics.

Analogously to the dynamics of independent electrons in a sample with small-size defects, the memory effects for the electron fluid
in a defectless sample in a magnetic field  are due to the ``extended collisions'' between electron-like Fermi-liquid quasiparticles
[see Fig.~S1(a) and Fig.~1 in the main text].   Indeed, if the mean free path of quasiparticles relative to their collisions
is of the order of the length of the cyclotron circle $2 \pi R_c$, two quasiparticles can  suffer
 several successive  collisions with slow  subsequent changes of their relative impact scattering parameter  [see Fig.~S1(a)]:
$
   \boldsymbol {\varrho} (t') =
    \mathbf{r} _{0,1} (t') - \mathbf{r} _{0,2} (t')
   \, $, $
t' \approx t   - N_rT
  \, .
$
  Here  $N_r=1,2,..$ is a number of successive  rotations in an extended collision.

 \begin{figure}[t!]
\centerline{\includegraphics[width=.8\linewidth]{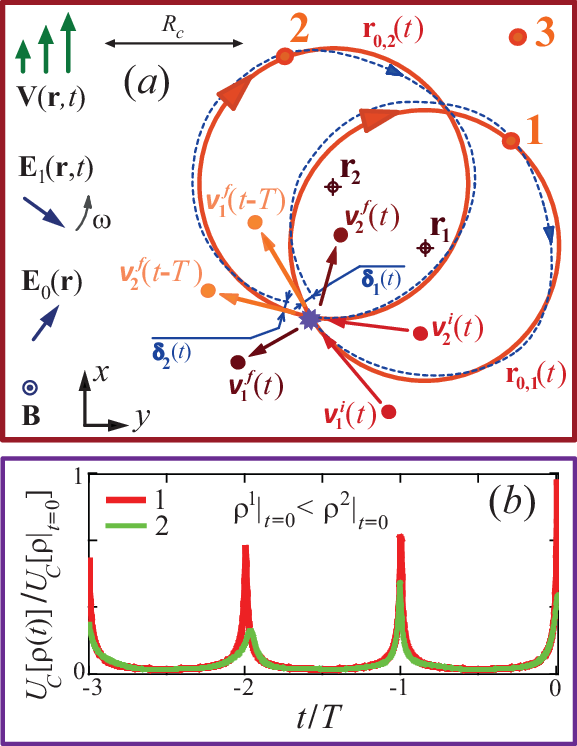}}
\caption{
($a$) Sketch of the microscopic mechanism of the memory effect in collisions of electron-like quasiparticles in a 2D  electron Fermi-liquid
 in a classical magnetic field $\mathbf{B} = B \mathbf{e}_z$. Motion of the fluid, averaged by dynamics of quasiparticles,
is described   by the hydrodynamic velocity $\mathbf{V} (\mathbf{r},t) $.  It contains the components which are linear by the
 applied electric fields, $\mathbf{V}_0 (\mathbf{r})   \sim  \mathbf{E}_0 $    and   $ \mathbf{V}_1 (\mathbf{r},t) \approx
\partial \mathbf{u} (\mathbf{r},t) / \partial  t    \sim \mathbf{E}_1(t) $, as well as the  nonlinear by $\mathbf{E}_{0}$ and $\mathbf{E}_{1} (t)$
components.  Red solid lines are the quasiparticle trajectories at  $\mathbf{E}_0, \mathbf{E}_1 , \mathbf{V }\equiv 0 $,
which are  exact cyclotron circles. Blue dashed lines are the trajectories  accounting the elastic force, microscopically  related
to  space-dependent perturbations of   the quasiparticle energy spectrum and leading to the term $\sim V_{ij} (\mathbf{r},t)$ in
Eq.~(\ref{main_eq_gen_S}),  as well as to the forces from  the full dc and the ac fields~$\mathbf{E}_0(\mathbf{r},t) $ and~$ \mathbf{E}_1
 (\mathbf{r},t)  $~(\ref{f}).  The  profiles of $\mathbf{V}(\mathbf{r},t)$  and $\mathbf{u}(\mathbf{r},t)$ are changing  weakly
 by $\mathbf{r}$ on the scale of $R_c$. The value of the relative impact scattering parameter, $ \boldsymbol{\varrho } (t')=
  \mathbf{r}_{0,1}(t') - \mathbf{r}_{0,2}(t') $,     at the moment $ t' = t-T $ differ from its value
at  $t'=t$ on the difference  $\boldsymbol{\delta } (t)   = \boldsymbol{\delta }_{1}(t) - \boldsymbol{\delta }_{2}(t) $
 of the trajectory  mismatches by a period: $\boldsymbol{\delta }_{p}(t) = \mathbf{r}_{0,p}(t) - \mathbf{r}_{0,p}(t-T)  $,
 $p=1,2$. The values~$\boldsymbol{\delta} _p (t)$ are estimated as   the mean shifts of the whole fluid in the centers~$\mathbf{r}_p$
by a period: $ \boldsymbol{ \delta} _p (t) \sim [\mathbf{u}  (\mathbf{r}_p,t) - \mathbf{u} (\mathbf{r}_p,t-T ) ] $.
\\
$_{} \;\: (b$) The screened Coulomb energy $U_C( \varrho(t) )$ with the character screening radius $a_B$
of two probe interacting quasiparticles in the time moments $t<0$ preceding  the current
collision at $t=0$ at some nonzero ac field $\mathbf{E}_1 (t) $ with $\omega = 2.4 \, \omega_c$~(schematic).
 Green  and red  curves correspond to a larger and a smaller
distances $ \varrho |_{t=0} \sim a_B$ between the two electrons
 [the value $\boldsymbol{\varrho}(t) =\mathbf{r}_{0,1} (t) -\mathbf{r}_{0,2} (t)  $ is the two-particle
 scattering parameter, $a_B $ is the character radius of Debye screening in 2D case being the Bohr radius].
  It is seen that the interaction energy for a
 given pair of quasiparticles has maximums at the moments $t_{N}  = -N T$, where $N=0,1,2,$ and so on. The local maximums
$U_C( \varrho(t_N) )$ are nonmonotonic with the increase of $N$ due to the effect from the ac field $\mathbf{E}_1(t)$
in the electron trajectories~$ \mathbf{r}_{0,1/2} (t)$
 [see panel~(a)].
}
\end{figure}

Such events lead to the dependence of the relaxation rate  of the momentum flux,
 $  \partial \hat{\Pi}  / \partial  t$, in the moment $t $
on the characteristic of the fluid in the moments $t'= t - N_rT $ of previous  scattering  events.
The physical reason for the dependence of the fluid relaxation rate at the present moment~$t$ on the fluid characteristics
at previous moments, $t'<t$, consists in a rather long collisionless motion of quasiparticles before moment~$t$. Indeed, when describing
the motion of a fluid by a one-particle
  distribution function and the corresponding quantities~$V (\mathbf{r},t)$ and~$\Pi(\mathbf{r},t)$, their
average values, used in the theory, are actually formed after last collisions of quasiparticles, that is, at times $t'<t$, significantly
 preceding moment~$t$. Accordingly, in the interval $t - (N_r+1)T \lesssim t' \lesssim  t - N_r T $
 the evolution of the fluid is quasi-deterministic. Such motion type is described by
 the retarded terms resulting from averaging of many  deterministic electron trajectories.

The corresponding  phenomenological  dynamic equations  of the fluid, extending the viscoelastic equations from
  Refs.~\cite{el,Semiconductors} in order  to account  only double extended collisions  ($N_r=1$), can be
formulated as follows:
\begin{equation}
\label{main_eq_gen_S}
 \left\{
 \begin{array}{l}
 \displaystyle
   \frac{\partial n }{ \partial t }      +  n_0 \, \mathrm{div }\mathbf{V} = 0
   \\
   \\
   \displaystyle
    \frac{  \partial V_i  } {  \partial t  } \,  =\,
          \frac{ e }{m} \,  E_i (\mathbf{r},t)  + \omega_c \epsilon _{ikz} V_k - \frac{1}{m} \,
  \frac{\partial \Pi_{ij}  }{ \partial x_j }
\\
\\
 \displaystyle
    \frac{ \partial \Pi_{ij}  / \partial t }{  1 \: +\: \delta F_{2,ij} (\mathbf{r},t)/(1+ F_2 ) }
    \, = \, 2 \omega_c \,  \epsilon _{ikz} \Pi_{kj} \, -
\\\\
  \displaystyle
  - \frac{\Pi_{ij} }{ \tau_{2}  }
   - \frac{m(v_F^{\eta})^2}{4} \,  V_{ij}   - \Gamma _{ijkl}(\mathbf{r},t)  \,  \Pi_{kl} (\mathbf{r},t-T)
\end{array}
\right.
.
\end{equation}
Here   $ n_0 $ is the quasiparticle characteristic being analogous   to the equilibrium density  $n_0 = n-\delta n$ in the case
of a Fermi gas;
$e$ is the electron charge;  $m$ is the electron renormalized mass;
the electric field
 $\mathbf{E} (\mathbf{r},t)$ contains the  components from  dc and  ac  external fields induced by the electric bias and
 the microwave radiation as well as from the internal dc and ac field related to the non-equilibrium charge density
 $ e \, \delta n (\mathbf{r},t)$:  $ \mathbf{E} (\mathbf{r},t) = \mathbf{E}_0 (\mathbf{r}) +  \mathbf{E}_1 (\mathbf{r},t) $,
\begin{equation}
\label{f}
\begin{array}{c}
 \mathbf{E}_0 (\mathbf{r}) = \mathbf{E}_0 ^{ext} +  \mathbf{E}_{0}^{int} (\mathbf{r})
 \:,
 \\\\
 \mathbf{E}_1 (\mathbf{r},t) = \mathbf{E}_1  ^{ext}  (t) +  \mathbf{E}_{1}^{int} (\mathbf{r},t)
 \:;
 \end{array}
\end{equation}
 $\epsilon _{ikm}$ is the antisymmetric unit tensor; $V_{ij} =V_{ij}  (\mathbf{r},t) $ is the tensor of the gradients
 of the velocity $\mathbf{V}$:
\begin{equation}
V_{ij} =  \frac{ \partial V_i }{ \partial x_j } + \frac{ \partial V_j }{ \partial x_i  }
\: ;
\end{equation}
$\tau_{2}$  is the shear stress relaxation  time without the memory effects;
the parameter $v^{\eta}_F \gg v_F  $ defines the
 viscosity $\eta_0$ of a highly viscous fluid at zero magnetic field~\cite{vis_res_2}:
\begin{equation}
\eta_{\,0} = \eta|_{B=0} = \, ( v_F^\eta ) ^2
 \tau_2  / 4
 \:;
\end{equation}
$ \hat{\Gamma} _{ijkl} (\mathbf{r},t)$ is the tensor describing the retarded relaxation  of $\Pi_{ij}$ due to the extended collisions
[see Fig.~S1(a)]; $T=2\pi/\omega_c$ is the cyclotron  period; and $ \delta F _{2,ij}   ( \mathbf{r} ,t ) $  are the perturbations
 of the coefficient $F_2$ of the Landau function. The  values  $ \delta F _{2,ij}  $ describe the effect of the viscous motion
  of the fluid on the quasiparticle energy spectrum in a non-linear by $V_i$ and $P_{ik}$ approximation.

The Navier-Stokes-like non-stationary equations of the 2D electron liquid at a strong interparticle interaction without the memory effects
 were derived in Refs.~\cite{el,el2,el3} for zero magnetic field and  in Ref.~\cite{Semiconductors} for a nonzero
classical magnetic fields.  Those equation are Eqs.~(\ref{main_eq_gen_S}) with omitted  the $F_{2,ij} $- and $\Gamma_{ijkl}$-terms.
  Their criteria of applicability can be formulated as follows~\cite{vis_res_2,Semiconductors}:
\begin{equation}
 \label{cond}
 \begin{array}{c}
   F_{1} \gg  1    \, , \;\;\;\;  \mathrm{and  } \;\;\;\;   F_m \ll F_1  \;\;\; \mathrm{at} \;\;\;  m \geq 3
\,;
\\
 \\
R_c \ll \Delta x
\quad  \mathrm{or  }\quad
v_F/ \omega \ll \Delta x
\,,
\end{array}
\end{equation}
where   $F_m $ are the angular harmonics  of the Landau interaction function~$F_{\mathbf{p},\mathbf{p}'}$;  $ \Delta x $
is the characteristic spacescale  of flow inhomogeneities; $R_c = v_F/ \omega_c $ ia the cyclotron radius of electron-like
quasiparticles; and $v_F$ is the Fermi velocity. Provided conditions~(\ref{cond}) are fulfilled,
 the parameter $F_2$ can be moderate, small, or large.

In view of a phenomenological qualitative type of our theory,   we do not presented in Eq.~(\ref{main_eq_gen_S}) the exact Fermi-liquid
renormalizations of all the parameters $v_F^\eta$, $m$, $\omega_c$, $\tau_2$.   They were studied and presented
in Ref.~\cite{Semiconductors}
at the absence  of memory effects. In this connection,
we make some related simplifications:
do not distinguish between the two cyclotron frequencies in equations for $ \partial \mathbf{V} / \partial t$
and $ \partial \hat{\Pi} / \partial t $; make the change:
\begin{equation}
  \delta F_{2,ij} \, / \, (1 +F_2 )
  \; \to \;
   \delta F_{2,ij}
\:,
 \end{equation}
and keep only  the renornalization of the viscosity  amplitude via  $v_F^\eta  \gg v_F $ [as it is substantial for applicability
of Eqs.~(\ref{main_eq_gen_S})  for description of the shear  stress waves].

Apparently, the consideration of Refs.~\cite{vis_res_2,Semiconductors} and extending them equations~(\ref{main_eq_gen_S})
 are qualitatively valid also for a moderately non-ideal electron
 liquid, in which the Landau parameters $F_m$ are greater than  the critical values $F_{m,c} \sim 1 $, when the  shear-stress mode
becomes possible in zero magnetic field~\cite{Silin,Inti_et_al}.

In magnetic field,  the strong quantum many-particle correlation effects, namely a reconstruction   of the Fermi-liquid ground state
and quasiparticles of 2D interacting electrons,    should appear.    The electron systems becomes substantially different
from the conventional Fermi-liquid. This may be a quantum form of description of  interparticle pair correlations related
 to the extended
 collisions with classical mechanics picture [see Fig.~S1].    For example, the effects  of reconstruction of the ground and excited states
 for 2D electrons in quantizing magnetic fields were studied in
 Refs.~\cite{KFS,KFS2}.   Similar reconstruction  in moderate magnetic fields
  can change the structure of  the macroscopic Navier-Stokes-like equations
of the type of~(\ref{main_eq_gen_S}) and substantially   change the  applicability condition  $F_1 \gg 1  $~(\ref{cond}).

Finally, we note that the resulting formulas for the flow characteristics,  which we will obtain below from Eq.~(\ref{main_eq_gen_S}),
 remain reasonable even at $F_{1} \ll 1$. In this limit,
 all parameters in Eqs.~(\ref{main_eq_gen_S})  correspond to the Fermi gas,
 and their solutions varies on scales $\Delta  x$, which is
 the edge of applicability of
 any hydrodynamic-like equations:
\begin{equation}
    \Delta x \, \sim \, R_c \, \sim \, v_F/\omega \,.
\end{equation}
This may point out that,  even for weakly non-ideal Fermi gas, the  consideration based  on the distribution function of quasiparticles
 and solution of the kinetic-equation, at least at some flow parameters,
 leads to the results being similar to ones following from Eqs.~(\ref{main_eq_gen_S}).
 Indeed, for example, for stationary flows   in magnetic fields, electron hydrodynamics
 is realized at any inter-particle scattering rates and values of~$F_1$ in the bulk regions
  of sufficiently wide samples, $W\gg 2R_c$~\cite{a,Holder}.
Moreover, if a system does not contain objects with sizes smaller than the minimum microscopic length, say, $R_c$ or $l_2=v_F\tau_2$,
then the kinetic and hydrodynamic descriptions
often give qualitatively similar flow distributions with minimal scales  $R_c$ or $l_2$
(see discussion in Ref.~\cite{c} and references therein).

Below in this Section~2 we construct  the relaxational contribution to the memory effects,  described   by
the term $ \hat{\Gamma} _{ijkl} (\mathbf{r},t) \, \Pi_{kl} (\mathbf{r},t-T) $. The elastic   contribution to the memory effects,
 described   by the term $ \sim \delta F _{2,ij}  ( \mathbf{r} ,t)$,    will be studied  in next Section~3.

Analogously to the coefficients $\alpha$ and $\beta$ in Eq.~(\ref{Gamma_imp}),
  the memory coefficients  $\delta F_{2,ij} $ and $\Gamma_{ijkl}$ in Eq.~(\ref{main_eq_gen_S}) should be   proportional to
the probability    to make a cyclotron rotation of  a quasiparticle in magnetic field without collisions with other quasiparticles:
\begin{equation}
\label{P1}
     P = e^{ -T / \tau_q }
    \:,
\end{equation}
where the value  $\tau_q$  is  the interparticle  departure scattering time.  The shear scattering time~$ \tau_2 $ in Fermi systems
 is usually much longer  than the time $ \tau _ q $~\cite{Alekseev_Dmitriev,Novikov}, therefore there exists a wide diapason
of  magnetic fields:
\begin{equation}
 \label{ddd}
1/ \tau_2 \, \ll \, \omega_c  \, \lesssim    \, 2 \pi/ \tau _q
\,,
\end{equation}
where  the probability  $P$~(\ref{P1}) is not too close to unity, $(1-P)\sim 1 $. For such   $\omega_c  $, in rough approximation,
 we can replace the factor  $(1-P)   $ by unity [it   could appear in the memory terms in Eqs.~(\ref{main_eq_gen_S}) and
    describes the probability of any scattering for a particle during one cyclotron rotation; compare Eqs.~(\ref{main_motion_eq_imp})
 and  (\ref{main_eq_gen_S})].  Apparently, in diapason~(\ref{ddd}) the quantization of the density of states
and appearance of many returns, $N_r \gg 1$,  in extended collisions  (and/or  more complex not-pair correlation effects)
are not substantial.
Therefore, we will consider it as the range of applicability by magnetic field of the developed  theory of the fluid dynamics.

The characteristic radius $a_B$ of the interparticle interaction potential is much smaller than the cyclotron radius.
In this case, each  extended collision consists of two (or several) successive collisions of two quasiparticles in a small region
 [a blue star in Fig.~S1(a)]   and of almost collisionless  motions of these two  quasiparticles  far from the  region of scattering.
 This motion  is mainly determined by the magnetic field, but is also   substantially affected   by the  full electric field
$\mathbf{E}  (\mathbf{r},t)   $~(\ref{f})  as well as by the elastic force, related to the space-dependent change of
the quasiparticle energy  spectrum~\cite{el,Semiconductors}  and leading to the term
\begin{equation}
  - \, [ \, m ( v_F^\eta)^2  / \, 4\, ]\:  V_{ij} (\mathbf{r},t)
\end{equation}
in Eq.~(\ref{main_eq_gen_S}).  The elastic force and the  components $\mathbf{E}_{0}^{int} $ and $\mathbf{E}_{1}^{int} $ of
field~(\ref{f}) are inhomogeneous by $\mathbf{r}$ and are  determined by the formation of the flow in the whole sample. These inhomogeneities
lead to the difference of the full forces  acting on two quasiparticles  ``1'' and ``2'' [see Fig.~S1(a)]. This difference defines the dependence
of the probability of the extend collisions on the flow magnitude and shape in the current moment and in the past.  For example, the probability
to return for two quasiparticles is much greater for a stationary homogeneous flow than for a fast ac  flow, in which
the memory about positions of two quasiparticles can be  lost  with a large probability in one  cyclotron rotation [see Fig.~S1(a)].

To describe such memory effects quantitatively, one should use the Boltzmann-like kinetic equation with a generalized collision operator
of the inter-quasiparticle scattering, being nonlocal by time. Such collision operator, leading to the motion equation of the type
of Eqs.~(\ref{main_eq_gen_S}), could be derived within some sophisticated procedure from the classical or the quantum Liuville equations
for the interacting Fermi-liquid electron-like quasiparticles. Herewith some other retarded terms in the resulting effective transport
equations, except the term $ - \Gamma_{ijkl} (\mathbf{r},t) \Pi_{kl} (\mathbf{r},t-T)$  accounted in  Eq.~(\ref{main_eq_gen_S}),
may  be possible, for example:
$ A_{ij} E_{j}(\mathbf{r},t-T) $, $B_{ijkl} E_{j}(\mathbf{r},t-T) \Pi_{kl}(\mathbf{r},t-T)  $,
and so on~\cite{S4}. However we hope   that the term $ - \Gamma _{ijkl} (\mathbf{r},t) \Pi_{kl}(\mathbf{r},t-T)$ is sufficient to
account the main part of the memory effects in interparticle collisions and below describe its origin and estimate its magnitude.

Quantitatively,  the extended inter-particle collisions are characterized by the following values.
Two successive  collisions are controlled by the impact relative scattering parameter of particles~``1'' and ``2''  [see Fig.~S1(a)]:
\begin{equation}
 \label{rel_im_sc_par}
  \boldsymbol{\varrho}  (t') =  \mathbf{r}_{0,1} (t' ) - \mathbf{r}_{0,2} (t' ) \:, \quad  t' = t, \; t-T \:.
\end{equation}
During one cyclotron period $t-T < t' <t $, two scattered quasiparticles are moving along the trajectories $\mathbf{r}_{0,1} (t)$ and
$\mathbf{r}_ {0,2}(t)$  with the centers near the points  $\mathbf{r}_1$ and $\mathbf{r}_2$. Their  impact scattering parameter $  \boldsymbol
{\varrho}  (t')$~(\ref{rel_im_sc_par})  differs in the moments $t' = t$  and $t' = t-T $ on the value
$\boldsymbol{ \delta}  (t)  = \boldsymbol{ \delta} _1 (t) - \boldsymbol{ \delta} _2 (t)  $, where
$\boldsymbol{ \delta} _1 (t) $ and $\boldsymbol{ \delta} _2 (t)  $    are the mismatches of the
 trajectories $\mathbf{r}_{0,1} (t)$ and
  $\mathbf{r}_{0,2}(t)$ by a cyclotron period:
\begin{equation}
\label{differences_of_mismatches_of_the_quasiparticle_positions}
\begin{array}{c}
\displaystyle
 \boldsymbol{ \delta }(t) =
 \boldsymbol{\varrho}  (t) -  \boldsymbol{\varrho}  (t-T) = \boldsymbol{ \delta} _1 (t) - \boldsymbol{ \delta} _2 (t) \:,
 \\
 \\
 \boldsymbol{ \delta} _p (t) =  \mathbf{r}_{0,p} (t) - \mathbf{r}_{0,p} (t-T )\,, \quad p=1,2\,.
 \end{array}
\end{equation}
Nonzero values of the mismatches $ \boldsymbol{ \delta} _p (t) $ are  induced by the action of the elastic force $V_{ij} (\mathbf{r},t)$
 and the electric fields~(\ref{f}), as it was discussed above [see also  Fig.~S1(a)].

In accordance with the essence of the extended collisions, the tensor $ \Gamma _{ijkl} (\mathbf{r},t)$ in the retarded relaxation
 term in Eq.~(\ref{main_eq_gen_S})  is determined by  the mismatches
$\boldsymbol{ \delta }(t) $~(\ref{differences_of_mismatches_of_the_quasiparticle_positions})  of the scattering parameters~$ \boldsymbol{
\varrho}  (t')$~(\ref{rel_im_sc_par}), averaged  by all the quasiparticle pairs near the given point~$\mathbf{\mathbf{r}}$.  The
averaged  mismatch~$ \langle \boldsymbol{  \delta }  (t) \rangle $ is apparently  expressed via the  macroscopic  variables of the fluid in
the past, generally speaking, via all variables:  $n$, $\mathbf{V}$, $\hat{\Pi}$. Thus the  tensor  $ \Gamma _{ijkl} (\mathbf{r},t)$
 depends on   these  variables as an operator:
\begin{equation}
    \label{}
     \Gamma _{ijkl} (\mathbf{r},t) = \hat{\Gamma} _{ijkl} [n(\mathbf{r}',t'),\mathbf{V}(\mathbf{r}',t'),\hat{\Pi}(\mathbf{r}',t')]
     \:,
\end{equation}
where $ |\mathbf{r}-\mathbf{r}'| \lesssim R_c $  and $t-T\lesssim t'< t$.

We consider high-frequency regime when that interparticle are rare as compared with cyclotron period: $\omega,\omega_c \gg 1/\tau_2$.
 In this case, a collision of one of two particles during  an extended collision with a third particle  is a relatively rare event, thus $P \sim 1 $.
  Thus the only  macroscopic variable determining the mismatches $\boldsymbol{\delta}_1(t)$ and   $\boldsymbol{\delta}_2(t)$ is the
displacement vector $\mathbf{u} (\mathbf{r}',t')$  in  the centers $\mathbf{r}' = \mathbf{r}_{1}, \, \mathbf{r}_{2} $ of the particle trajectories~``1'' and~``2'' (see Fig.~S1):
\begin{equation}
  \boldsymbol{\delta}_p(t)
  \sim  \mathbf{u} (\mathbf{r}_p,t) - \mathbf{u} (\mathbf{r}_p,t-T)  \:, \quad p=1,2
  \:,
\end{equation}
where $\mathbf{r}_p$ are the centers of the two neighbor unperturbed trajectories.
For   the  resulting dependence of $\hat{\Gamma}$ on the shape and the magnitude of the flow  we have:
\begin{equation}
    \label{Gamma_u}
 \begin{array}{c}
  \hat{\Gamma} _{ijkl} [ \mathbf{u} (\mathbf{r}',t') ]
 =  \Gamma _{ijkl} \Big  (    \big \langle  \, \boldsymbol {\delta u}  (\mathbf{r}_1 , \mathbf{r}_2,t) \,
 \big  \rangle _{\mathbf{r}_1 , \mathbf{r}_2 }  \Big)
  \end{array}
 \:,
 \end{equation}
where the brackets  $\langle   \, . \, \rangle _{\mathbf{r}_1 , \mathbf{r}_2 } $ denote  the averaging by  $ |\mathbf{r}-\mathbf{r}_p|
\lesssim R_c $, $p=1,2$, and the vector $\boldsymbol {\delta u} $ is:
\begin{equation}
\label{expr}
     \begin{array}{c}
  \boldsymbol {\delta u} (\mathbf{r}_1 , \mathbf{r}_2,t) =
   [\mathbf{u}( \mathbf{r}_1,t)-\mathbf{u}( \mathbf{r}_1,t-T)]
 -      \\
     \\
          -
     [\mathbf{u}( \mathbf{r}_2,t)-\mathbf{u}( \mathbf{r}_2,t-T) ]
   \end{array}
 \:.
 \end{equation}
Here the displacement $\mathbf{u} (\mathbf{r},t)$   can consist of the ac   near-elastic as well as the  dc  dissipative parts, both
related  to the ac and  dc components of  $\mathbf{V} (\mathbf{r},t)$ by Eq.~(\ref{conn_s}).

In the hydrodynamic regime, the characteristic spacescale of the inhomogeneity of the flow, $W$,  is
much larger than $R_c$, therefore we have:
\begin{equation}
    \label{d_u}
     u_m(\mathbf{r}_1,t') - u_m(\mathbf{r}_2,t') \approx   (x_{1 n} - x_{2 n}) \,
       \frac{ \partial u_m(\mathbf{r}_1,t') }{ \partial x_n }
    \:,
\end{equation}
where the summation by $n$ is supposed. Thus the expression $\boldsymbol {\delta u}  (\mathbf{r}_1 , \mathbf{r}_2,t) $~(\ref{expr}),
averaged by $\mathbf{r}_{1}$ and $\mathbf{r}_{2}$,     is expressed via
  the shift (the ``mismatch'') $\Delta_{mn}(\mathbf{r},t)$  of the strain
 tensor $\varepsilon_{mn}(\mathbf{r},t)$ in a cyclotron period. For the resulting $\hat{\Gamma}$ we obtain:
\begin{equation}
\label{Gamma_gen}
\begin{array}{c}
\hat{\Gamma} _{ijkl} [ \, \mathbf{u}(\mathbf{r}',t') \,] (\mathbf{r},t)
 =
\Gamma _{ijkl} \big(  \, \Delta _{mn} (\mathbf{r},t)\, \big)
\,,\;\;
\\
\\
 \Delta _{mn} (\mathbf{r},t)  = \varepsilon _{mn}  (\mathbf{r},t)  - \varepsilon _{mn}  (\mathbf{r},t-T)\:.
\end{array}
\end{equation}
Expansion of $\Gamma _{ijkl}$ by  $\Delta _{mn}$ up to the quadratic order is:
\begin{equation}
\label{Gamma_ex_app}
\begin{array}{c}
\displaystyle
\Gamma _{ijkl} (  \Delta _{mn})
= \Gamma ^{(0)}_{ijkl}
  \, + \,  \alpha^{mnop} _{ijkl } \, \Delta _{mn} \, \Delta_{op}
     \:,
\end{array}
\end{equation}
where the tensors $\Gamma ^{(0)}_{ijkl}$ and $\alpha^{mnop} _{ijkl }$ are determined by the parameters  of the fluid and depend
 on temperature and magnetic field.  The  linear term $ a^{mn}_{ijkl } \, \Delta _{mn}$ is absent in Eq.~(\ref{Gamma_ex_app}) due to
the symmetry of the relaxation  processes relative to the inversion  of the direction  of flows.

From Eq.~(\ref{conn_s}) we obtain the expression for the mismatch of the strain tensor via the velocity gradient~$V_{mn} (\mathbf{r},t')$:
\begin{equation}
   \label{Delta_ij}
 \Delta _{mn} (\mathbf{r},t) \approx
 \int _{t-T} ^t dt' \; V_{mn} (\mathbf{r}, t'  \, )
 \:.
 \end{equation}
Here, similarly to Eqs.~(\ref{f}), the tensor  $V_{mn}   $ can contain both  the ac component corresponding to an almost elastic  dynamics
 and the dc component corresponding to a dissipative viscous motion.

One should expect that the tensors $\Gamma ^{(0)}_{ijkl} $ and  $\alpha^{mnop} _{ijkl }$ are proportional to the probability $P$~(\ref{P1}) [or, possibly,
$P^2$] to make a full rotation for a quasiparticles in a pair without collisions with other quasiparticles
and  to some interparticle   scattering rate, apparently,  the rate of the relaxation of the shear stress
 $ 1/\tau_2$.

According to Fig.~S1(a), the relative value of the correction to the probability of an extended collision from the deviation of
the trajectories $\mathbf{r}_{0,p} (t)$ from exact cyclotron circles is proportional to the squared
 ratio~$| \boldsymbol  {\delta }(t)|/a_B \ll 1$:
\begin{equation}
 \label{rel_v}
 \frac{
\Gamma_{iklm} -  \Gamma_{iklm}  ^{(0)}
}
{\Gamma_{iklm} } \, \sim  \,  \Big(\,  \frac{ | \boldsymbol  {\delta }(t) | }{ a_B } \,\Big)^2
 \,.
 \end{equation}
Here the  value $|\boldsymbol  {\delta }(t)| = |\boldsymbol  {\delta }_1(t) - \boldsymbol  {\delta }_2(t)|$
is the  characteristic magnitude of the difference  of the mismatches of the   two quasiparticles trajectories
near the point $\mathbf{r}$ at the moments $t$ and $t-T$. Its magnitude is [see the definition
 and the estimation of~$\boldsymbol  {\delta }(t)$
in Eqs.~(\ref{differences_of_mismatches_of_the_quasiparticle_positions})-(\ref{d_u})]:
\begin{equation}
 \label{es2}
  | \boldsymbol
    {\delta }(t)| \sim || \hat{\Delta }(\mathbf{r},t) || \, R_c
\end{equation}
 The value $ a_B$ in Eq.~(\ref{rel_v}) is the Bohr radius,
 being the estimate for  the size  of the interparticle interaction potential~$U_C(r)$
  (the Rytova-Keldysh screened   2D electrostatic potential).

In this way, from Eq.~(\ref{Delta_ij}) and the inequality $| \boldsymbol  {\delta }(t)|/a_B \ll 1$ we obtain of the condition
of applicability of
the memory terms:
\begin{equation}
\label{cr_E_Dx}
V_{char} T R_c / \Delta x \, \ll \, a_B
\,,
\end{equation}
where $V_{char} $ is the character flow amplitude and $\Delta x$ is its character  spacescale.
Equation~(\ref{cr_E_Dx}) is the condition  on the the electric fields~$E_{0,1}$, controlling the magnitude
of~$V_{char} $.

According to
the definition  of the tensors  $\Gamma ^{(0)}_{ijkl}$  and~$\alpha^{mnop} _{ijkl }$ in  Eq.~(\ref{Gamma_ex_app}), we obtain from
Eq.~(\ref{rel_v}),~(\ref{es2}):
\begin{equation}
\label{est_for_mem_tens}
 \begin{array}{c}
    \displaystyle
    \Gamma ^{(0)}_{ijkl} = \, A_{ijkl}
    \:
    \frac{ e^{-T/\tau_q} }{ \tau_2 }
    \:,
    \\
    \\
    \displaystyle
         \alpha^{mnop} _{ijkl } = \,  B^{mnop} _{ijkl }
         \:
 \Big( \, \frac{   \displaystyle R_c }{   \displaystyle a_B } \, \Big)^2
   \:
   \frac{ e^{-T/\tau_q} }{ \tau_2 }
    \:,
\end{array}
\end{equation}
where $A_{ijkl}  $ and $B^{mnop} _{ijkl } $ are the numeric constants independent on the fluid parameters.  We note that for the case of
 the electron scattering on soft impurities, analogous formulas  for the memory constants $ \alpha$ and $\beta $ in the tensor~(\ref{Gamma_imp})
were derived in Ref.~\cite{Beltukov_Dyakonov} and lead to the estimations of $ \alpha$ and $\beta $ analogous to Eq.~(\ref{est_for_mem_tens}).

Formulas~(\ref{est_for_mem_tens}) lead to a definite  dependence of the   retarded  relaxation terms on the magnetic field
(via $T $ and $R_c$) and on the temperature $T_e$ of the electron fluid via the times $\tau_2$ and $\tau_q$. As we mentioned above,
 the departure time $ \tau_q$ is usually much smaller than the stress relaxation time $\tau_2$,  $    \tau_q \ll \tau_2    $, but they
 has similar temperature dependencies  for 2D degenerate
electrons  at any strength of interparticle interaction~\cite{Novikov,Alekseev_Dmitriev}.  Up to the  logarithmic factor
$\ln(\varepsilon_F / T_e )$,  being weakly dependent on the temperature $T_e$, both the rates  $1/\tau_q$ and $1/\tau_2$ are
the quadratic functions of~$T_e$~\cite{Novikov,Alekseev_Dmitriev}:
\begin{equation}
 \label{tau_temp_dep}
    \frac{ \hbar }{ \tau_{q,2} } = C_{q,2} (T_e)  \, \frac{ T_e^2 }{ \varepsilon_F }
   \:.
\end{equation}
Here $\varepsilon_F$ is the Fermi energy of the electron fluid and $C_{q,2} (T_e) $ are the dimensionless coefficients, which
are  determined  by the interparticle  interaction  parameter $r_s$, related to the electron density $n_0$, and  can weakly depend on
temperature  via the logarithm  $\ln(\varepsilon_F / T_e )$ (see discussion in Ref.~\cite{Alekseev_Dmitriev}).

\newpage

\subsection{ 3. Memory effects in elastic part of interparticle interaction  in highly viscous 2D electron fluid }
 The memory effect due to the formation of pairs of quasiparticles can also lead to non-dissipative  ``elastic'' retarded terms
in Eqs.~(\ref{main_eq_gen_S}).   Microscopically, elastic retarded terms are related to a collisionless motion of quasiparticles
in pairs during several cyclotron periods, a formation of a statistical distribution of quasiparticles
before the current moment (after collisions preceding
the collisionless motion of pairs),
 and  the dependence of the energy of pairs, containing the long-range term
 $U_C(|\mathbf{r}_{0,1} (t)  - \mathbf{r}_{0,2} (t)  |)$ as well as the short-range terms with
 the Landau function~$ F_{ \mathbf{p}_{0,1} (t) , \mathbf{p}_{0,2} (t) }$, on the electric
 and stress fields  during the almost free rotation between collisions  [see Fig.~S1(a,b)].

In this section we develop a phenomenological description of this effect within
the Navier-Stokes-like hydrodynamic equations~(\ref{main_eq_gen_S}).  Namely, we construct
 the qualitative  form of the corrections:
\begin{equation}
    \delta F_{2,ij} (\mathbf{r},t ) = \delta F_{2,ij} [ \, \mathbf{V}(\mathbf{r} ' ,t ' )\,
    ,
     \,  \hat{\Pi } (\mathbf{r} ' ,t ' ) \, ]
    \: ,
\end{equation}
to the Landau parameter $F_2$,  appearing in Eqs.~(\ref{main_eq_gen_S}).

The energy of the interparticle interaction in the fluid and the resulting elastic forces
 depend on correlations
in the positions $ \mathbf{ r }_{0,\, p}(t)$ and  $ \mathbf{ v }_{0,\, p}(t)$  the velocities
 of all  quasiparticles with numbers $p$ in time and space.   Evolution of correlations between the positions
and velocities of particles  in pairs are controlled by the cyclotron rotation, leading to the approach  and the removal
of quasiparticles one from another. Such change of the distances between the two quasiparticles in a pair
 is accompanied by the increase and the decrease of the magnitudes of
 of the electrostatic long-range screened Coulomb interaction $U_C (|\mathbf{ r }_{0,\, 1}(t) - \mathbf{ r }_{0,\, 2}(t) |)$
  and  the Fermi-liquid
short-range many-particle interaction controlled by the Landau terms with
$F_{\mathbf{ p }_{0,\, 1}(t) , \,  \mathbf{ p }_{0,\, 2}(t) } = F_ { \mathbf{ p }_{0,\, 1}(t) - \mathbf{ p }_{0,\, 2}(t) }$.

 For a known evolution of the distributions and correlations of quasiparticles,
  the energy of the interparticle interaction,
 containing, in particular, the long-range term $\sum _{ p_1 , p_2 } U_C(|\mathbf{r}_{0,p_1} - \mathbf{r}_{0,p_2}  |)$,
  is determined at each time moment  $t$
  primarily by the pairs of  quasiparticles located
at minimal distances one  from each other:
\begin{equation}
   \min _ { {\small
   \begin{array}{c}
   \, \mathrm{all \, pairs} \, (p_1, p_2)
   ,
   \\
     t'<t \,
     \end{array}
     } }\,
    [ \, |\mathbf{r}_{0, \, p_1} (t')  - \mathbf{r}_{0, \, p_2} (t')  | \, ] \sim a_B
\: .
\end{equation}
At the moments of time $ t '= t- N_r T $
 these quasiparticles
approach each other at distances of the order of the radius of their interaction, $\sim a_B$. In Fig.~1(b)
we illustrate
the interaction  energy $U_C (|\mathbf{ r }_{0,\, 1}(t) - \mathbf{ r }_{0,\, 2}(t) |)$
 of two quasiparticles in a pair in retarded time moments.

 If during one period
many pairs of quasiparticles  did not destroyed due to collisions with ``third''  quasiparticles,  their configurations
in the moments of time  before  one or several cyclotron revolutions  ago, $t'=t-N_rT$,    are
 similar to the configuration at the current moment of time~$t$ [see Figs.~S1(a,b)].     So the interaction energy of
the quasiparticles, the correlation   between  which is important
at $t$,      depends mainly on the distributions of quasiparticles   at the moments  $t$
 and  $ t '= t- N_r T $ ($N_r  \geq 1 $), when also had especially closely approached one to another, on the distances $\sim a_B$.

 Within the Fermi-liquid theory for 2D electrons, the effect of the distribution of quasiparticles
  on their energy spectrum is described by the wavevector-dependent effective  Landau function,
  containing both the short-range as well as the long-range electrostatic contributions.
 We account the perturbation  of the elastic part of the interparticle interaction  via a nonlocal
(in time) perturbation of the second harmonic of the Landau function $ F_2 $, as
 it is related with the non-equlibrium shear stress related characteristics of the fluid. Indeed, the parameters $F_0$ and $F_1$
  determines compressibility
 and the quasiparticle mass in the Fermi liquid~\cite{LP},
 while the parameter $F_2$ determine the viscoelastic cyclotron frequency and the magnitude of the viscosity coefficients~\cite{Semiconductors}.

For simplicity, we consider only   the corrections to $F_2$ for  the flow in a long [see Fig.~1(a)
in the main text] and account only the   current-time and the retarded contributions with one recollision, $N_r=1$:
\begin{equation}
 \label{two_c_Fxxxy}
   \delta F _{2,xx/xy}
      =
   \delta F _{2,xx/xy} ^{(0)} + \delta F _{2,xx/xy} ^{(1)} \:,
\end{equation}
where the values $ \delta F _ {2, xx / xy} ^ {(0)} (y, t)  $ are expressed via the values  $ V (y, t) $
and~$ \Pi_ {xx/xy} (y, t ) $,  while the values $ \delta F _ {2, xx / xy} ^ {(1)} (y, t)  $ refer to  this values
in the retarded moment,  $t'=t-T$:  $ V ( y, t-T) $ and $ \Pi_ {xx/xy} (y, t-T) $.
 Here we imply that such perturbation of the parameter $F_2$ are relatively small:
 \begin{equation}
\big| \,  \delta F _ {2, xx / xy} ^ {(0), (1)} (y, t)  \big| \ll 1
  \, .
\end{equation}
The difference between the $ xx $- and $ xy $-components of the corrections in $ F_2 $  is due to a nonzero
shear stress leading to a breaking  of symmetry  of the Fermi surface and
the Landau function~$ F (\varphi) $  with respect to the angle between the particle momenta $\mathbf{p}_{1,2}$:
$ \varphi  =
 \angle (\mathbf{p}_{1} , \mathbf{p}_{2} )$.
Below we omit the current-time-contribution $  \delta F _{2,xx/xy} ^{(0)}$ in Eq.~(\ref{two_c_Fxxxy})
  as it does not lead to specific effects in the non-linear components of the dc flow
 and account only the nontrivial retarded term,
$ \delta F _ {2, xx / xy} ^ {(1)}  (y,t)$.

The power of thermal energy dissipated at a given point $y$ of the viscous fluid flow in the moment $t'$ is:
\begin{equation}
  \label{W_in_flow}
     \mathcal{W} (y,t') = - \, ( \, V_{ik} \, \Pi_ {ik}   \, ) \, |_{y,\,t'}\, / \, 2  \:,
\end{equation}
where summation over same indices is supposed.   The Landau functional of the total fluid energy and, thus,
 the Landau parameters $F_m$ are expressed via the integral of powers $\mathcal{W} ^k$.
Similarly, in a viscous inhomogeneous flow  the relationship between the total fluid  energy
and the inhomogeneous part of the distribution function   of quasiparticles, $ \delta f _{\mathbf{p}} $,
contains not only the linear and quadratic Fermi-liquid terms, but also high-order by $ \delta f _{\mathbf{p}} $ terms.
Correspondingly, the perturbation of the quasiparticles spectrum  as well as of the Landau parameters
 acquire correction  proportional  to  powers of $ \delta f _{\mathbf{p}} $,  therefore to the powers of
   $V_{ik} $, and~$ \Pi_ {ik} $.

By analogy with Eq.~(\ref{W_in_flow}), for the perturbation of the second harmonic of the Landau function $F_2$ in a Poiseuille flow,
 in which only the derivative $ \partial V_x / \partial y$ is nonzero, we write:
\begin{equation}
 \label{abcd}
 \begin{array}{c}
  \displaystyle
    \delta F _{2,xx} ^{(1)} ( y,t ) =   \Big[ (\, a  \:  \Pi_{xx}
                            + b \: \Pi_{xy}   \, ) \, \frac{\partial V }{\partial y } \Big]  \, \Big|_{\,y,\,t'=t -T}
   \, ,
   \\
   \\
   \displaystyle
   \delta F _{2,xy} ^{(1)}  ( y,t ) = \Big[ (\, c \:  \Pi_{xx}
                            + d \: \Pi_{xy}   \, ) \, \frac{\partial V }{\partial y } \Big] \, \Big|_{\, y,\,t'=t -T}
   \, .
   \end{array}
\end{equation}
In this work we do not construct the full form of the non-liner perturbations of the parameter~$F_2$, but
propose and use only the simplest relation~(\ref{abcd}) for a Poiseuille flow the presence of magnetic field.  Within
the proposed above mechanism of perturbations of the elastic part of the interparticle interaction energy due to
 formation of pairs,  the coefficients $ a, b, c, d $ are proportional to the probability~$P(B)$~[Eq.~(\ref{P})]
for a quasiparticle to make a complete cyclotron circle  without collisions with a third quasiparticle:
$  a, \, b, \, c, \,  d \, \sim \,  e ^ {- \, T \, /\,  \tau_q} $.

It is noteworthy that appearance of the coefficients $(1-\delta F_{2,ij}) $  near the derivative~$\partial \hat{\Pi}/  \partial t$
in Eq.~(\ref{main_eq_gen_S}) for a given flow frequency $\omega$ is effectively equivalent to the
renormalization of the coefficients  $2 \omega_c$   in the right-hand side of the equation for~$\partial \hat{\Pi}/  \partial t$.
 That is, the  memory effects  in the interparticle interaction energy  for an electron fluid is mainly
 a non-linear elastic (non-dissipative)  effect, being the flow-dependent shift  of the frequency $2 \omega_c$ of
the own  dynamic of the shear stress of the fluid:
\begin{equation}
   2 \omega_c
   \to
     2 \omega_c + \Delta \omega_{c,\,l}
   \,, \quad
  \Delta \omega_{c,\,l}   =\sum _l M _{lij} \,\delta  F_{2,ij}
\end{equation}
where the tensor $M _{lij} $ stems from the form of Eqs.~(\ref{main_eq_gen_S}) for a given flow.

For the just described  elastic memory effects, the inequality   $ | \boldsymbol  {\delta }(t)|/a_B \ll 1$,
which guarantee sufficiently small amplitudes of the change of the scattering parameter after one return in the extended collisions
 due to the external and internal field [see Figs.~S1(a,b)],
also should be fulfilled, thus the condition of applicability of the memory terms (\ref{cr_E_Dx}) also takes place.

 We emphasize that we have developed
   a ``minimal'' phenomenological  model of a non-linear dynamic of an electron fluid, in which some proper
    phenomenologically justified terms
 are accounted to describe photoresponse of the fluid. Others more complex types of interparticle correlations
 and  corresponding memory terms
 can be possible in macroscopic dynamic equations of the type of~Eqs.~(\ref{main_eq_gen_S}).

\newpage

\subsection{ 4.  Linear responses of highly viscous electron fluid \\  on dc and ac electric fields  }
In order to study the photoresistance effect within the proposed model,  first of all, we need to find the linear responses of the
fluid on a dc and an ac electric fields $\mathbf{E}_0$ and $\mathbf{E}_1(t)$.

Both these fields can be written as  $ \mathbf{E} ( t ) = \mathbf{E} ( \omega ) \,  e^{-i\omega t} + c.c. $ with $\omega = 0 $ and
$\omega \neq 0 $.  The linear responses should be calculated by linearized equations~(\ref{main_eq_gen_S})  applied
 to the $\omega  $ harmonics of the  hydrodynamic velocity $\mathbf{V} (\mathbf{r},t ) =\mathbf{V}(\mathbf{r},\omega) e^{-i\omega t}
+ c.c. $, the density perturbation $ \delta n  (\mathbf{r},t ) = \delta n  (\mathbf{r},\omega)  e^{-i\omega t} + c.c. $, and the
shear stress tensor $\hat{\Pi} (\mathbf{r},t) = \hat{\Pi} (\mathbf{r},\omega)  e^{-i\omega t} + c.c.$, corresponding to the harmonics
of~$ \mathbf{E} ( t )$.

First, we formulate the resulting equations for the amplitudes $ \mathbf{V} (\mathbf{r},\omega) \sim  \hat{\Pi} (\mathbf{r},\omega)
\sim \mathbf{E} (\omega)$ at a general geometry of a flow.

A solution of the last of equations~(\ref{main_eq_gen_S}) without the memory effect term $-\Gamma_{iklm} (\mathbf{r},t) \Pi_{lm}
(\mathbf{r}, t-T) $ leads to the following linear relation between the time harmonics   of  the  momentum  flux    $ \Pi_{ij} = \Pi_{ij}
(\mathbf{r},\omega)$ and of the gradients of the harmonics of the velocity $ V_{ij} ( \mathbf{r},\omega ) =  \partial V_i /
 \partial x_j  + \partial V_j  /\partial x_i $~\cite{vis_res_1}:
\begin{equation}
\label{conn_Pi_dVdx}
 \Pi _{ij} = m \,[  \, \eta_{x  x }  (\omega) \, V_{ij}
  + \epsilon _{ikz}  \eta_{xy}(\omega) \, V_{kj}  \,]\:,
\end{equation}
where  the ac viscosity coefficients $\eta_{xx} = \eta_{xx}(\omega) $ and $ \eta_{xy}= \eta_{xy}(\omega) $ have the form~\cite{vis_res_1}:
\begin{equation}
\label{eta_xx__eta_xy}
\begin{array}{c}
\eta_{xx}
\\
\eta_{xy}
\end{array}
\Big \}
 =
 \frac{(v^{\eta}_F)^2 \tau_2 /4}
   {1+ ( 4 \omega_c^2-\omega^2 ) \tau_2^2 -2 i \omega \tau_2 }
\, \Big\{
 \begin{array}{c}
1- i \omega \tau_2
\\
2 \omega _c \tau_2
\end{array} .
\end{equation}
The second of equations (\ref{main_eq_gen_S}) with $\hat{\Pi} (\mathbf{r},\omega)  $ from Eq.~(\ref{conn_Pi_dVdx}) is  transformed
into the Naiver-Stokes equation  for the harmonic of the velocity  $ \mathbf{V} = \mathbf{V}(\mathbf{r},\omega)$:
 \begin{equation}
\label{Naiver_Stokes}
 \begin{array}{c}
    \displaystyle
 -i \omega \mathbf{V} = \frac{e }{m} \, \mathbf{E} (\omega)
  + [ \,   \mathbf{V} \times  \boldsymbol{\omega } _c \, ]
 +
 \\
 \\
  \displaystyle
 + \, \eta_{xx} (\omega) \:  \Delta _L\mathbf{V}
 +  \, \eta_{xy}  (\omega)  \: \Delta_L \mathbf{V} \times \mathbf{e}_z
 \:,
 \end{array}
\end{equation}
where   $\Delta _L= \partial ^2 / \partial x ^2+ \partial ^2 / \partial y ^2 $ is the Laplace operator.

The role of  the retarded relaxation term in the last of equations~(\ref{main_eq_gen_S}) for the linear response is as follows. If we
choose  the main part $\hat{\Gamma}^{(0)}$ of the relaxation tensor $\hat{\Gamma}(\mathbf{r},t)$  in the simplest form:
\begin{equation}
   \label{tau_shtrix}
    \Gamma_{ijkl} ^{(0)} = \delta_{ik} \delta_{jl} \, \frac{1}{ \tau_2'} \:,
\end{equation}
we arrive to  a redefinition of the relaxation rate $1/\tau_2$ which enters Eq.~(\ref{eta_xx__eta_xy}) as compared with its value
in the absence   of the memory effects:
\begin{equation}
   \label{redef}
   \frac{ 1}{\tau_2}
    \, \to \,
   \frac{   1}{\tau_2 }
   \,  +\,
   \frac{  e^{2\pi i \omega /\omega_c } }{ \tau_2'}
\:.
\end{equation}
For the time $\tau_2 ' $ here we should use estimate~(\ref{est_for_mem_tens}): $1/ \tau_2' \sim e^{-T/\tau_q} / \tau_2$.

Further in this work  we do not take into  account  redefinition~(\ref{redef}) of $\tau_2$ by the two reasons.  First, in view of
estimate~(\ref{est_for_mem_tens}), the second term in this formula  is much smaller than the first one on the
factor~$ e^{-T / \tau_q} $, which is substantially  smaller  than unity at $\omega_c\sim \omega \gg 1/\tau_2, 1/\tau_q $. Second,
 our analysis shows  that redefinition~(\ref{redef}) leads only  to  some subtle distortion of
the sinusoidal dependence of a factor in the resulting photoconductivity on the ratio $\omega / \omega_c$. For the current work,
being  a first proposal of the hydrodynamic mechanism of MIRO, such detail  is not  substantial.

We remind that the parameter  $v^{\eta}_F$ in Eq.~(\ref{eta_xx__eta_xy})  for a strongly non-ideal electron Fermi liquid with
large Landau parameters, $F_m \gg 1 $,  is much greater than the actual Fermi velocity $v_F$~\cite{vis_res_2,Semiconductors}.

Second,  we use equations  (\ref{eta_xx__eta_xy}) and (\ref{Naiver_Stokes}) to calculate the the linear response of the electron
fluid  on the  applied dc and the ac   fields, $\mathbf{E}_0$ and $\mathbf{E}_1(t)$, in a Poiseuille-flow like geometry: a  long
straight sample with rough edges [see Fig.~1(a) in the main text].

We consider that the sample width $W $ is much  smaller than
 the characteristic plasmon  wavelength~$l_p$:
\begin{equation}
 \label{W}
 W \ll l_p
 \:, \qquad
 l_p = s/\omega
  \:.
\end{equation}
Here   we use the estimate of $l_p$ for a quantum well with a metallic gate. The value  $s$ is the plasmon velocity in such structure
which is usually much larger than $v_F$ and $v_F^\eta$. In defectless samples of such widths the dc and ac linear responses  are formed
mainly by  a dissipative viscous flow and by standing magnetosonic waves, respectively, whereas the plasmonic   contribution to the ac
 flow component is suppressed to the extent of the small parameter~$ v_F^\eta/s \ll 1 $~\cite{vis_res_2}.

In such flow geometry, only  the $x$ component of the velocity $\mathbf{V}$ and  the $xy$ component of the velocity gradient tensor~$V_{ij}$ are
substantial in both the ac and dc components~\cite{vis_res_2}.  For the ac flow component, the density perturbation $\delta n (y,t) $ and the $y$
component of the velocity  $V_{y,1} (y,t) $ are relatively small quantities being proportional to  $ (v_F^\eta/s)^2 \ll 1 $~\cite{vis_res_2}.
However, such $\delta n $  determines  the not small ac  component  $ E^{int}_{y,1}   (y,t)  $ of the internal Hall field, which screens
the $y$ component of the radiation fields $E_{y,1}(t)$ [see Eqs.~(\ref{f})]. For the dc flow component, $ V _{y,0} \equiv 0 $ and
the density perturbation $\delta n (y) $ determines the dc part of the Hall field $ E^{int}_{y,0} (y) $.

Thus, in order to find the velocity $\mathbf{V} = \mathbf{V}  _0 + \mathbf{V} _1 $  we need only the $x$-component of
 equation~(\ref{Naiver_Stokes}). For each component $\mathbf{V}  _0 $ and $  \mathbf{V} _1 $  it takes the form:
\begin{equation}
  \label{Naiver_Stokes_Puassel}
    \displaystyle
 -i \omega V = \frac{e }{m} \, E_x (\omega)
 + \, \eta_{xx} (\omega) \:  \frac{ d^2 V }{dy^2}
 \:.
 \end{equation}
The both fields  $ E^{int}_{y,0} (y) $  and $ E^{int}_{y,1} (y,t) $ are calculated from   the $y$-components of
equation~(\ref{Naiver_Stokes})~\cite{je_visc}.

The resulting response of the fluid in such sample on a dc electric field $\mathbf{E}_0 ^{ext} = E_0 \mathbf{e}_x$ is the dc Poiseuille
flow with a parabolic profile. From Eq.~(\ref{Naiver_Stokes_Puassel}) at zero frequency, $\omega = 0$, with the diffusive boundary
conditions:
\begin{equation}
  \label{z}
   V \big|_{y= \pm W /2 } = 0
   \:,
\end{equation}
we obtain for the dc component $V _0 $  of the velocity $V  $~\cite{je_visc}:
\begin{equation}
  \label{V_0}
   V_0 (y)= \frac{eE_0}{2 m   \eta _{xx} } \,  \Big[ \Big(\frac{W}{2}\Big)^2 - y^2 \, \Big ]
  \:,
\end{equation}
where the dc diagonal viscosity coefficient is:
\begin{equation}
\label{e}
\eta _{xx}  =  \eta _{xx}  (0)=  \frac{ ( v_F^{\eta} ) ^2 \tau_{2} / 4 }  {1+4 \omega_2 \tau_{2}^2 }
\:.
\end{equation}
The corresponding averaged sample conductivity:
\begin{equation}
\label{I}
  \sigma _0 = \frac{ I_0}{E_0W } \, ,  \quad   \;\; I_0 = e n_0 \int _{-W/2}^{W/2} d y  \, V_0 (y)
  \:,
\end{equation}
and the averaged resistivity $ \varrho _0 = 1/\sigma _0 $ strongly depend on a magnetic field  via the viscosity coefficient $\eta_{xx} $.

 \begin{figure}[t!]
\centerline{\includegraphics[width=0.8\linewidth]{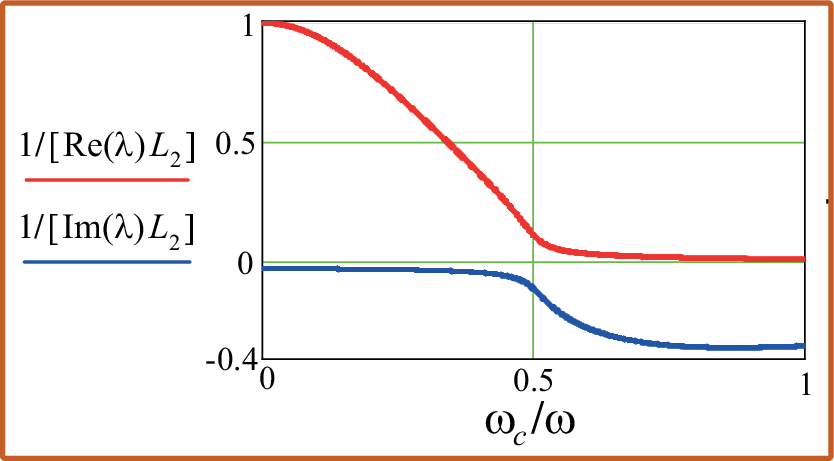}}
\caption{
Reciprocal real and imaginary parts of the eigenvalue $\lambda$, describing the decay length and   the wavelength of the transverse
magnetosonic waves. Thes values are plotted  as functions of the ratio $\omega_c/ \omega  \sim B $
  at a high   ac frequency, $   \omega \tau_2 = 20$. }
\end{figure}

Next, we calculate the linear  response of the fluid in the same long sample on the left and  the right ac circularly polarized radiation
fields:
\begin{equation}
  \label{r_f}
  \mathbf{E}^{ext} (t) =
   \mathbf{E}_{1} ^\pm e^{-i\omega t} +c.c. \:, \quad \mathbf{E}_1 ^\pm =
   \frac{E_1}{2}\left(
   \begin{array}{c}
   1
   \\
   \pm i
   \end{array}
   \right) \:.
\end{equation}
As in was mentioned above, at condition~(\ref{W}) the response $\mathbf{V}_1 =V_1 \mathbf{e}_x$ is  mainly  formed by the transverse magnetosonic
waves.  According to Eq.~(\ref{Naiver_Stokes_Puassel}), the resulting velocity is almost independent on the sign ``$\pm$'' of the circular polarization
of field~(\ref{r_f}):   $V^\pm_{1}(y,t)  \approx V_{1} (y,t)  $. From Eq.~(\ref{Naiver_Stokes_Puassel}) with the boundary condition~(\ref{z})
we obtain $ V_{1} (y,t) =  V_1 (y) \, e^{ -i \omega t } + c.c. $, where:
\begin{equation}
 \label{V_1}
   V_1 (y) = \frac{ e E_1  }{2 m  } \, \frac{i}{\omega}
     \, \Big [  \, 1- \frac{ \cosh( \lambda y ) }
           { \cosh( \lambda W / 2  ) } \, \Big  ]
 \:,
\end{equation}
and $\lambda = \lambda (\omega , \omega_c)$ is the eigenvalue corresponding to the transversal  magnetosonic waves. It is seen from
Eq.~(\ref{Naiver_Stokes_Puassel})  that:
\begin{equation}
\label{la}
  \lambda = \sqrt{  \frac{  - i \,\omega  \:\; }{ \eta_{xx}(\omega) } }
 \:\;.
\end{equation}
The reciprocal  imaginary  and  real parts of this eigenvalue,  $1/\mathrm{Im}  \, \lambda $  and $1/\mathrm{Re}  \, \lambda $,
 provide  the wavelength and the  decay length of the magnetosonic waves, respectively. Their dependencies on $\omega_c$ at a fixed
 frequency $ \omega \gg 1/\tau_2$ are drawn  in Fig.~S2.

Far above from the viscoelastic resonance  ($\omega  - 2\omega _c \gg 1/\tau_2$) we obtain from Ref.~(\ref{la}) the following estimates
for the length of decay and the wavelength of magnetosonic waves:
\begin{equation}
 \label{lambda_above}
    \frac{1}{ \mathrm{Re }  \, \lambda } \sim  L_s = v^\eta_F \tau_2
   \: ,\quad\quad
 \frac{1}{ \mathrm{Im }  \, \lambda  } \sim l_s = \frac{v_F^\eta }{ \omega }
  \:,
\end{equation}
which lead to the relation   $1/ \mathrm{Re} \lambda  \gg 1/ \mathrm{Im} \lambda  $. Thus, in this regime the solution $V_1(y,t) \sim e^{-
 i\omega t+ \lambda y}$   is well-formed weakly decaying waves.

Far below the viscoelastic resonance ($ 2\omega  _c- \omega  \gg 1/\tau_2$), visa versa, we have:
\begin{equation}
 \label{lambda_below}
      \frac{1}{  \mathrm{Re }  \, \lambda }
      \sim
      l_s \, , \qquad
      \frac{1}{  \mathrm{Im } \, \lambda }
      \sim
      L_s
     \:,
\end{equation}
thus $1/ \mathrm{Re} \lambda  \ll  1/ \mathrm{Im} \lambda  $. In this case the velocity profile $V_1(y,t)\sim e^{- i\omega t+ \lambda y} $
is formed by exponentially decaying  non-oscillating  eigenmodes, and the imaginary part of $\lambda $ is not substantial.

It follows from Eqs.~(\ref{V_1})-(\ref{lambda_above}) that above the resonance,  $  \omega  > 2 \omega_c $,   the standing magnetosonic waves
 $V_1(y,t) \sim e^{-i\omega t  + \lambda y }$  are formed. They  are localized in the whole sample    at the intermediate sample widths:
\begin{equation}
  l_s \ll W \ll L_s
  \:,
\end{equation}
and in the near-edge regions, $ W/2-|y| \lesssim L_s $, in the samples with the large widths (provided $L_s \ll l_p$):
\begin{equation}
  L_s \ll W \ll l_p
   \:.
\end{equation}
In the last case, the flow  in the central part of  the sample, $W/2-|y| \gg L_s $, is a trivial response, $V_1 \sim ieE_{1}^{\pm} /(2m\omega)$,
of a dissipativeless homogeneous electron  media on  $E_{1,x} (t)$ [see Eq.~(\ref{V_1})].

In the samples with the widths $W \gg l_s $ and at the frequencies below the resonance,  $ \omega  < 2 \omega_c $, the ac response $V_1(y,t)$
is located in the narrower near-edge regions, $ W/2-|y| \lesssim l_s $ and have non-oscillation exponential profile [see Eq.~(\ref{lambda_below})].

In the narrowest samples:
\begin{equation}
 \label{narrowest}
   W \ll l _s
  \:,
\end{equation}
as it follows from Eqs.~(\ref{V_1})-(\ref{lambda_below}),  the velocity profile $V_1(y)$ is parabolic  in both
the cases  $ \omega  > 2 \omega_c $
and $ \omega  < 2 \omega_c $, and its amplitude $V_1(0)$ decreases with the decrease of $W$ as $ \sim |\lambda |^2 W^2$.
We note that last equation~(\ref{narrowest}) is consistent with the condition $W= \Delta x \gg v_F /\omega $~(\ref{cond})
 of the macroscopic  hydrodynamic description of the flow as for a highly correlated fluid one should imply that $v_F^\eta \gg v_F$
 (see Section~2 and Refs.~\cite{Semiconductors}\cite{vis_res_2}).

In Fig.~S3 we plot the energy   absorbed by the Poiseuille flow $\mathcal{W} _{tot}
  = \int _{t-T} ^{t }   \int _{-W/2} ^{W/2 }  dt dy \mathcal{W}(y,t)$, where $\mathcal{W}(y,t)$
is given by Eq.~(\ref{W_in_flow}). The evolution of the flow described above with a change in the relations between the frequencies $\omega$
 and  $\omega_c $, on the one hand,   and the sample width $W $ and the eigenvalue $\lambda $, on the other hand, is reflected in the change
 in the dependence $\mathcal{W} _{tot}  (\omega_c ) $. In particular, oscillations in $\mathcal{W} _{tot}  (\omega_c ) $ at $\omega_c <\omega/2$ for
not very wide samples are associated with the appearance of standing magnetosonic waves.

Now we can calculate the shift (mismatch) of the $xy $ component of strain tensor, $ \Delta _{xy} (\mathbf{r} , t) $ in a
cyclotron period,  that enters the retardation relaxation term $- \Gamma _{ijkl}[\hat{\Delta } (\mathbf{r},t)] \Pi_{kl} (\mathbf{r},t-T)$
in the motion equation~(\ref{main_eq_gen_S}) for the  Poiseuille flow. In the presence of both the  dc  and ac fields
$\mathbf{E}_0$,  $\mathbf{E}_1 (t)$, as it takes place in the experiments of photoresistance, the shift $ \Delta _{xy} (\mathbf{r} , t)
 \equiv  \Delta (y , t) $ in the linear approximation  by $E_0$ and $E_1$ contains the two contributions:
\begin{equation}
  \Delta  (y,t) = \Delta_0 (y)+ \Delta _1 (y,t)
  \:,
\end{equation}
from the dc dissipative and the ac almost elastic  components of the linear response: $V_{lin}(y,t) = V_0(y) + V_1(y,t) $.
Equations~(\ref{Delta_ij}), (\ref{V_0}), and (\ref{V_1}) yield:
\begin{equation}
 \label{Delta_0}
   \Delta _0 (y) =
   \frac{ 2   \pi }{\omega_c} \,
   \frac{ d V_0 }{d y}
\end{equation}
and $\Delta _1 (y,t ) = \Delta _1 (y ) \,e^{ -i \omega t} + c.c.  $, where
\begin{equation}
\label{Delta_1}
 \Delta _1 (y )  = \frac{ i}{\omega } \:
  \Big[ \, 1- e^{ 2   \pi i \,  \omega /\omega_c } \, \Big]
     \: \frac{  d V_1 }{ d y }
   \:.
\end{equation}
It is noteworthy that this formula contain the factor $ e ^ { 2  \pi i \,  \omega / \omega_c  } $ being periodic by the reciprocal
 magnetic field.

 \begin{figure}[t!]
\centerline{\includegraphics[width=.95\linewidth]{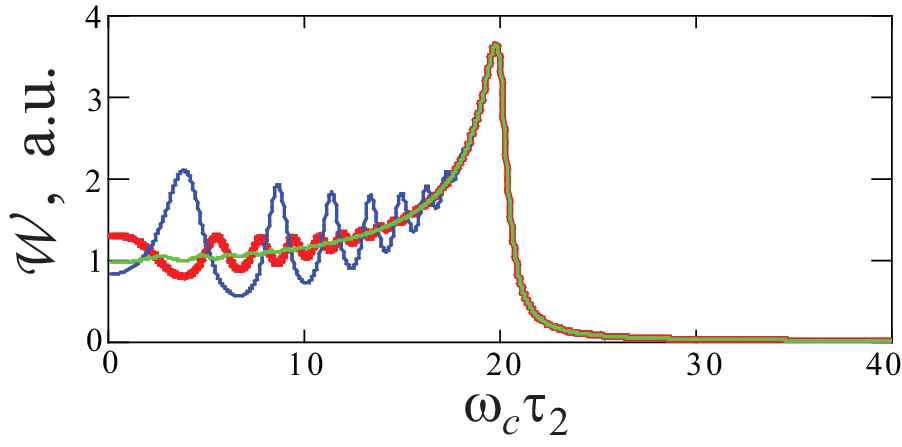}}
\caption{
Energy absorbed by the flow of a highly viscous electron fluid in a long sample from an incident high-frequency field as a function
 of magnetic field.  The calculation is made for the ac frequency  $\omega \tau_2 = 40$  and the sample widths  $ W/L_s  = 2, 4, 8$
for  blue, red, and green curves, respectively.
 }
\end{figure}

\newpage

\subsection{ 5. Non-linear response of fluid
  on dc and ac electric fields: memory effects
 in interparticle scattering}
In this section we calculate the nonlinear correction to the dc flow  component $V_0(y)$~(\ref{V_0}), induced by the ac component $V_1(y,t)$~(\ref{V_1})
 and the memory relaxation term, $-\Gamma_{ijkl } \Pi_{kl}$, in Eqs.~(\ref{main_eq_gen_S}), and the resulting mean
 photoconductivity~$\Delta  \delta \varrho _{\omega } $ of the sample.

We continue to consider the flow in a long sample  with the width $W$ which is much smaller than the characteristic plasmon wavelength
$l_p$ [see Eq.~(\ref{W})].  Similarly as for the linear component, only  the $x$ component of the velocity $\mathbf{V}$ and  the $xy$
components of the tensors $V_{ij}$ and  $\varepsilon _{ij}$ present  in the non-linear component [see Fig.~1(a) in the main text].
In such geometry, equations~(\ref{main_eq_gen_S})  take the simplified form:
 \begin{equation}
\label{main_eq_partic}
\left\{ \:
\begin{array}{l}
\displaystyle
  \frac{\partial V }{\partial  t } = \frac{e}{m}\, E  ^{ext} (t)
  - \frac{1}{m} \, \frac{ \partial  \Pi_{xy} }{ \partial y}
\\
\\
\displaystyle
\frac{ \partial  \Pi_{xx} }{ \partial t } = 2 \omega_c \Pi_{xy}
  - \frac{ \Pi_{xx} }{ \tau_{2} }
  -
   \\
\\
\displaystyle
 \qquad \qquad  \qquad \qquad \;
-   \Gamma  (\Delta)   \,  \Pi_{xx}(y, t-T)
\\
\\
\displaystyle
\frac{ \partial  \Pi_{xy} }{ \partial t } = - 2 \omega_c \Pi_{xx}
    - \frac{ \Pi_{xy} }{ \tau_{2}  } -
 \\
\\
\displaystyle
 \qquad \;\;
 - \,\,    \frac{m(v^{\eta}_F)^2}{4} \, \frac{\partial V }{\partial  y } \, -   \Gamma  (\Delta)
  \,   \Pi_{xy }(y,t-T)
\end{array}
  \right.
   ,
\end{equation}
where $ E ^{ext} (t) =E_0 + E_{x,1}(t)$ and we used the simplest form of  the retarded relaxation tensor $\Gamma_{ijkl}  =\Gamma \delta _{ik}
\delta _{jl} $, analogous  to Eq.~(\ref{tau_shtrix}). For the dependence of  $\Gamma $ on $\Delta$, according to Eq.~(\ref{Gamma_ex_app}),
 we write:
\begin{equation}
\label{Gamma_partic}
\Gamma (  \Delta)  =  1/\tau ' _{2}
    -  \alpha \, \Delta ^2
    \:, \qquad
    \alpha >0
\:.
\end{equation}
The positiveness of $\alpha$  corresponds to the decrease  of the probability  for a particle ``1'' to scatter again  on a particle ``2''  with
the increase  of the amplitudes  of  the velocity $ V (y,t)$, of the strain $ \varepsilon _{xy} (y,t)$, and, thus,  of the trajectories
mismatches~$\boldsymbol{\delta} _{1,2}(t)$  during a cyclotron period  (see Fig.~S1). Note that the same sign of the analogous parameters $\alpha $
and $\beta $ in the tensor~(\ref{Gamma_imp}) of the memory effect in the scattering of non-interacting electrons on defects  was obtained
in Ref.~\cite{Beltukov_Dyakonov} for weak localized scatters with a smooth potential.

The mismatch  $\Delta = \Delta (y,t)$  in Eqs.~(\ref{main_eq_partic}) and (\ref{Gamma_partic}) is expressed via the integral of
$ \partial V(y,t') / \partial y$ by Eq.~(\ref{Delta_ij}).

The photoconductivity and the photoresistance at small ac powers are given by the correction $I_\omega$  to the dc current $I_0$~(\ref{I}),
linear by the ac power $\mathcal{W }\sim E_1^2$. Thus, together with the linear responses   $I_\omega \sim  V_0  \sim  E_0 $ and $ V_1 \sim   E_1 $,
one needs to calculate the time-independent nonlinear component $V_\omega (y) $ of the  dc flow of the order of~$ \sim E_0E_1^2$:
\begin{equation}
 \label{V_series}
  V (y,t)= V_0(y) + V_1 (y,t)  + V_ \omega (y)  \: , \quad
\end{equation}
and the corresponding contribution $\hat{\Pi}_\omega  \sim E_0 E_1^2 $ to the momentum flux tensor:
\begin{equation}
 \label{Pi_series}
   \hat{\Pi}  (y,t)= \hat{\Pi}_0(y) + \hat{\Pi}_1 (y,t)
     + \hat{\Pi}_\omega  (y)
   \: .
\end{equation}
Here $\hat{\Pi}_0 $ and $\hat{\Pi}_1 $  are the linear contributions in  $\hat{\Pi}$, related to $V_0$ and $V_1$ by~Eq.~(\ref{conn_Pi_dVdx}).

The equations for $\hat{\Pi}_\omega $ and $V_ \omega $  are derived by  the substitution of the dc and the ac linear components, $\hat{\Pi}_0$,
$\Delta_0$ and   $\hat{\Pi}_1$,   $\Delta_1$,   into the nonlinear parts of the retarded relaxation terms in Eqs.~(\ref{main_eq_partic}) with
$\Gamma(\Delta)$~(\ref{Gamma_partic}) and by the integration of the resulting equations for $\partial \mathbf{V} / \partial  t $  and
 $ \partial \hat{\Pi} / \partial  t $ by one cyclotron period. After such procedure, we obtain:
\begin{equation}
\label{Pi_zv_main_eq0}
 \left\{ \:
\begin{array}{l}
\displaystyle
    \frac{ d \Pi  _{\omega  , xy}}{ d y }  =  0
\\
\\
\displaystyle
    \frac{ \Pi _{\omega , xx} }{ \tau_2 }
     - 2 \omega_c \Pi _{\omega ,  xy} =
   \\
\\
\displaystyle
 \qquad \qquad
    =\alpha \,
   \langle \,  \Delta ^2 (y,t) \, \Pi _{xx}(y,t-T) \, \rangle_{\omega }
\\
\\
\displaystyle
     2 \omega_c \Pi _{\omega ,   xx} + \frac{ \Pi _{\omega ,   xy} }{ \tau_2 }  =
    - \frac{m(v^\eta_F)^2}{4 } \frac{d V_ \omega  }{ d y} \,+
 \\
\\
\displaystyle
  \qquad \qquad
     + \, \alpha \,
   \langle \, \Delta ^2 (y,t) \, \Pi _{xy}(y,t-T) \, \rangle_{\omega }
\end{array}
 \right.
 .
\end{equation}
Here the angular brackets with the subscript ``$\omega$'' denote the operation of taking  the contribution of the order of  $\sim E_0E_1^2$,
averaged  by the last period of the ac field:
\begin{equation}
\label{br_om0}
    \langle f(t) \rangle_{\omega }
      =
      \frac{1}{T} \int _{t-T_\omega} ^ t d t' \: f_{\omega } (t')
   \:,
\end{equation}
where $T_\omega = 2\pi /\omega  $ is the period of the ac field and  $f_\omega (t) = f_{12}(t) E_0 E_1^2 $ denotes the term ``12'' in the decomposition
of the function $f(t)$ in power series by the amplitudes of the dc and the ac fields: $ f(t) = \sum _{m,n=1}^\infty  f_{mn}(t) E_0^mE_1^n $.

From the last two equations in system~(\ref{Pi_zv_main_eq0})   we obtain the expression for the nonlinear part $\hat{\Pi}_{\omega}$ of the
momentum flux tensor. In particular, for its  $xy$-component we have:
\begin{equation}
\label{Pi_star_res0}
\begin{array}{c}
\displaystyle
   \Pi  _{\omega , xy}  (y)= -m \, \eta_{xx} \frac{d V _{ \omega  } }{ dy}
    +
\\
\\
\displaystyle
    +
    \frac{\alpha \tau_2}{1+4\omega_c^2 \tau_2 ^2 } \,
      \big[\,
    \langle \, \Delta ^2 (y,t) \, \Pi _{xy}(y,t-T) \, \rangle_{\omega }
    -
\\
\\
\displaystyle
     \qquad \quad -
   \,
    2 \omega_c \tau_2  \langle \,  \Delta ^2 (y,t) \, \Pi _{xx}(y, t-T) \, \rangle_{\omega }
   \, \big]
\:.
\end{array}
\end{equation}
Here each nonlinear terms in the brackets  $\langle \:..\: \rangle_\omega $~(\ref{br_om0})  contain the two parts with different  combinations of
the  dc, $V_0 \sim \Pi_{0,ij} \sim \Delta_0 $,  and ac, $V_1 \sim \Pi_{1,ij} \sim \Delta_1$, linear responses:
\begin{equation}
\label{contrs_in av0}
\begin{array}{c}
    \langle  \, \Delta ^2 (y,t)  \, \Pi _{ij}(y,t-T)  \, \rangle_{\omega } =
     \\
     \\
      = \langle \,  \Delta _ 1 ^2 (y,t)  \, \rangle
   \: \Pi _{0, ij} (y)
   \, +
\qquad
\\
\\
 \qquad
      + \,
   \langle  \, \Delta _ 1  (y,t)  \: \Pi _{1, ij} (y,t-T)   \, \rangle
   \:   \Delta _0(y)
\:.
\end{array}
\end{equation}
Here $ij  \, = \, xx, \, xy$ and  the angular brackets without a subscript denotes the operation of  averaging over the last period of the ac
 field:
\begin{equation}
  \langle g(t) \rangle = \frac{1}{T} \int _{t-T_\omega} ^ t d t'
   \: g(t')
 \:.
\end{equation}

The linear components of the momentum flux tensor in Eq.~(\ref{contrs_in av0})  have the form [see Eq.~(\ref{conn_Pi_dVdx})]:
\begin{equation}
\label{first}
   \Pi_{0,(xx/xy)} (y) =- m \, \eta_{xy/xx}   \frac{ d V_{0} }{ dy }
\end{equation}
and
\begin{equation}
\label{second}
   \Pi_{1,(xx/xy)} (y,t) =- m\, \eta_{xy/xx}   \frac{ d V_{1} }{  d y  }
    \,  e^{-i\omega t } + c.c.
\:.
\end{equation}
In view of Eq.~(\ref{Delta_1}), the squared shift of the deformation~$\Delta _1^2 $ in Eq.~(\ref{contrs_in av0}), averaged by $t-T_\omega<t'<t$,
is:
\begin{equation}
\label{av_1}
 \langle \Delta _ 1 ^2 (y,t)  \rangle =
   \frac{4}{ \omega^2} \sin^2 \Big(\frac{\pi \omega }{\omega  _c }\Big)
    \:  \Big| \, \frac{dV_1}{dy} \Big|^2
   \:.
 \end{equation}
Due to Eqs.~(\ref{Delta_1}) and (\ref{second}), the expression under the angular brackets  in the last term in  Eq.~(\ref{contrs_in av0}),
averaged  by $t-T_\omega<t'<t$,  takes the form:
\begin{equation}
\label{av_20}
  \begin{array}{c}
\displaystyle
   \langle \, \Delta _ 1  (y,t)   \,\, \Pi _{1, (xx/xy)} (y,t-T)  \,  \rangle
     =
\\
\\
\displaystyle
=
\frac{ 2}{\omega} \: \mathrm{Re}  \big[ \,
  i\,   \eta_{xy/xx}(\omega)  \, \big( e^ {2 \pi i\omega /\omega_c } - 1
 \big)\,
\big]
   \,  \Big| \, \frac{dV_1}{dy} \Big|^2
   \:.
     \end{array}
\end{equation}

Substitution of formulas~(\ref{Pi_star_res0})-(\ref{av_20}) for $\Pi  _{\omega  , xy} $ into the first of equations~(\ref{Pi_zv_main_eq0}) leads
to the final  equation  for  the nonlinear part of the flow profile $ V_\omega  = V_\omega (y) $:
\begin{equation}
 \label{eq_for_v_zv_fin0}
    \frac{d ^2 V_\omega }{ d y^2} = G(y) \:,
\end{equation}
where $G(y) $ is given by formulas~(\ref{Delta_0}), (\ref{Delta_1}), and~(\ref{Pi_star_res0})-(\ref{av_20}). Equation~(\ref{eq_for_v_zv_fin0})
should be solved with the diffusive  boundary conditions on the component $V _ \omega (y)$: $ V _ \omega  | _ { y = \pm W/2 } = 0 $. The result
 takes the form:
\begin{equation}
\label{result_SI0}
  V_{\omega}(y)  \, = \, \frac{e ^3 E_0 E_1 ^2  \alpha }{ m^3 (v^\eta_F)^2}
  \,
   \frac{u + 2 \, \mathrm{Re} \,v }{\omega^4}
   \:  J(y) \:,
\end{equation}
where $J(y)$ is the dimensionless factor determining the profile of the nonlinear flow component:
\begin{equation}
  \label{J_y0}
  J(y) =
   \frac{| \lambda   |^2   }{|\cosh (\lambda W/2  ) |^2}
   \int  _{|y|} ^{W/2}   d \tilde{y}  \;  \tilde{y}
     \;|  \sinh (\lambda \tilde{y}  ) | ^2
\end{equation}
and the dimensionless  values $u$ and $v$ contain the dependence on the frequencies $\omega$ and $\omega_c$:
\begin{equation}
\label{u}
  u =
  4 \sin ^2 \Big( \frac{\pi \omega}{\omega _c }\Big)
 \, ( -1 + 4 \omega_c ^2 \tau_2^2 )
 \:,
\end{equation}
\begin{equation}
\label{v}
 \begin{array}{c}
\displaystyle
 v =
  \frac{ 2\pi \omega}{\omega _c}
   \Big[\sin\Big( \frac{2\pi \omega}{\omega _c } \Big)
  -
   2i  \sin^2 \Big( \frac{\pi \omega}{\omega _c } \Big)  \Big]
\times
\\
\\
\displaystyle
  \times
    \frac{(-1 + i \omega \tau_2 + 4 \omega_c ^2 \tau_2^2 )(1+ 4 \omega_c^2 \tau_2^2)}
    {1+ ( 4 \omega_c^2-\omega^2 ) \tau_2^2 -2 i \omega \tau_2 }
\:.
 \end{array}
\end{equation}

A direct calculation of the factor $J (y) $~(\ref{J_y0}) yields:
\begin{equation}
\label{J_res}
\begin{array}{c}
\displaystyle
  J (y)=
  \frac{\nu^2 + \mu^2}{ 8 \,  \nu^2 \, \mu^2 \,   \big| \cosh [\, (\mu - i\nu ) W/2\, ]\big |^2}
    \,\times
\\
\\
\displaystyle
  \times\,
  \Big\{ \, \mu^2   \big[
     - \cos( \nu W) + \cos(2 \nu y)   \, -
 \qquad \quad
  \\
\\
 \displaystyle
 -
  \,  \nu W  \sin(\nu W)   + 2 \nu y  \sin(2 \nu y) \, \big]
  +
 \quad
 \\
 \\
 \displaystyle
 \quad
+ \,  \nu^2
  \big[
 -\cosh( \mu W) + \cosh(2 \mu y) \, +
  \\
 \\
  \qquad  \quad
 +
  \: \mu W  \sinh( \mu W)  -  2ty  \sinh(2 \mu y)
 \, \big]   \, \Big\}
 \:.
\end{array}
\end{equation}
where  $\mu = \mathrm{Re} \, \lambda  $ and $ \nu = -\mathrm{Im} \, \lambda    $. It is seen from Fig.~S2 that both the values $\mu$  and $\nu$
are positive at any $\omega $ and $\omega _c$.    In the limiting cases of the narrow and the wide samples,  $W \ll \min(1/\mu,1/\nu)$ and
$ W \gg \max(1/\mu,1/\nu)$, we obtain from Eq.~(\ref{J_res}):
\begin{equation}
\begin{array}{c}
  \displaystyle
    J (y)= \frac{ ( \nu^2+\mu^2 ) ^2 }{4} \Big[ \Big( \frac{W}{2} \Big)^4  - y^4  \Big]
\end{array}
\end{equation}
and
\begin{equation}
\begin{array}{c}
\displaystyle
  J (y)= \frac{ W \,  ( \nu^2+\mu^2 ) }{ 4\mu }
 \,
  \big[ \,
   1   -   e^ {-2\,\mu\,( \, W/2 \, - \, |y| \, )}  \,
     \big]
 \:.
\end{array}
\end{equation}
In this way, in the very narrow samples the profile of the flow perturbation  $ V_{\omega}(y) $ is a fourth-order parabola, while in
the very  wide samples it is almost flat in the central region, $W/2-|y| \gg 1/\mu $, and exponentially goes to zero in the near-edge
regions,  $ W/2-|y| \lesssim 1/\mu $.  For the samples with the intermediate widths, $ \min(1/\mu,1/\nu) < W < \max(1/\mu,1/\nu)$, the
 component~$V_\omega(y)$ contains comparable  decaying  and oscillating parts [see Eq.~(\ref{J_res})].

  \begin{figure}[t!]
\centerline{\includegraphics[width=0.85 \linewidth]{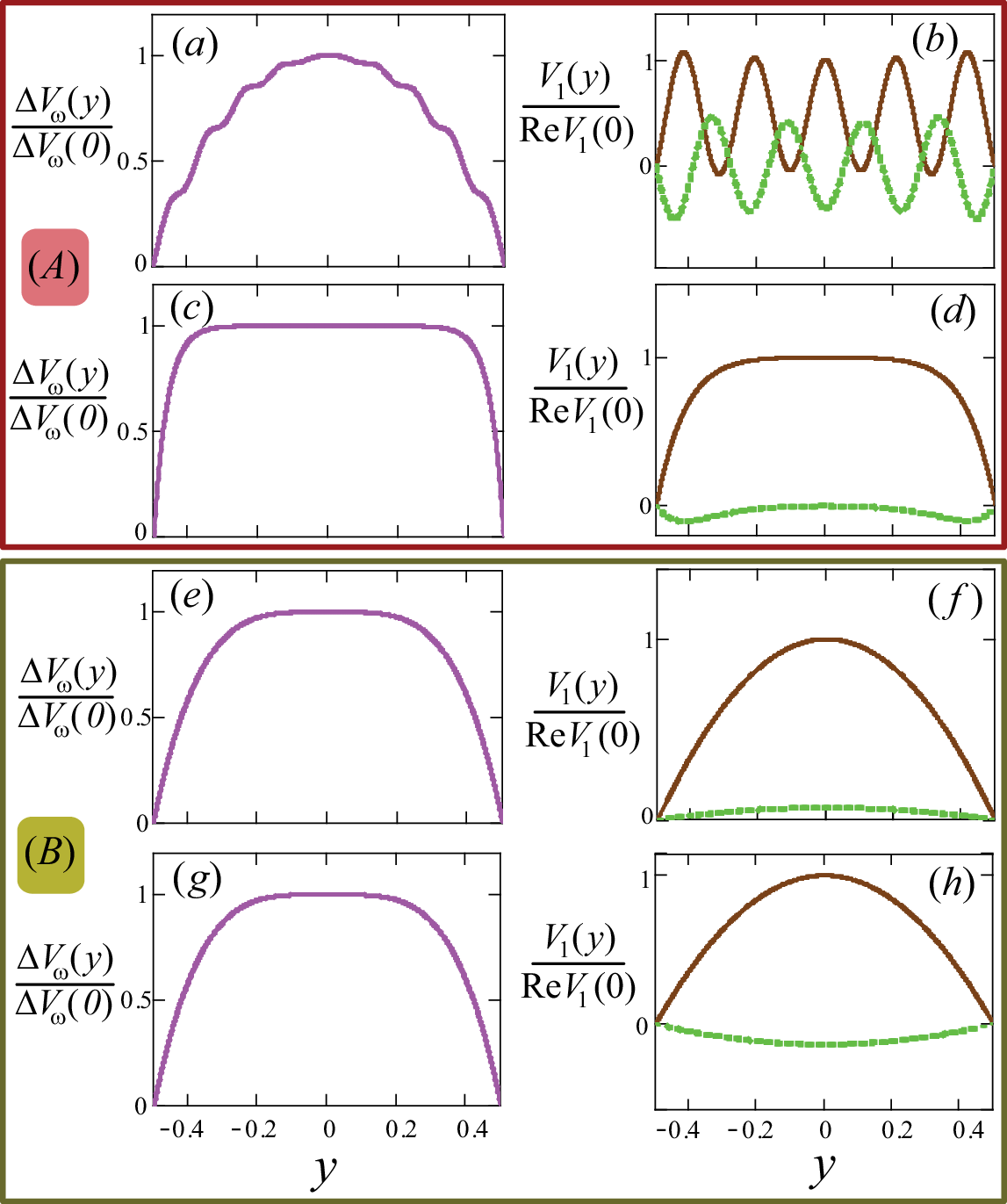}}
\caption{
Profiles of the flow  components  $V_{\omega}(y)$ [left panels] and  $V_1(y)$ [right panels; brown  and green lines refer  to  $\mathrm{Re }
 \, V_1(y)$  and   $\mathrm{Im} \, V_1(y) $, respectively]. All curves are plotted for the ac frequency $\omega \tau_2 = 40$. Panel ($A$) shows
the results for a relatively wide sample, $W/L_s =1$,  in the magnetic fields  $\omega_c \tau_2 = 0.34 $  ($a$,$b$) and $\omega_c \tau_2 =
0.52 $ ($c$,$d$). At $\omega_c \tau_2 = 0.34 $ the system is above the viscoelastic resonance and  the wavelength and the decay length of
the magnetosonic  waves are:  $2\pi/(L_s \mathrm{Im }\lambda )= 0.21$ and   $1/(L_s \mathrm{Re}\lambda ) = 1.004$, whereas at
$\omega_c \tau_2 = 0.52 $ the system turns out   below the viscoelastic   resonance that correspond   to the decay length  $1/(L_s \mathrm{Re} \lambda)
 = 0.031$.  Panel~($B$) presents the results for a narrow sample, $W/L_s =0.01$,   at the magnetic fields  $\omega_c \tau_2 = 0.34 $  [($e$,$f$); above
the resonance;  $\mathrm{Im} \lambda $ and  $ \mathrm{Re} \lambda $ are the same as in panels ($a$,$b$)] and to $ \omega_c \tau_2 = 0.6 $
[($g$,$h$); below the resonance;  $1/(L_s\mathrm{Re}\lambda) = 0.038$].
 }
\end{figure}

The profiles  $V_\omega(y)$  and $V_1(y) $ for the medium and  the narrow samples are drawn in Fig.~S4.

From the resulting expressions~(\ref{result_SI0})-(\ref{J_res}) and estimate~(\ref{est_for_mem_tens}) for the coefficient in the memory term,
$ \alpha = B\,  (R_c/a_B)^2   e^{-T/\tau_q} / \tau_2 $, one can find the magnetooscillations of the  velocity  $V _{\omega} (y) $ at any $y$.
Note that $R_c =  v_F /\omega_c $  contains the actual Fermi velocity, unlike the viscosity coefficients  containing the parameter $v_F^\eta $,
$ v_F^\eta \gg v_F$. At the strong magnetic fields and  the high ac frequencies, $\omega _c \sim \omega \gg 1/\tau_2$, and far from the resonance,
 $ | \omega  - 2 \omega _c  | \gg  1 / \tau_2 $, for  the point $y=0$,   corresponding to the typical values
of $V_{\omega}(y)$ (see Fig.~S4), we have:
\begin{equation}
 \label{V_omega_0}
      V_{\omega} (0)
          =
        \sin^2 \Big ( \frac{ \pi \omega}{ \omega_c} \Big )  \,  V_{\omega} ^{max,1}
          +
        \sin\Big ( \frac{ 2\pi \omega }{ \omega_c} \Big ) \, V_{\omega} ^{max,2}
\end{equation}
where the amplitudes of these two oscillating contributions are:
\begin{equation}
  \label{V_omeg_est}
    V_{\omega} ^{max,1} = 16\,  \frac{e ^3 E_0 \, E_1 ^2  \, v_F^2    }{ m^3 \, (v^\eta_F)^2 \, a_B^2   }
    \, \frac{ \tau _ 2 }{\omega^ 4}
        \: e^{ -T/\tau_q} \:       J_c
     \: ,
\end{equation}
\begin{equation}
 \label{V_omeg_est_2}
   \begin{array}{c}
    \displaystyle
    V_{\omega} ^{max,2} =
       - \frac{ 4 \pi \,  \omega \omega_c  }{\omega^2 - 4 \omega_c^2   } \:
    V_{\omega} ^{max,1}
   \: ,
\end{array}
\end{equation}
and the dimensionless factor $J _c \equiv J(0)$ is:
\begin{equation}
\label{J}
     J _c
        =  \frac{ | \lambda | ^2 }  { | \cosh (\lambda W/2  ) |^2 }  \int \limits  _{0} ^{W/2}  d \tilde{y} \; \tilde{y}
  \;| \sinh (\lambda \tilde{y}  ) | ^2
\:.
\end{equation}

From Eq.~(\ref{J_res}) we obtain that above the viscoelastic resonance, $\omega> 2 \omega_c $, when $ \mu \ll \nu $, the factor
$J$ for different sample width  is estimated as:
\begin{equation}
\label{above}
J _c
  \sim
  \nu^2 W^2
\left\{
 \begin{array}{l}
 \displaystyle
    \displaystyle
 \frac{ 1 }{W  \mu}
      \, \;  , \;   \;  \quad  W   \gg  1/\mu
 \\
 \\
 \displaystyle
 \frac{ 1 }{ \cos^2 ( \, \nu W / 2 \, )} \, ,
 \; \;
  1/\nu \ll W \ll  1/\mu
 \\
 \\
  \displaystyle
   \nu^2 W^2 \, , \;  \; \quad  W \ll  1/\nu
 \end{array}
\right.
   \:.
\end{equation}
This formula for the intermediate sample widths, $ 1/\nu \ll W \ll  1/\mu$, is valid for the sample width far from the following condition:
the coincidence of the sample width  $W$ with a half-integer number of the wavelengths of transversal magnetosound~\cite{vis_res_2}:
\begin{equation}
 \label{c}
    \nu W  = \pi  + 2 \pi l
   \:,
\end{equation}
where $ l=0,\pm1, \pm2, ... $. At this condition acoustic-like resonances related to standing magnetosonic modes inside the sample occur in
the flow.   For the sample widths, given by Eq.~(\ref{c}), the denominator of $J_c$ in Eq.~(\ref{above}) becomes zero.  Therefore in the vicinities
of such  $\omega $  and $\omega_c$  one should use  the exact expression for the denominator: $|  \cos [\,(\nu  + i \mu )\, W / 2   ]  \, |^2$.
Thus at small values of the deviation parameter  $\Delta w_l = \nu W/2 - (\pi/2 + \pi l )  $, $| \Delta w_l   | \ll 1 $, and provided
$\mu W \ll 1 $ we have:
\begin{equation}
  \label{res}
   J _c \sim \frac{ \nu  ^2 W ^2 } { \Delta w_l ^2  + (\mu W /2 ) ^2 }
   \:.
\end{equation}
It follows from this formula that the resonances at the peculiar values of $\nu W $~(\ref{c}) can be more or less sharp depending
on parameters $\omega \tau_2$ and~$W/L_s$.

Below the viscoelastic resonance, $\omega < 2 \omega_c $, when  $ \mu \gg \nu  $ (see Fig.~S2),  one obtains from  Eq.~(\ref{J_res}):
\begin{equation}
\label{below}
  J_c   \sim  \mu W
\left\{
 \begin{array}{l}
 \displaystyle
    1 \, \;  , \;   \quad  W   \gg  1/\mu
 \\
 \\
  \displaystyle
   \mu^3 W^3 \, , \; \quad W \ll 1/\mu
 \end{array}
 \:.
\right.
 \end{equation}

From  Eqs.~(\ref{lambda_above}),  (\ref{lambda_below}),  (\ref{J})-(\ref{below})  we estimated  the  amplitudes $V^{max,1/2}_\omega$~(\ref{V_omeg_est}),
 (\ref{V_omeg_est_2})  in an exact form for different regimes:   above the viscoelastic resonance ($\omega>2\omega_c$)  for the
wide [$W \gg 1/\mu (\omega_c)$], the medium [$1/\nu (\omega_c) \ll W \ll 1/\mu (\omega_c)  $],  and   the narrow [$W \ll 1/\nu (\omega_c) $]
samples  as well as below the viscoelastic resonance  ($\omega<2\omega_c$) for the wide [$W \gg 1/\mu (\omega_c)$] and the narrow [$W \ll 1/\mu
(\omega_c)$] samples.

\begin{figure}[t!]
\centerline{\includegraphics[width=0.98\linewidth]{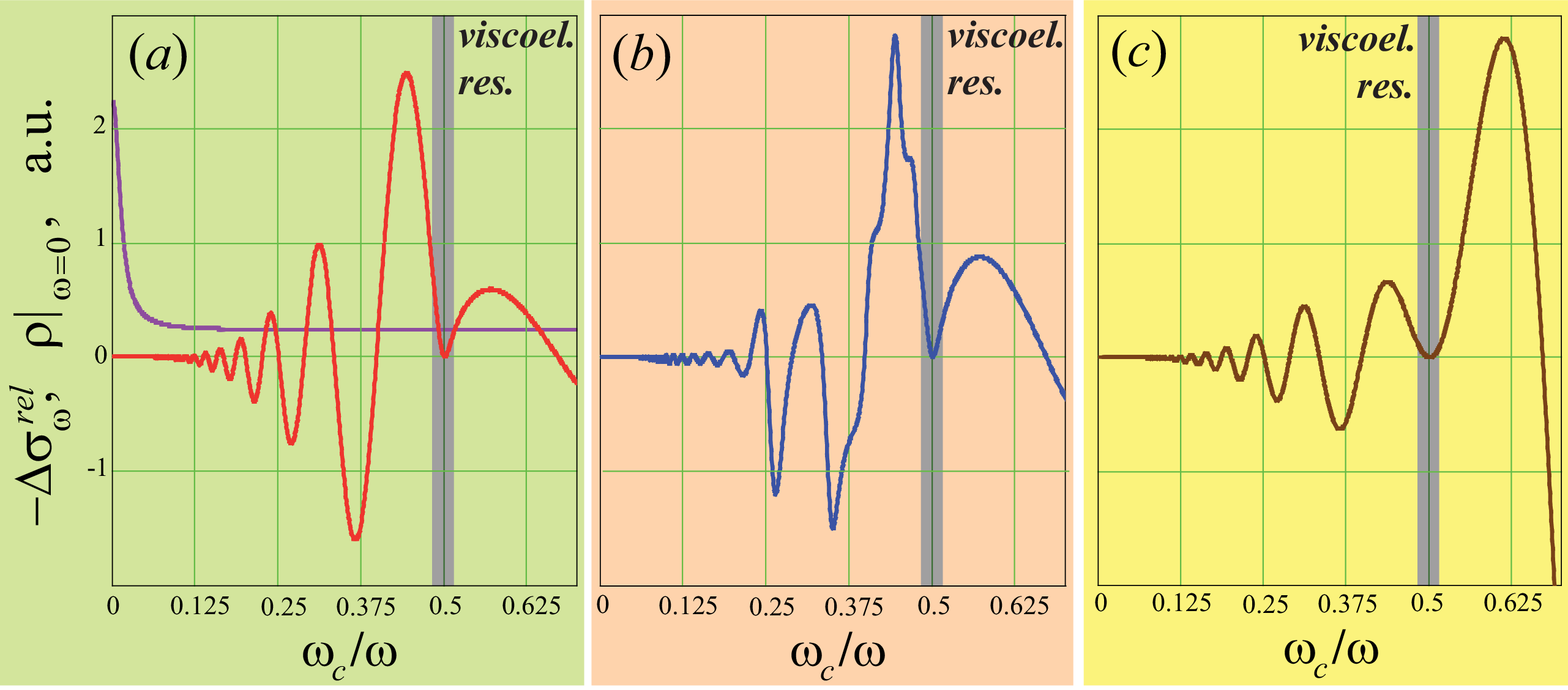}}
\caption{
 Photoconductivity (the relaxation contribution) with the inverse sign,  $  - \Delta \sigma_\omega ^{rel}$,
 for Poiseuille flows of an electron fluid
 in long defectless samples. The parameters of the system are $ \omega \tau_{2} = 40 $; $\tau_q / \tau_2 = 0.7$;     $  W / L_s =6   ,\,
 1.19   ,\, 0.03 $   [red, blue, and brown curves in panels ($a$-$c$), respectively].  Grey stripes schematically denote the region
 near  the viscoelastic resonance, $\omega = 2 \omega_c$, where the  fluid dynamics is mainly  dissipative (in the zero approximation
 by $ \omega \tau_{2}  \gg 1 $)   and   the elastic contribution in photoconductivity related with the perturbations of
  the Landau parameters $\delta F_{2,ij}$ becomes very sub (see Eq.~(\ref{main_eq_gen_S}) and next Section~6). In panel ($a$) we also plotted
 the dc magnetoresistance $\varrho|_{\omega = 0 }  (\omega _c)$ (violet curve) for  a sample with the same parameters of the electron
fluid and some density of defects.   Such magnetoresistance consists of the two contributions~\cite{je_visc}:
  the hydrodynamic one $\varrho_{xx,hydr}(B)$, being  proportional to $\eta_{xx} \sim (1+4\omega_c^2 \tau_2^2)^{-1}$ and dominating
  at $\omega_c \tau_2 \lesssim 1 $ ($ \omega_c / \omega \lesssim 0.1  $), and the Ohmic one,  $\varrho_{xx,Ohm}(B) = const$ being
  proportional to the rate of  the scattering of quasiparticles on impurities,  $1/\tau_{imp}$,
  and dominating at $\omega_c \tau_2 \gg 1 $.
 }
\end{figure}

We do not present all  the resulting formulas for $V^{max,1/2}_\omega$ for these cases because of  their cumbersomeness. Let us discuss in detail
 only the result for the most interesting regime: above the viscoelastic resonance, $\omega>2\omega_c \gg 1/\tau_2$, for the medium sample
width, $1/\nu \ll W \ll 1/\mu $, when the magnetooscillations acquire irregular shape [see ~S5].  In this case and  far from
 the magnetosonic resonances~(\ref{c}), the velocity amplitudes~(\ref{V_omeg_est}) and (\ref{V_omeg_est_2}) in this regimes take the form:
\begin{equation}
  \label{res_est_med}
    V_{\omega} ^{max,1} =  (\, \omega^2  - 4 \omega_c^2   ) \, V^0_{\omega}
    \, ,
\end{equation}
\begin{equation}
  \label{res_est_med_2}
     V_{\omega} ^{max,2} = - 4 \pi \,
     \omega \omega_c \, V_\omega^0
 \:,
\end{equation}
where
\begin{equation}
 \label{ampl}
 V_{\omega} ^0 =   \frac{e ^3 E_0 \,  E_1 ^2 \, v_F^2    }{ m^3 \, ( v_F^\eta ) ^4  \, a_B^2  } \:
    S(\nu) \: \frac{  W^2 \, \tau _ 2  }{ \omega^4 }
       e^{  -T/\tau_q }
\end{equation}
and  $S (\nu) \sim 1/\cos^2(\nu W/2)$, when the value  $\nu W$ is not too close to $(1+2N_r) \, \pi$ with integer $N_r$.

Now we can find   the dependencies of the radiation-induced correction $\Delta \sigma _{\omega} $  to the dc conductivity, $\sigma  = \sigma  _0
 + \Delta \sigma _{\omega} $,  on temperature, ac frequency,  and magnetic field far from  the magnetosonic resonances~(\ref{c}).  The value
 $ \Delta \sigma _{\omega} $  is defined as:
\begin{equation}
   \label{Delta_sigma}
  \Delta \sigma_\omega = \frac{I_\omega  }{E_0W} \:, \qquad  I_\omega
    = e n_0 \int _{-W/2} ^ {W/2} dy \: V_\omega (y)\:.
\end{equation}
Equations~(\ref{J})-(\ref{V_omega_0})   and~(\ref{res_est_med})-(\ref{ampl})   yield for the photoconductivity~$\Delta \sigma_\omega  $ in
 the considered   case,  $1/\nu \ll W \ll 1/\mu $, when the standing magnetosonic waves are well-formed:
\begin{equation}
\label{UV}
\Delta \sigma_\omega = (U+V) \, \Delta \sigma_\omega ^{max}\:,
\end{equation}
where the factors  $U=U( \omega / \omega  _c )$ and $V=V( \omega / \omega  _c )$  are:
\begin{equation}
\label{UV2}
\begin{array}{c}
\displaystyle
U  =
  \Big[ 1 -   \Big( \frac{ 2 \omega _c}{ \omega }\Big)^2   \Big] \,
     \sin^2 \Big( \, \frac{ \pi \omega }{ \omega  _c  } \, \Big)
   \:,
  \\
 \\
\displaystyle
V  =
   - \frac{ 4 \, \pi \,  \omega_c  }{\omega   }
         \, \sin  \Big( \, \frac{ 2 \pi \omega }{\omega  _c } \, \Big)
       \:,
\end{array}
\end{equation}
and the amplitude $\Delta \sigma_\omega ^{max} = \Delta \sigma_\omega ^{max} (  \, \omega , \, \omega_c \tau_q \, ) $  is:
\begin{equation}
 \label{dep_on_T_ome}
  \Delta \sigma_\omega^{max}
    \sim  \frac{e^4 E_1^2  \, n_0  \, v_F^2    }{m^3 \, ( v_F^\eta ) ^4 \,  a_B^2    } \,
    \frac{ W^2\tau_2  }{ \omega^2 }
    \: {\Large e}
    ^{ -  2 \pi / ( \omega_c  \tau_q) }
   \:.
\end{equation}
Here the relaxation times   $\tau_q=\tau_q(T_e) $ and $\tau_2 = \tau_2(T_e) $ are given by   Eqs.~(\ref{tau_temp_dep}).  It follows from
this result  that the oscillation of photoconductivity   are suppressed  with the increase of the temperature $T_e$ and
 the ac frequency~$\omega$.

One can see from Eq.~(\ref{res}) that in the vicinities of the  magnetosonic  resonances~(\ref{c}) the value  $\Delta \sigma_\omega^{max}
 \sim J_c $  increases as compared with Ref.~(\ref{dep_on_T_ome}) in the large factor~$\nu ^2  / \mu ^2 \gg 1  $.

In Fig.~S5  we present the averaged photoconductivity  $ \Delta \sigma _{\omega}  =
 \Delta \sigma _{\omega} ^{rel} $~(\ref{Delta_sigma}) multiplied on $-1$,
as  this value is presented as photoresistance in linear by the ac power $\mathcal{W}$ approximation is:
\begin{equation}
   \Delta   \varrho _{\omega }
   =  - \Delta \sigma _{\omega}   /\sigma_0^2
   \: .
\end{equation}
The curves are plotted at a fixed  ac frequency  $\omega$ and the sample widths $W$  smaller,   comparable, and larger than
the characteristic decay length~$ L_s $. The dependence $ -\Delta \sigma _{\omega}(\omega _c ) $ exhibits  oscillations by
the inverse magnetic field. The shape of   the obtained oscillations  is sinusoidal in  the limiting cases  $W\gg L_s$ and  $W\ll L_s$,
while for intermediate sample widths, $ W \sim L_s$,  it is irregular. The last property is related to the appearance of the magnetosonic
 resonances in the ac linear response~$V_{1}(y,t)$ at the sample width satisfying conditions~(\ref{c}). These resonances manifest themselves
most clearly when $W \sim L_s$ and, thus a not too many wavelengths fit in the sample,   but the damping of waves is relatively weak.

\newpage

\subsection{ 6.   Non-linear response of  fluid
on dc and ac electric fields: memory effects
in elastic part of interparticle interaction}
In this section, we calculate the ``elastic'' contribution in the nonlinear correction  $ V _ {\omega }  (y)$
to the dc velocity $ V_0(y) $, which originates from the nonlinear by $\mathbf{V}(\mathbf{r},t)$ retarded  part
of the energy of quasiparticles   which is expressed  via the perturbations $\delta F _{2,ij} $~(\ref{abcd}) of
the Landau parameter $F_2$.  This contribution is additive
with  the relaxational  contribution in $ V _ {\omega } $ from retarded relaxation
 due to the memory effect in extended collisions, calculated in
 previous Section~5.  The resulting contribution in photoconductivity  $\Delta \sigma _ \omega$ has the  properties being
partially different from the ones of the relaxational ones.

The equations for $\hat{\Pi}_\omega $ and $V_ \omega $  are derived by  the substitution of the dc and the ac linear components, $\hat{\Pi}_0$,
$\Delta_0$ and   $\hat{\Pi}_1$,   $\Delta_1$,   into the nonlinear parts of the retarded elastic terms in Eqs.~(\ref{main_eq_gen_S}) with
$\delta F_{2,ij} $~(\ref{abcd}). As we mentioned in Section~2, for brevity we omit the amplitude of the unperturbed Landau parameter
writing just  $F_2=0$ in the left part of Eq.~(\ref{main_eq_gen_S}).

For the geometry of Poiseuille flow,  the resulting equations for $\partial \mathbf{V} / \partial  t $  and
 $ \partial \hat{\Pi} / \partial  t $  are similar to Eqs.~(\ref{main_eq_partic}) accounting the retarded
 relaxation via the terms $\sim\hat{\Gamma}$.  In order to calulate the dc non-linear contributions
  to the velocity and the stress tensor, one should again integrate the motion  equations
by one cyclotron period, as it was done in Section~5 for the inelastic contribution due to the term $\Gamma_{ijkl} \Pi_{kl} $. We obtain:
\begin{equation}
\label{Pi_zv_main_eq}
 \left\{ \:
\begin{array}{l}
\displaystyle
\frac{1}{m} \frac{d \Pi  _{\omega  , xy}}{ d y } =0
  \\
  \\
   \displaystyle
\frac{ \Pi _{\omega , xx} }{ \tau_2 }
  - 2 \omega_c \Pi _{\omega ,  xy}
  \, =
    \\
    \\ \displaystyle
 \qquad\quad \quad \quad
  =\,
 \Big\langle \delta  F_{2,xx}^ {(1)} (y,t )\, \frac{  \partial \Pi _{xx}(y,t)}{\partial  t } \, \Big \rangle_{\omega }
   \\
   \\
   \displaystyle
2 \omega_c \Pi _{\omega ,   xx} + \frac{ \Pi _{\omega ,   xy} }{ \tau_2 }  =
 - \,  \frac{m _0 \, (v^\eta_F)^2}{4 } \frac{d V_ \omega  }{ d y}
  \, +
   \\
   \\
    \displaystyle
 \qquad\quad \quad \quad  \;
+  \,    \Big \langle   \delta F_{2,xy}^ {(1)} (y,t ) \,
 \, \frac{  \partial \Pi _{xy }( y,t) }{ \partial  t }  \, \Big \rangle_{\omega }
\end{array}
 \right.
 .
\end{equation}
Here all the notations correspond to the ones in Section~5.

From the last two equations in system~(\ref{Pi_zv_main_eq})  we obtain the expression for the nonlinear part $\hat{\Pi}_{\omega}$ of the
momentum flux tensor. For the $xy$-component we have:
\begin{equation}
\label{Pi_star_res}
\begin{array}{c}
\displaystyle
\Pi  _{\omega , xy}  (y)= -m_0 \, \eta_{xx} \frac{d V _{ \omega  }  (y)}{ dy}
\, +
\\
\\
\displaystyle
+ \,
 \frac{ \tau_2}{1+4\omega_c^2 \tau_2 ^2 } \,
\big[\,
 \Big \langle   \delta F_{2,xy}^ {(1)} (y,t ) \,
 \, \frac{  \partial \Pi _{xy }( y,t) }{ \partial  t }  \, \Big \rangle_{\omega }
 -
\\
\\
\displaystyle
  \qquad \quad -
\,
 2 \omega_c \tau_2
  \Big \langle   \delta F_{2,xx}^ {(1)} (y,t ) \,
 \, \frac{  \partial \Pi _{xx}( y,t) }{ \partial  t }  \, \Big \rangle_{\omega }
\, \big]
\:.
\end{array}
\end{equation}
Here the nonlinear terms in the brackets  $\langle \:..\: \rangle_\omega $~(\ref{br_om0}) again contain the two contributions with
the different  combinations of the  dc and ac linear responses:
\begin{equation}
\label{contrs_in av}
\begin{array}{c}
 \displaystyle
 \Big \langle   \delta F_{2,xx}^ {(1)} (y,t ) \,
 \, \frac{  \partial \Pi _{ xx}( y,t) }{ \partial  t }  \, \Big \rangle_{\omega }=
 \\
   \\
   \displaystyle
 = \Big[ \, a\,  \Big  \langle \,  \Pi _ {1,xx}  (y,t-T ) \,   \frac{ \partial \Pi _ {1,xx}  (y,t ) }{ \partial t } \,\Big  \rangle
\, +
\\
\\
\displaystyle
+ \,  b \,   \Big  \langle \,  \Pi _ {1,xy}  (y,t-T )   \, \frac{ \partial \Pi _ {1,xx}  (y,t ) }{ \partial t } \, \Big \rangle    \,
\Big]
\,   \frac{d V _0(y)}{ d y }
\end{array}
\end{equation}
and
\begin{equation}
\label{contrs_in av_2}
\begin{array}{c}
\displaystyle
 \Big \langle   \delta F_{2,xy}^ {(1)} (y,t ) \,
 \, \frac{  \partial \Pi _{ xx}( y,t) }{ \partial  t }  \, \Big \rangle_{\omega }=
 \\
   \\
   \displaystyle
 = \Big[ \, c \, \Big   \langle \,  \Pi _ {1,xx}  (y,t-T )  \,  \frac{ \partial \Pi _ {1,xy}  (y,t ) }{ \partial t } \, \Big  \rangle
\, +
  \\
  \\
  \displaystyle
+ \, d \,  \Big  \langle \,  \Pi _ {1,xy}  (y,t-T ) \,   \frac{ \partial \Pi _ {1,xy}  (y,t ) }{ \partial t } \, \Big  \rangle    \,
\Big]
\,   \frac{d V _0(y)}{ d y }
\:.
\end{array}
\end{equation}
At the considered regime of high frequencies and magnetic fields,  $\omega \sim \omega_c \gg 1/\tau_2$,   the viscosity coefficients
are related as:
\begin{equation}
-i \omega \eta_{xy} (\omega)  \approx  2\omega_2  \eta_{xx} (\omega)
\:.
\end{equation}
After substitution of the expression
for $\hat{\Pi}_1$ via $\mathbf{V}_1 $ and averaging over the interval $t-T<t'<t$, the values in Eqs.~(\ref{contrs_in av})
 and~(\ref{contrs_in av_2})  takes the form:
 \begin{equation}
 \Big  \langle \,  \Pi _ {1,ij}  (y,t-T )  \,
    \frac{ \partial \Pi _ {1,kl}  (y,t ) }{ \partial t } \, \Big  \rangle =
     Q _{ij,kl} \, R (y)
 \end{equation}
 where
\begin{equation}
\label{av_1}
  \begin{array}{c}
  \displaystyle
  Q _{xx,xx}
 \\ \\
  Q _{xy,xy}
 \end{array}
 \Bigg\}
 \,
  =
  \,
-  \frac{  2}{\omega}
   \,  \sin \Big( \frac{ 2 \pi \omega }{\omega_c} \Big)
   \Bigg\{
  \begin{array}{c}
  \displaystyle
   4 \, \omega_c ^ 2
 \\\\
    \omega ^ 2
 \end{array}
   \:,
 \end{equation}
\begin{equation}
\label{av_2}
  \begin{array}{c}
  \displaystyle
  Q _{xx,xy} =  Q _{xy,xx} =
4 \,  \omega_c \,  \cos \Big( \frac{ 2 \pi \omega }{\omega_c} \Big)
   \:,
     \end{array}
\end{equation}
and
\begin{equation}
\label{P_y}
R (y)=
      | \eta_{xx} (\omega)| ^2 \:  \Big| \, \frac{dV_1 (y)}{dy} \Big|^2
  \end{equation}

Substitution of formulas~(\ref{Pi_star_res})-(\ref{P_y}) for $\Pi  _{\omega  , xy} $ into the first of equations~(\ref{Pi_zv_main_eq}) leads
to the final  equation  for  the nonlinear component  $ V_\omega  = V_\omega (y) $ of the dc velocity:
\begin{equation}
 \label{eq_for_v_zv_fin}
    \frac{d ^2 V_\omega }{ d y^2} = H(y) \:,
\end{equation}
where $H(y) $ is given by Eqs.~(\ref{Pi_star_res})-(\ref{av_2}). Equation~(\ref{eq_for_v_zv_fin}) should be solved with the diffusive  boundary
conditions on the component $V _ \omega (y)$: $ V _ \omega  | _ { y = \pm W/2 } = 0 $. The result  takes the form:
\begin{equation}
\label{result_SI}
   V _{\omega}(y)  = \frac{e ^3 E_0 E_1 ^2   }{ m_0^3 (v^\eta_F)^2  }
   \: \frac{   |\eta_{xx} (\omega)| ^2   }{ \omega \, \eta_{xx} } \:  \Phi  (\omega , \omega _c)
   \,  J(y ) \:,
\end{equation}
where the factor $ J(y) = J(y, \lambda )  $,  determining the shape of $V_\omega(y)$, is the same [Eq.~(\ref{J_y0})], as for the inelastic contribution
 and  the factor  $ \Phi ( \omega , \omega_c ) $ has the form:
\begin{equation}
\label{u2}
\begin{array}{c}
  \displaystyle
\Phi  = 4 \:\Big \{
\,
 \Big[  - a \, \omega_c \tau_2   +    2 d    \,\Big( \frac{\omega_c }{\omega  } \Big)^2 \Big] \,
 \sin \Big( \frac{ 2 \pi \omega }{\omega_c} \Big)
 +
 \\
 \\
   \displaystyle
 + \,
  \Big[-  2  b \, \omega_c \tau_2  +   c\, \Big] \, \frac{ \omega_c }{ \omega }   \,
 \cos \Big( \frac{ 2 \pi \omega }{\omega_c} \Big)
 \,
  \Big \}
  \:.
  \end{array}
\end{equation}
We remind that the coefficients $ a, b, c, d $ are proportional to the probability $P$~(\ref{P1}) for two quasiparticles in a pair
 to make a complete  cyclotron rotation without a collision with  a third quasiparticle.

The resulting dependence of the elastic contribution to the sample photoconductivty,
 $\Delta \sigma _{\omega}^{el}  = \int _{-W/2} ^{W/2} dy \, V^{el} _{\omega}(y)  $, [here $ V^{el} _{\omega}(y) = V  _ {\omega}(y)   $
 is given by Eq.~(\ref{result_SI})]
  on magnetic field at fixed $\omega$ exhibits
 MIRO-like oscillations at $\omega_c < \omega/2$ and a huge peak near $\omega_c = \omega/2$  [see Fig.~4(a) in the main text]. By the
last property,  the elastic contribution  drastically  differs from the inelastic contribution,
 which has no peak at $\omega_c = \omega/2$ [see Fig.~3(a) in the main text].

\newpage

\subsection{ 7. Discussion of results  and model }
 {\em 7.1.~Independence of photoresistance on the sign of circular polarization of radiation }

In experiments~\cite{Smet_1,Ganichev_1}  it was demonstrated that     the dependence of  photoresponses
$\Delta \sigma _\omega$   and $\Delta \varrho _\omega$ on the polarization  helicity ``$\pm$''
  is absent in ultra-high quality GaAs quantum well samples with low defect densities and
very large experimental values of mobilities
  (see discussion in the main text).  In previous Sections~5 and~6, we have shown that for purely hydrodynamic flows
  in narrow samples, $W \ll l_p$~(\ref{W}), the viscoelastic  contribution dominates in the ac flow, the component $E_{1,y} ^{ext}$
 of the external electric field is screened inside
the sample, and    the velocity $\mathbf{V}_1\, ||\, \mathbf{e}_x$ becomes independent of $E_{1,y} ^{ext}$ and the
helicity of polarization. Thus within our theory photoresponses $ \Delta \sigma _\omega$   and $\Delta \varrho _\omega$
also are independent on the polarization  helicity ``$\pm$''.

We remind that for  Ohmic  flows  in  wide  bulk  samples  the  situation  is  totally  different.    In  theory~\cite{Beltukov_Dyakonov}  and
other  theories for disordered samples it is implied that the ac electric field acting  on independent electrons in Ohmic samples is just
the external radiation field~$\mathbf{E}_1(t)$.   Unlike the current hydrodynamic model, the memory term  due to extended collisions
of electrons with localized  defects in the equation for  $\partial \mathbf{V}/\partial t$ contains not the mismatch of
the deformation,~$\hat{\Delta } (\mathbf{r},t)$, but  the mismatch  of the electron position,~$\mathbf{r}_0(t)$, after
one cyclotron rotation, $\boldsymbol{\Delta }(t) = \mathbf{r}_{0}(t)- \mathbf{r}_0(t-T)$.  Such mismatch~$\boldsymbol{\Delta } \neq 0  $
arises due to  the force  $e\mathbf{E}_0 +e\mathbf{E}_1(t)$ from external electric  fields, directly  acting on electrons between
their collisions with defects. Obviously, $\boldsymbol{\Delta }(t) $ and the resulting photoconductivity strongly  depend  on
the helicity~$\pm$ of the polarization of~$\mathbf{E}_1(t )$~\cite{Beltukov_Dyakonov,rev_M}.

Thus, the independence of the photoresistance of the sign $\pm$ of circular polarization is a necessary (but not sufficient)  evidence
of the formation of a high-frequency hydrodynamic flow  of a viscous electron fluid. The dependence  on $\pm$ can appear in sufficiently
wide samples,  $ W \gtrsim l_p$, owing to the formation of  a substantial  plasmonic component in $V_1$, or at a sufficiently large strength
of disorder when the main contribution  to photoconductivity becomes related to the scattering of single quasiparticles on defects or
 sample edges.

In Ref.~\cite{rev_M} the following idea about the most typical origin of the absence of the dependence of  MIRO on the helicity of the circular
polarization of an incident  radiation field   $\mathbf{E}_1 ^{ext} (t)$  was formulated.   The difference of the  polarization of
the actual ac field $\mathbf{E}_1(\mathbf{r},t) = \mathbf{E}_1^{ext}(t) +\mathbf{E}_1^{int} (\mathbf{r},t)   $, which is ``felt''
by 2D electrons,  and the polarization  of the  incident field  $\mathbf{E}_1^{ext}(t)$  can be  induced  by some elements of
an experimental  setup. This can result in   an independence (more exactly, a weak dependence) of MIRO on the sign of the polarization
of $\mathbf{E}_1 ^{ext} (t)$. However, it was not indicated in Ref.~\cite{rev_M} that this modification of the polarization can be
 a consequence of the screening of one of the components of the incident field  $\mathbf{E}_1^{ext} (t)$ by the internal
 ac field~$\mathbf{E}_1^{int} (\mathbf{r},t) $ originating from the redistribution of the very 2D electron fluid
 density~$ \delta n (\mathbf{r},t) $ and, thus,  can be  inextricably related with the mechanism  of MIRO.
  It is noteworthy that in Ref.~\cite{Savchenko} it was experimentally demonstrated that particular  setups of experiments
     (shapes of samples,
   designs of ac radiation sources)
actually can sufficiently change the sensitivity of the MIRO effect relative to the polarization of the external
radiation~$\mathbf{E}_1^{ext}(t)$.  However, in experiment~\cite{Savchenko}  an almost complete independence
of photoresistance on the helicity of the circularly polarized radiation  \{such as it was observed, for example,
 in experiment~\cite{Smet_1}; see Fig.~3(b2) in the man text\} was not achieved~\cite{I_A_Dmitriev_private_communication}.

In this way, an experimental setup and other external factors, apparently,  can strongly weaken the dependence of photoresistance on the sign~$\pm$,
 but not almost absolutely   eliminate it, as it was observed in Ref.~\cite{Smet_1} and other works.

In Refs.~\cite{Mikhailov,Chepelianskii_Shepelyansky} theories of another type  for 2D electron magnetotransport  were developed,
in which the photoresistance is  associated with the ac-field-induced redistribution of the electron density  near sample
contacts~\cite{Mikhailov} or around localized defects inside a sample~\cite{Chepelianskii_Shepelyansky}. Although these  theories lead to
  magnetooscillations of photoresistance with a weak dependence on the sign of the circular polarization of the incident radiation,
  there exist substantial problems  in the comparison of other predictions of those theories with experiments on MIRO.  For example,
there is a discrepancy between the shapes of the predicted oscillations and of the oscillations observed     in experiments.
The  latter ones  are often  sinusoidal, whereas in Refs.~\cite{Mikhailov,Chepelianskii_Shepelyansky} the sinusoidal shape of
 magnetooscillations was not obtained.
\\
\\
\indent {\em 7.2.~Mixed hydrodynamic-Ohmic flows }

Below, in this and next Subsections~7.2 and~$7.3$, we  discuss how the particular violation of the ``maximal'' conditions of
 applicability of the developed purely hydrodynamic model (the absence of bulk disorder and a  very large strength  of
 the interparticle interaction) may  be relaxed so that the obtained above key  results  on photoresistance of a highly viscous fluid
remain qualitative  valid.

First, in this subsection, we discuss the case of a disordered sample  in which the hydrodynamic  contributions  to the ac linear in  $E_1$
and  the dc nonlinear in  $E_1$ components of the flow components still  persist, despite of the dominance of the Ohmic contribution in
  the main dc flow component, linear in  $E_0$. We will estimate the parameters corresponding to such ``mixed'' regime.

At high magnetic fields, $\omega_c \tau_2 \gg 1 $, the magnitude of a purely hydrodynamic dc flow  unlimitedly increases,  being proportional
to $\omega_c^2$ [Eqs.~(\ref{V_0})  and (\ref{e})].  Thus, the purely hydrodynamic  dc averaged resistance, $  \varrho_{0} = E_0 W /I_0 $,
strongly decreases with the increase of~$\omega_{c}$ and~$W$:
\begin{equation}
  \varrho_{0} \, \propto \, 1 \, / \,  (W \omega_c)^2
  \: .
\end{equation}
Such divergence is limited by the formation of an Ohmic flow in central region of the sample where the scattering of quasiparticles
  on disorder (and/or phonons) becomes to dominate and to determine the magnitude of  the current and the resistance~$ \varrho_{0} $
\{see Refs.~\cite{je_visc,Alekseev_Dmitriev}  and Fig.~S5(a)\}.

The    Ohmic   contribution  in the dc current $I_0$ dominates  if the sample width is  much larger than the Gurzhi
length~$l_G = l_G(\omega_c)  $:
\begin{equation}
  \label{n}
    W \gg l_G \:, \quad l_G  = \sqrt{ \tau_{tr}  \eta _{xx}  (\omega_c) }
  \:,
\end{equation}
where $ \tau_{tr}   $ is the total momentum relaxation time   due to   the scattering of quasiparticles on bulk disorder and acoustic
phonons. In this case, the Ohmic part of the dc flow is located  in the central sample region:
\begin{equation}
 \label{reg_b}
  - W/2 + l_G \lesssim y \lesssim W/2 - l_G
    \:,
\end{equation}
or, more precisely,   $W/2 -|y| \gg l_G$.   The resulting  hydrodynamic-Ohmic  magnetoresistance has the form \{see Fig.~S5(a) and
discussion in Ref.~\cite{je_visc}\}:
\begin{equation}
  \label{r}
     \varrho  _{0}  (B) \sim \frac{\eta_{xx} (B) }{ W^2 }  + \frac{1}{\tau_{tr}}
  \:.
\end{equation}

One can show that at the considered high-frequency regime, $ \omega_c \sim \omega \gg 1/\tau_2 $, the inequality~(\ref{n})  is consistent
with  inequality~(\ref{W}),  which ensures the dominance of  the viscoelastic pact  in the ac component,  provided the following condition is fulfilled:
\begin{equation}
 \label{n2}
  s \,  \gg    \, v_F^\eta  \sqrt { \frac{ \tau_{tr}}{ \tau_2} }
 \:.
\end{equation}
 Here $s \sim \sqrt{n_0}$ is the velocity of plasmons for the structures with 2D electrons near a metallic gated.
Although for literal applicability of the presented above hydrodynamic description of
the viscoelastic component one needs $v_F^\eta  \gg v_F$,
 the last  two parameters, apparently, do not differ too much at realistic electron densities   $ n_0 \sim 10^{11}  - 10^{12} $~cm$^{-2}$
 in GaAs quantum wells~\cite{recentest_,Alekseev_Dmitriev} (see discussion in the main text).
 Herewith the plasmon velocity $ s  \sim n_0^{1/2}$
is usually far greater than $v_F$~\cite{Semiconductors}.  Thus  condition~(\ref{n2}) can be typically  fulfilled in such GaAs structures.
It follows from the above conditions that for the frequencies  $\omega  \sim \omega  _c$ in the interval:
\begin{equation}
   \frac{ v_F^\eta  \sqrt { \tau_{tr} } }{ W \sqrt {  \tau_2 } }
   \ll
   \omega
  \ll
  \frac{  s}{W}
  \:,
\end{equation}
the dc current component $I_0$ is predominantly due to the Ohmic flow  $V_0 (y) \approx  const $   in the central (bulk)  region~(\ref{reg_b}),
 whereas the ac flow  component  $V(y,t)$ is still formed mainly by the magnetosonic waves   in the whole sample.

For such flows, only in the near-edge regions:
\begin{equation}
  \label{near_edg}
       W/2 -|y| \lesssim l_G \:,
\end{equation}
both the dc and ac flow components  $V_0(y)$ and  $V_1(y,t)$ are controlled by viscosity  and viscoelasticity. In the bulk
region~ (\ref{reg_b}) the dc component of the flow is homogeneous, $V_0(y) = const $, thus values~(\ref{contrs_in av0}), (\ref{contrs_in av})
and~(\ref{contrs_in av_2}), determining the relaxational and the elastic parts of $V_\omega(y)$, are zero, and the bulk region provides no
hydrodynamic   contributions to photoresistance.

Now we are able to  estimate the magnitude of the   hydrodynamic contribution to photoconductivity  $\Delta \sigma _ \omega $  from
the near-edge  regions~(\ref{near_edg}),  originating only from the memory effect in the interparticle scattering (see Section~2).

For a rough estimate,   we can use  the derived    above formulas for $\Delta  \sigma _ \omega $  of a Poiseuille flow, changing
in  them the sample width $W$   on the characteristic  width of the near-edge regions, $ l_G$~(\ref{n}).   Herewith we neglect
 the contribution  to photoconductivity from the bulk region~(\ref{reg_b}), which is controlled by disorder.  Our estimations, based on
  Eqs.~(\ref{dep_on_T_ome})  for $\Delta \sigma _\omega |_{W=l_G} $  and on the formulas from Refs.~\cite{rev_M} and~\cite{Beltukov_Dyakonov}
for   photoconductivity due to the memory effects in the scattering on disorder,
  show that such assumption can be realistic.

In this way, when the Ohmic contribution in the dc conductivity $\sigma_0$ and  the  hydrodynamic contribution   $\Delta \sigma _\omega $
 of the near edge-layers~(\ref{near_edg}) in the nonlinear photoconductivity dominate,   the averaged photoresistivity in the linear
by ac power regime, $ \Delta \varrho_\omega \sim E_1^2 $, takes the form:
\begin{equation}
 \label{dr}
 \Delta \varrho_\omega = -  \Delta \sigma_\omega / \sigma_0^2
 \:,
\end{equation}
where $\Delta \sigma_\omega$ is calculated by Eqs.~(\ref{UV})-(\ref{dep_on_T_ome}),~(\ref{result_SI}), applied to the near-edge
layers~($W=l_G$), while the averaged dc conductivity  $\sigma_0$ of a  sample  is close to the Ohmic  conductivity  of
the bulk region~(\ref{reg_b}):
 \begin{equation}
\label{rr}
  \sigma  _0 = e^2 n_0 \tau_{imp,tr}/m
\:.
\end{equation}
Note that the last value has no dependence  on magnetic field,  the electron temperature, and the sample width.

Equation (\ref{dr}) leads to an estimate for the dependence of the photoresistivity  on magnetic field~$B \propto \omega_c  $,
ac frequency~$\omega$, and the electron
temperature~$T_e$.

For example, for the inelastic relaxation contribution  $\Delta \sigma_\omega ^{rel} $ due to the extended collisions
of quasiparticle pairs in the interval of the magnetic fields  corresponding to the inequality $ 1/\nu  \ll l_G (\omega_c)\ll 1/\mu$
[and, thus, to the applicability of Eq.~(\ref{dep_on_T_ome})],   equations~(\ref{UV})-(\ref{dep_on_T_ome}),
(\ref{n}), (\ref{dr}), and~(\ref{rr})~yield:
\begin{equation}
\label{Delta_varrho_omega}
\begin{array}{c}
 \displaystyle
   \Delta \varrho_\omega^{\, rel}
      =
   \sin^2\Big( \frac{\pi \omega}{\omega_c} \Big)
    \, \Delta \varrho ^{(1)} _{\omega }
     +
   \sin\Big( \frac{ 2\pi \omega }{ \omega_c } \Big  )
    \, \Delta \varrho ^{ (2) } _{\omega }
  ,
 \end{array}
\end{equation}
where
\begin{equation}
  \label{delta_ro}
   \Delta \varrho ^{(1)} _{\omega}
 = -
  \displaystyle    \frac{ A _O}{\omega^2 \omega^2_c } \,
    \Big(\, 1  - \frac{4 \omega_c^2}{\omega^2 }   \Big)
    \:  {\displaystyle  e}^{  -  A_E   T_e^2 / \omega_c}
\end{equation}
and
\begin{equation}
  \label{delta_ro_2}
   \Delta \varrho ^{(2)} _{\omega}
  =
  \displaystyle    \frac{ 4  \pi A _O}{\omega^3 \omega_c } \,
        \:  {\displaystyle  e}^{  -  A_E    T_e^2 / \omega_c}
   \:.
\end{equation}
Here the  coefficients  $A_O$ and $A_E$ are independent of $\omega$,  $\omega_c$,  $T_e$  and are easily calculated
from Eqs.~(\ref{dep_on_T_ome}), (\ref{n}),  and~(\ref{rr}). It is seen from Eqs.~(\ref{delta_ro}) and (\ref{delta_ro_2})
that at $1 < \omega/\omega _c<4\pi $  the first  amplitude, $ \Delta \varrho ^{(1)} _{\omega}$, is numerically smaller than
the second one, $   \Delta \varrho ^{(2)} _{\omega}$.

We note that there may be many sources of other temperature dependencies of photoresistance as compared with the last
result~(\ref{Delta_varrho_omega})-(\ref{delta_ro_2}). One of the them, proposed in Ref.~\cite{theor_2}, is related with heating
 of the 2D electron system by microwave radiation. Others can be related to, for example, with the   thermoelectric effects
(without or  with the heating of the electron system).  The last effects for the case of hydrodynamic viscoelastic flows can be studied
on base   of the theories of thermoelectric magnetotransport and heat release without memory effects  in space-inhomogeneous electron flows
(see, for example, Refs.~\cite{Semina,Titov})   as well as of the electron energy relaxation in quantum wells due to the interaction
with acoustic phonons  (see, for example, Refs.~\cite{Karpus_1,Yassievich}).

Beside this, the ballistic contributions to the observed photoresponses,  induced by the scattering of quasiparticles  on the edges
 and/or macroscopic defects can reduce the temperature dependencies of all types   (for the consideration of the ballistic contribution
in dc hydrodynamic magnetotransport  see,  for example, Refs.~\cite{Holder,a,c}).
\\
\\
\indent  {\em  7.3.~Case of weak interparticle interaction  }

In this subsection, we discuss the possibility of weakening of the condition~(\ref{cond}) of  a very large strength of the interparticle interaction,
 which allows using simple viscoelastic equations~(\ref{main_eq_gen_S}) to describe high-frequency dynamics of the fluid.

In Refs.~\cite{vis_res_2,Semiconductors} transverse shear-stress waves in the presence of magnetic field
 were studied  within the hydrodynamic Navier-Stokes-like equations (See Sections~2 and~4)
for a strongly non-ideal electron Fermi liquid, in which  $  F_1 \gg 1 $ and $ v_F \ll v_F ^\eta $~(\ref{cond}).
 We remind  that the hydrodynamic description of flows by the variables $n$, $\mathbf{V}$, and $\hat{\Pi}$ implies the predomination
 of the zero, the first, and the second harmonics  by the velocity variable $\mathbf{v}$ in the quasiparticle distribution function
$f(\mathbf{r},\mathbf{v},t)$.  Possibly,  the results of Refs.~\cite{vis_res_2,Semiconductors}  remains qualitatively valid also
for the electron fluid   with moderate  interparticle interaction, when~$F_m \sim 1 $.

Indeed, it is known that shear stress transversal sound can be excited in an electron fluid  in zero magnetic field not only
at the very low densities~$n_0$ corresponding to  $r_s \gg  1 $  and to large  Landau  parameters: $  F_m \gg 1 $,  but also
at moderate densities, when $F_m \sim 1$~\cite{LP}. Here  $r_s$ is the dimensionless parameter characterizing the relative magnitude
of the  interparticle Coulomb interaction energy. Namely, the  transverse waves in 3D and 2D Fermi cases  in zero magnetic field
   can be excited if the Landau parameters~$F_m$ are greater than some critical  values $F_{m,c} ^{2D/3D} \sim 1 $, herewith
   $ F_{m,c} ^{3D} >F_{m,c} ^{2D} $~\cite{LP,Inti_et_al}. In the latter case, an ac flow is described not by
the hydrodynamic variables $n$, $\mathbf{V}$, and $\hat{\Pi}$,  but by  the whole distribution function $f(\mathbf{r},\mathbf{v},t)$
of the Fermi-liquid quasiparticles. A decomposition of such $f(\mathbf{r},\mathbf{v},t)$ by the angle $ \varphi $ of
the quasiparticle  velocity~$\mathbf{v}$  contains many comparable angular harmonics  $e^{i m \varphi }$.

However, we note that the interaction parameter $r_s$ is near unity for many of GaAs quantum wells (for example,
for the structures studied in experiments~\cite{exp_GaAs_ac_1,exp_GaAs_ac_2,recentest_}), thus
the Landau parameter $F_1$ turn out to be not large, possibly, much smaller than unity.
 Therefore  a question of  the microscopic structure of the highly correlated fluid
and the maximally general condition of applicability of the Navier-Stokes-like equations of the form
 of equations~(\ref{main_eq_gen_S}) remains
open.

As it is discussed in the main text, one ``straightforward'' solution of the problem of finding the applicability  conditions
 of the developed theory
to these structures with $r_s \sim 1 $ can  be as follows.

It is possible that, even at $r_s \sim 1 $, for many ranges of sample widths $W$, magnetic fields $B$, and frequencies $\omega$,
 the hydrodynamic equations~(\ref{main_eq_gen_S}) remain the proper effective equations: that is,
      for many relationships between $R_c$,  $\omega$, and at $ W \gg R_c, \, v_F/\omega$
the electron fluid flows
  calculated within the kinetic equation can turn out
  to be qualitatively similar to the flows calculated within the hydrodynamic theory.
  This is to be true when
there is no source of small-scale disturbances with sizes $\Delta x \ll R_c$ and far from the frequencies of the ``geometric resonances'',
 related to commensurability of the sample size~$W$ and the size of trajectories~$2R_c$. The latter effect in a weakly non-ideal electron gas apparently plays the role of resonances due to standing shear waves in a strongly non-ideal electron liquid.

 For example, the validity of the last statement was demonstrated in Ref.~\cite{c}
 for the near-edge regions of a Poiseuille flow of a 2D electron fluid at $\omega_c \tau_ 2 \ll 1 $ in a wide bulk sample with
  the width~$ W \gg l_2 $.
   The  similarity between the kinetic and the hydrodynamic description of such flow is due to the fact
    that harmonics $f_m$ of the distribution function~$f$ can decrease moderately
quickly with $m$, therefore the contributions of the hydrodynamic part of~$f$ ($m \leq 2$) and the ballistic part ($m>2$)
   can have similar shape and give comparable contributions to the flow characteristics.
 As another example, we note that in  theoretical works~\cite{a,Holder} it was demonstrated   that,
in narrow high-quality samples where $W\ll l_2$, the hydrodynamic regime of 2D electron transport
begins to form at $B>B_c$, starting immediately  after the critical magnetic field $B_c$,
 when the cyclotron diameter $2R_c(B_c) $ becomes equal to the sample width~$W$.

We  hypothesize that
there can  be other ``non-trivial'' additional reasons of applicability  of the hydrodynamic
 equations~(\ref{main_eq_gen_S}) for weakly interacting 2D electrons.
  It can be possible  that, even at moderate and small parameter $r_s$ and corresponding  small $F_m$,
the emergence of pair correlations (described in the classical-mechanics approach as the extended collisions, see Fig.~S1),
 can be responsible  for the formation of the highly correlated electron system with the a viscoelastic  and the memory effects,
  being similar by its macroscopic properties to the   Fermi liquid with large $F_m$.

Indeed, first, within semiclassical kinetic approach, a  rigorous microscopic description of the paired electrons (see Fig.~S1)
 cannot  be reduced to
the dynamics of the one-particle  distribution function described by the Boltzmann   equation.
 A possible way of description of highly correlated systems of quasiparticles is based  on the dynamic
equations for the  two-particle correlation function.      For example, such equations were derived in
Ref.~\cite{Gantsevich_Gurevich_Katilius} for non-degenerate Boltzmann  gas in zero magnetic field.
 Similar equations for time evolution
of the two-particle correlation function of quiasiparticles can be derived for an electron Fermi liquid at $r_s \sim 1 $ in
a non-zero magnetic field. Possibly, they may lead macroscopic hydrodynamic  equations of the type of Eq.~(\ref{main_eq_gen_S})
with  the parameter~$v_F^\eta \neq v_F $ and some possible addition terms.

Second,  appearance in a magnetic field of  the paired electrons [see Fig.~S1(a)] may dramatically enhance
the degree of quantum coherence in the electron fluid.  As a result, a reconstruction  of the Fermi-liquid ground state
and quasiparticles of 2D interacting electrons  may appear.
 The reconstruction of the many-particle 2D electron  states in magnetic field within the Hartree-Fock approximation was studied in
 Refs.~\cite{KFS,KFS2}.  Beyond this approximation, different type of correlations in the electron system,
  in particular, related to the quantum analog of the paired electrons in the extended  collisions,  can be possible.
  In this connection, the flows
associated with the proper 2D electron-like quasiparticles  and the ground state,
 both formed by the weakly interacting ``bare electrons'' in classically-strong magnetic fields,
 may be described within some  effective
  Navier-Stokes-like  hydrodynamic equation similar to~Eqs.~(\ref{main_eq_gen_S}).

\newpage

\subsection{ 8.  Additional comparison with experiments }
{\em 8.1.
 Simultaneous observation of peak at doubled cyclotron frequency and  giant negative magnetoresistance }

First of all, we note that a very pronounced  giant negative magnetoresistance and the irregular shape of MIRO, including
the large peak near~$\omega =  2\omega  _c$, are usually observed in the same highest mobility  samples up to the lowest  temperatures
 (see  experimental works~\cite{Smet_1,Ganichev_1,exp_GaAs_ac_1,exp_GaAs_ac_2,recentest_} and Figs.~3,~4
in the main text).

This fact, apparently, point out on the realization of the hydrodynamic transport regime in both the dc and ac components of
 a flow  in 2D electron systems in the same samples.  In particular, the giant negative magnetoresistance, which was initially
predicted  for the flows of an ordinary viscous electron fluid~\cite{je_visc,Gurzhi_Shevchenko}, is also a characteristic property
 of a highly correlated viscous fluid. In this connection, we cite  the experimental data on magnetoresistance and  magneto-photo-resistance
and present the results of our calculations for these two physical values on the neighbour  panels of Figures 2-4.
\\
\\
\indent  {\em 8.2.
Dependencies of oscillating part of photoconductivity and photoresistance on electron temperature}

The  temperature and magnetic field dependencies of the amplitude of  MIRO   observed in  Ref.~\cite{Zudov_T} for a high-mobility GaAs
 quantum very well correspond  to Eq.~(\ref{delta_ro_2}), including the absence of a temperature dependence  in the pre-exponent
factor~$ A_O $.  The formula used   in  Ref.~\cite{Zudov_T} for fitting of the MIRO  amplitude was just the same as Eq.~(\ref{delta_ro_2}).

Herewith  we note that, a substantial temperature dependence of the pre-exponent   amplitude of phonotoresistance oscillations was reported
in experimental works~\cite{S5,W}. This result may not contradict to the correspondence of theoretical result~(\ref{Delta_varrho_omega})-(\ref{delta_ro_2})  to experiment~\cite{Zudov_T} by the following reasons. First, experiment~\cite{W}
was  performed for the GaAs structures with double quantum wells, where all the discussed effects may be quite different.  Second,
 in Ref.~\cite{S5} rather different formulas for fitting the  MIRO curves were used, as compared  with  Eq.~(\ref{Delta_varrho_omega})
 and the formulas used in  Ref.~\cite{Zudov_T},  so the results of Ref.~\cite{S5} cannot be directly compared  with the results
of Ref.~\cite{Zudov_T}  and the predictions of our theory.

In this way,  the results   on the temperature dependence of  MIRO experimentally obtained in Ref.~\cite{Zudov_T} can   be explained by
the model of a mixed Ohmic-hydrodynamic flow, developed in Section~7.2 and  leading
 to  photoresistance~(\ref{Delta_varrho_omega})-(\ref{delta_ro_2}).
\\
\\
\indent  {\em 8.3.
Behavior of MIRO with varying of ac frequency }

With the increase of the frequency $\omega $, the shape of the oscillations of  $ \Delta \varrho_\omega   $  as the function of
the ratio $\omega /\omega _c$ is retained  in our theory [see Eqs.~(\ref{Delta_varrho_omega})-(\ref{delta_ro_2})].  The pre-exponent factor
in the larger contribution $ \Delta \varrho ^{(2)}   _ {\omega} $~(\ref{delta_ro_2}) at a fixed ratio  $\omega /\omega _c$  is rapidly
suppressed with $\omega $ as $\sim  1/ \omega^4 $:
 \begin{equation}
 \frac{    \Delta \varrho ^{(2)} _{\omega} }{   (\omega / \omega_c  ) \: e^{  -  B  \,  T_e^2 / \omega_c}  }
  =
  \displaystyle    \frac{ 4\pi A _O }{\omega^4 } \,
   \:.
\end{equation}

 Beside this, some destructive ``geometric'' effects from inhomogeneities of the sample edges and from possible macroscopic defects
inside the sample on the formation of the standing waves in the ac flow component become more and more substantial  with the increase of $\omega $.
Indeed,  the characteristic wavelength, $l_s  = v_F^\eta/\omega $, which should be grater than the sample inhomogeneities in order
to form  the standing magnetosonic waves,   decreases with $\omega $. So a suppression and a smearing  of the ac flow  component
and of its contribution  to magnetooscillations  of  $ \Delta \varrho_\omega  $  are expected  with the increase of~$\omega$.

This prediction about the frequency dependence of magneto-photo-resistance of the radio frequency
is in  a good agreement   with the measurements  of MIRO in a high-mobility GaAs quantum well  at different
frequencies~$\omega $ in Ref.~\cite{S5}.
\\
\\
\indent
{\em 8.4.
Comparison of  departure time $\tau_q$  with the shear stress relaxation time~$\tau_2$ }

The departure times $\tau_q$, which appear in Eqs.~(\ref{P}) and~(\ref{P1}), controls the amplitude of MIRO via the probability $P=e^{-T/\tau_q}$.
It is of interest to compare the experimental value of~$\tau_q $ with the experimental value of the shear stress relaxation
time~$\tau_2$, which determine  the magnitude and the magnetic field dependence of the dc viscosity coefficients $\eta_{xx,xy}|_{ \omega = 0 }$.

  \begin{figure}[t!]
\centerline{\includegraphics[width=0.99 \linewidth]{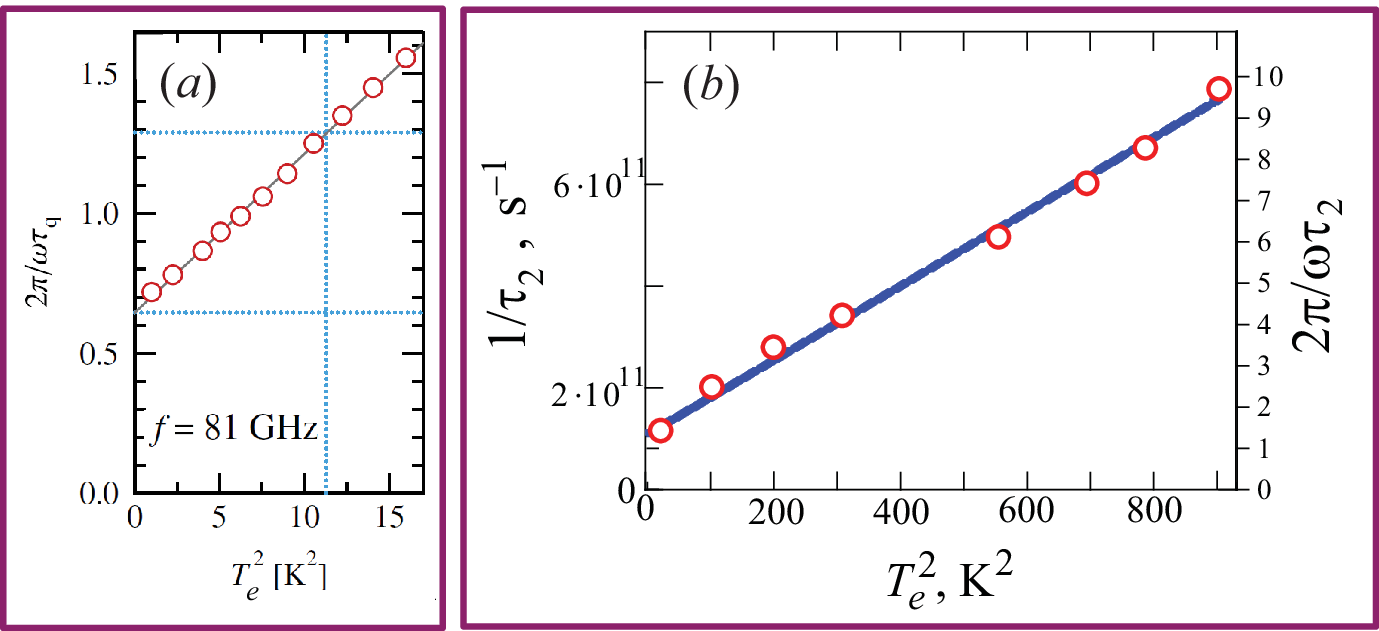}}
\caption{
(a)~Experimental data on the departure scattering time $\tau_q$ as a function of temperature $T_e$, obtained from  the measurements of
the amplitude of MIRO in a high-mobility GaAs quantum well with the 2D electron density $n_0= 2.8 \cdot 10 ^{11}$~cm$^{-2}$
($r _s = 1.05$). Thin solid line is  the quadratic fit $ a_q + b _q T_e^2$. The plot  is taken from Ref.~\cite{Zudov_T}.
(b)~Shear stress relaxation time $\tau_2$  at different $T_e$ obtained from fitting of the giant negative magnetoresistance
measured in Ref.~\cite{exps_neg_4}  in a high-mobility GaAs/AlGaAs quantum well with the same~$ n _0$ as for the sample
on which the data presented in panel~(a) was obtained.  Blue solid line is
 the quadratic fit $ a_2+ b_2 T_e^2$.   Right vertical scale shows the values $1/\tau_2$ in the same units as in panel~($a$) [i.~e.,
 divided by the ac frequency written in panel~($a$)].
  }
\end{figure}

In Fig.~S6 we present  experiential data for  the GaAs quantum  wells in the two  different samples with the close electron densities~$n_0$
 corresponding to  the intermediate interaction parameter  $r_s \approx 1.05$. In panel~($a$) we cite the data on the departure scattering
 time $\tau_q (T_e) $  obtained in Ref.~\cite{Zudov_T}  from the measurements of the temperature  dependence of  the MIRO
  amplitude $\Delta \varrho_\omega$.  This value  was supposed to be proportional to $P=e^{-T/\tau_q}$ with the departure rate  $1/\tau_q$
  contains the contribution  due to scattering on disorder, $1/ \tau_{q} |_{ T_e \to 0 }  =
  1/ \tau_{dis,q} $, and the interparticle contribution being quadratic
by temperature:  $1/ \tau_{ee,q} (T_e) \sim  T_e ^2$.

In panel~($b$) we plot experimental data on the shear stress relaxation time  $\tau_2(T_e) $
extracted by us from the giant negative  magnetoresistance of 2D electron fluid measured in Ref.~\cite{exps_neg_4} in a GaAs quantume well.
To obtain the experimental values of $\tau_2$, we used the fitting of the negative  magnetoresistance  by the procedure proposed in
Refs.~\cite{Alekseev_Dmitriev,je_visc}.   Both the rates  $1/\tau_q ( T_e )  $ and $1/\tau_2 (T_e)  $ are  indeed approximately quadratic
by the electron temperature $T_e$ and have some  residual values in the limit $T _e \to 0 $. It is seen that the temperature-dependent part of $1/\tau_q $
is in several times greater than the one of  $1/ \tau_2 $. This  is consistent with the calculations of these rates, $1/ \tau_{ee,q} (T_e) $
and $1/ \tau_{ee,2} (T_e) $,   within
the kinetic equation exactly accounting the interparticle scattering~\cite{Novikov,Alekseev_Dmitriev}.

Comparable experimental values of $\tau_q $ and $\tau_2 $ and their similar temperature  dependencies
 as well as the inequality $ \tau_q <  \tau_2  $ qualitatively
 evidence in favor of the hydrodynamic nature of the magnetoresistance and magnetoscillations of the photoresistance in
 the samples studied in Refs.~\cite{Zudov_T,exps_neg_4}. However, as it was  mentioned in Section~7.3, for a quantitative description of
the experimental data for samples with the interaction parameter   $r_s  \lesssim  1 $, a rigorous accounting of the correlations
due to formation of the quasiparticle pairs is needed.

We note that the values $2\pi/(\omega \tau_q)$ are of the order of unity [see Fig.~S6(a)].  Herewith after
 interpolation of the dependence $2\pi/(\omega \tau_q (T_e))$  in
 Fig.~S6(a) to higher $T_e$, this value becomes greater than unity.
 From comparison of the plot of $2\pi/(\omega \tau_2(T_e))$  in
 Fig.~S6(b) and the  interpolated dependence  in Figs.~S6(a)  we see that
the inequality $\tau_q\ll \tau_2$ is fulfilled at the temperatures $T_e \gtrsim 8$~K,
 in accordance with the condition~(\ref{ddd}) of applicability of our model of
 the  memory effects in the interparticle scattering.
\\
\\
\indent
{\em 8.5. Role of disorder in narrow samples }

Let us discuss the role of ``residual disorder'' in the relatively narrow samples,  $W \ll l_p$,  in the limit of lowest temperatures~$T_e$,
 when  any interparticle scattering  is switched~off.

At  $T_e \lesssim 1$~K,    the giant negative magnetoresistance as well as photoresistance of  GaAs quantum wells  retain  their
characteristic  properties,  while the scattering times  $\tau_q$ and  $\tau_2$ obtained from fitting of experimental data  attain some nonzero
residual values (see Fig.~S6 and Refs.~\cite{Zudov_T,Alekseev_Dmitriev,je_visc,Gusev_1,recentest_}).  According to the results of
 Refs.~\cite{Alekseev_Dmitriev,je_visc,Gusev_1}, it is believed that at   $ T_e \to 0  $ the electron transport is still hydrodynamic-like,
described by equations~(\ref{main_eq_gen_S}).  In particular,  in sufficiently narrow samples,
 $ W \lesssim l_G,\, l_p $ [ here  $l_G =  \sqrt{(\eta_{xx} \tau_{tr})|_{T_e =0}}$ ], the viscoelastic part of the flow    is to dominate.
 Herewith the relaxation of the momentum flux  $\hat{\Pi}$, including the retarded relaxation, in this regime is determined
by scattering of electron-like quasiparticles  on bulk disorder.

For applicability of hydrodynamics, one needs  that the shear stress relaxation time on disorder, $\tau_{imp,2}$,  is much  shorter than the momentum
relaxation time on the same disorder, $\tau_{imp,tr} \equiv \tau_{imp,1}  $:
\begin{equation}
\label{l}
   \tau_{imp,2} \ll  \tau_{imp,tr}
    \:.
\end{equation}
This inequality follows, for example, from the condition $l_G |_{B=0 , \,\omega =0} \ll v_F \tau_{imp,tr} $, which guarantees
 that the near-edge hydrodynamic hydrodynamic layers can be formed in wide samples, $ W \gg l_G $.
From the analysis of experimental data on the giant negative  magnetoresistance it was concluded that   this inequality, actually,
is often realized in high-mobility GaAs quantum wells with a good margin. (see, for example, Refs.~\cite{je_visc,Alekseev_Dmitriev,Gusev_1}).

According to the consideration of Refs.~\cite{je_visc,Semiconductors,vis_res_1,vis_res_2},  in this  hydrodynamic-like regime, the same
dependence  of the ac viscosity coefficients  $\eta_{xx/xy}(\omega)$  on the frequencies $\omega$, $\omega_c$ and the possibility of the excitation
of shear stress waves is to retain. However the memory effects in the scattering of quasiparticles  on disorder,
being similar in some
aspects to  the ones studied in Refs.~\cite{Beltukov_Dyakonov,Polyakov_et_al_classical_mem}, will affect the retarded relaxation
 of $\hat{\Pi}$ and the nonlinear  components of the flow (instead of the memory effects in the interparticle scattering  studied
 in this work).

In this way, a microscopic justification of the possibility of such hydrodynamic-like regime, including Eq.~(\ref{l})
 (possibly, related to the classical or quantum correlations induced the paired electrons),
 is needed for  deeper understanding of hydrodynamic magnetotransport in realistic high-mobility samples as well as
for a justification of the hydrodynamic explanation of  the discussed above experiments in the limit~$ T_e \to 0 $.
\\
\\
\indent
{\em 8.6. Role of disorder in wide samples }

As we discussed in Introduction of the main text,
 in sufficiently wide samples,  $ W \gg l_p \,  , \:  l_G $, photoresistance and, in particular,
MIRO, apparently,  has a bulk nature and
  are   usually  controlled by the purely  disorder mechanisms, related to the quantum or the  classical  dynamics of independent
2D electrons.
 These mechanisms have been extensively studied
 to date~\cite{theor_1,theor_1_1,theor_2,joint,Polyakov_et_al_classical_mem,Beltukov_Dyakonov,Chepelianskii_Shepelyansky,rev_M}.
 As it was mentioned in Introduction,  they explain many properties of MIRO in many experimentally studied samples. For example,
 the dependence of MIRO of the sign of circular polarization of radiation, corresponding to the magnetotransport
  theories in disordered samples, was clearly observed. In particular, in recent work~\cite{Savchenko} the dependence of photoresistance
on the polarization of radiation and on the geometry of the setup, which determines    the degree of polarization of radiation actually
 entering the sample, was studied in detail.

It is important that in the GaAs high-mobility quantum wells the macroscopic oval defects  are often present~\cite{d1,d2,d1_new}.
 The can appear in the growth process~\cite{d1,d1_new} or made artificially~\cite{d2}.
 Apparently, they do not
destroy  the hydrodynamic regime in some high-quality and sufficiently wide  samples (with a weak ``residual'' disorder between
the oval defects), but strongly modify the shape of the ac and the dc viscous  flows of the electron fluid.

 The oval defects lead
to the  decrease of the actual width of dc viscous 2D electron  flows in the linear by electric field regime~\cite{je_visc}.
 The effective flow width does not remain equal  to  the sample  width~$W$ when such defects are present,
  but becomes much smaller, $ W_{eff} \ll  W $~\cite{je_visc}, and is estimated as some value related to
the mean distance between the defects, $ \sim n_d ^ {-1/2} $ (here $n_d$ is the 2D defect density).
 It is natural to think that the oval defects can also
  significantly affect the high-frequency as well as  the dc nonlinear components
of the electron fluid flow in a similar way,
  reducing the effective width of conductive channels.
\\
\\
\indent
{ \em 8.7. Non-hydrodynamic versus hydrodynamic  mechanisms
 of magnetooscillations and peaks in photoresistance  of 2D~electrons }

Let us discuss recent work~\cite{peak_gr},   in which photoresistance and photovoltage of pure graphene samples in magnetic field were studied,
 together with recent work~\cite{recentest_},   in which magneto- and photoresistance of 2D electrons in ultra-high quality GaAs quantum wells
 samples of different width were examined.

In experiment~\cite{peak_gr} a larger peak near the doubled cyclotron frequency  $  \omega = 2\omega_c $ and smaller peculiarities near
 higher harmonics,  $  \omega = n\omega_c $, $ n \geq 3 $, were observed in photoresistance and photovoltage of high-quality graphene samples.
These peculiarities were explained as the manifestation of the conventional Bernstein modes (``charge cyclotron Bernstein  modes'')
 which are excited  in a 2D electron gas due to  nonlocality of the internal ac electric fields, related to a redistribution of
 the electron density. The last one was induced in experiment~\cite{peak_gr}  by the sharp contacts of the samples. A quantitative theory
of the  Bernstein waves in a 2D electron Fermi gas  in magnetic field  in the samples with inhomogeneous ac electric fields was developed
 in Ref.~\cite{peak_gr} in order to explain the observed photoresponses.

In experiment~\cite{recentest_}    the giant negative magnetoresistance and the peak in photoresistance near $\omega =2 \omega  _c $,
similar to the ones presented at Fig.~4 in the main text, were observed in high-quality GaAs quantum wells in samples with regions
having  different widths. For medium sample widths and at low radiation powers, the irregular shape of MIRO and
the giant $2\omega_c$-peak at were seen very well \{see Fig.~4(b) in Ref.~\cite{recentest_}\}.

We have the following comments on this experiments~\cite{peak_gr} and~\cite{recentest_}   and their explanation.

 From the one hand, although the excitation of the Bernstein modes
in  samples with sharp contacts  leads to strong peaks and peculiarities at $\omega = n\omega_c$ in photoresistance and photovoltage~\cite{peak_gr},
the calculated relative  magnitudes of the peaks at  $n \geq 3 $  (as compared with the $n=2$-peak magnitude) are sufficiently  larger
  than it was observed  in that work for  graphene  samples (compare Fig.~3(f) and~3(g) in Ref.~\cite{peak_gr} and see also Fig.~S3
 in its Supplemental material).

 From the other hand,  in Ref.~\cite{recentest_} it was experimentally shown that, in high-quality
 GaAs quantum wells samples of medium widths, the ratios of magnitudes of the  $   2\omega_c $-peak and the higher harmonic peaks, $n \geq 3 $,
apparently  being analogous to the one observed in Ref.~\cite{peak_gr}, are also very large, similar to these ratios for graphene samples studied
in Ref.~\cite{peak_gr}   (see Fig.~4(b) in Ref.~\cite{recentest_}).

Section~5 of Supplemental material   of Ref.~\cite{peak_gr} presents experimental results on the polarization dependence of
the photovoltage  $\Delta V_{ph} $ of a graphene sample. The measured dependence of $\Delta V_{ph} $  on magnetic field in the region
of the main harmonic of the cyclotron resonance,  $\omega=\omega_c $,  strongly depends on the sign $\pm$ of the circular polarization
of radiation,   while in the region near the doubled harmonic,~$\omega=2\omega_c $,  it is almost independent of the sign~$\pm$.
Herewith  there are no visible peculiarities on the photovoltage curves at  $\omega=3\omega_c $ and higher harmonics in Figure~S5
of this section. Such behaviour of the experimental data agrees very well with the predictions of the hydrodynamic theory
 developed in this work.

In this connection, we also note that the  sharp contacts of the  samples  studied in Ref.~\cite{peak_gr}   can lead not only to
the formation of an inhomogeneous ac electric  field,    leading  to the excitation of the Bernstein modes, but also to the formation
 of the viscoelastic contribution in spatially inhomogeneous  flows of the electron fluid, exhibiting the $2\omega_c$-resonance in the viscosity coefficient~$\eta_{xx}(\omega)$.

Let us discuss  other properties   of the experimental  results of Refs.~\cite{peak_gr,recentest_}.

When  the radiation power becomes very high, the magnitudes of all the  peaks $\omega = n\omega_c$  observed
in Ref.~\cite{recentest_} turn out comparable \{see Fig.~4(c,d) in Ref.~\cite{recentest_}\}. Apparently, this is due to
the destruction of the collective motion of the viscous electron fluid by the strong external electric forces, that allows
 only the magnetoplasmon and Bernstein  waves to be excited in the system of almost independent electrons (as these waves
are related to perturbations of charge density and the corresponding large Coulomb  forces).  As it is seen from  Fig.~4(c,d)
 in Ref.~\cite{recentest_}, such regime is strongly non-linear by the radiation field:   the photoresistance amplitude maximums
becomes comparable or larger than the values of the dc resistance, and   the zero resistance states are formed.

Note also that only the $\omega = 2\omega_c$-peak is observed without any peculiarities at~$\omega = 3\omega_c$
 and~$\omega = 4\omega_c$  in experiment~\cite{recentest_}  for   the narrowest 25~$\mu $m sample section at low ac field power,
when the viscoelastic contribution apparently dominates. This fact is well seen after vertical scaling of Fig.~4(b)
in  Ref.~\cite{recentest_}. Herewith for the wider 50~$\mu $m sample section, in which the magnetoplasmon component  in the ac flow
is to be larger,  both the  $\omega = 2\omega$-peak and a peculiarity at $\omega = 3\omega$ are seen after vertical
scaling.

Apparently, for the proper explanation of the magneto-photo-conductivity of the discussed samples  it is especially important
to verify the truth of the following statement  (or disprove it). The shapes of the magnetic field  dependencies of
 photoresistance~$\varrho_\omega(B)$ and photovoltage~$\Delta V_\omega(B)$ can differ  significantly from the ones  of the absorbtion power
 of radiation~$\mathcal{W}(B)$. Therefore, in order to explain the experimental dependencies  of photoresistance  and
photovoltage, it is necessary to calculate them within some nonlinear model of magnetotransport in a Fermi gas or a Fermi liquid,
 while calculation  of only the single  absorbed power~$\mathcal{W}(B)$  can be not  enough to describe the dependencies
of~$\varrho_\omega(B)$ and~$\Delta V_\omega(B)$. Namely, the last dependencies can differ very much from the shape of~$\mathcal{W}(B)$.
 In our opinion, this conclusion  is to be made from a comparison of these dependencies   obtained in  experiments~\cite{recentest_,peak_gr},
in the non-hydrodynamic theory of Ref.~\cite{peak_gr}, as well as in the hydrodynamic theory of Ref.~\cite{vis_res_2} and
of the present work (see Figs.~3,~4  in the main text and Fig.~S3,~S5 here).  Indeed the shape of the peak can differ from sample to sample,
but, as it was discussed in Section~8.1,  the appearance of a ``giant'' peak in the magneto-photo-resistance of the sample is accompanied
by the appearance of a giant negative magnetoresistance.

From all above facts, apparently, one should make the following conclusions.  On the one hand, for the samples with  relatively
strong  disorder density and/or in the regime of large electron temperatures (or strong ac power heating), excitation of the conventional
Bernstein modes  leads to the appearance of the peaks and peculiarities in the measured photoresistance  at $\omega = n\omega _c$
with the amplitudes,  those  decreases not  too fast with the increase of~$n$.  On the other hand,   observations
 in some high-quality samples  of the giant peak  in photoresistance near  $  \omega = 2\omega_c $,  being especially large
as compared with the peculiarities at higher cyclotron harmonics, $n \geq 3$, together with simultaneous observations  of
the irregular shape of MIRO and the giant  negative temperature-dependent magnetoresistance, evidence  in favor of hydrodynamic regime
of electron transport and formation
of ac flows of  a highly correlated  2D electron fluid.


\begin{thebibliography}{}
\bibitem{f}   
It is noteworthy that for 2D electrons in magnetic fields the non-perturbative  quantum correlation effects,
   namely a reconstruction  of the Fermi-liquid ground state
and quasiparticles,  can take place at any small~$r_s$. These effects   within
the Hartee-Fock approximation
were studied in
 Refs.~\cite{KFS,KFS2}; they also can substantially change the parameters  and  criteria  of applicability
of  the model of this  work.
\\
\end{thebibliography}
\end{document}